\documentclass[a4paper,twocolumn,11pt,accepted=2024-02-02]{quantumarticle}
\pdfoutput=1
\usepackage[utf8]{inputenc}
\usepackage[english]{babel}
\usepackage[T1]{fontenc}
\usepackage{hyperref}
\PassOptionsToPackage{compress}{natbib}
\usepackage[numbers]{natbib}
\usepackage{bbm}
\usepackage{physics}
\usepackage{mathtools}
\usepackage{subcaption}
\usepackage{float}
\DeclareMathOperator{\sgn}{sgn}
\DeclareMathOperator{\erfc}{erfc}

\begin{document}

\title{Continuous-time quantum walks for MAX-CUT are hot}

\author{Robert J.\ Banks}
\email{robert.banks.20@ucl.ac.uk}
\affiliation{London Centre for Nanotechnology, UCL, London WC1H 0AH, UK}
\author{Ehsan Haque}
\affiliation{Newham Collegiate Sixth Form Centre, 326 Barking Rd, London, E6 2BB, UK}
\author{Farah Nazef}
\affiliation{Newham Collegiate Sixth Form Centre, 326 Barking Rd, London, E6 2BB, UK}
\author{Fatima Fethallah}
\affiliation{Newham Collegiate Sixth Form Centre, 326 Barking Rd, London, E6 2BB, UK}
\author{Fatima Ruqaya}
\affiliation{Newham Collegiate Sixth Form Centre, 326 Barking Rd, London, E6 2BB, UK}
\author{Hamza Ahsan}
\affiliation{Newham Collegiate Sixth Form Centre, 326 Barking Rd, London, E6 2BB, UK}
\author{Het Vora}
\affiliation{Newham Collegiate Sixth Form Centre, 326 Barking Rd, London, E6 2BB, UK}
\author{Hibah Tahir}
\affiliation{Newham Collegiate Sixth Form Centre, 326 Barking Rd, London, E6 2BB, UK}
\author{Ibrahim Ahmed}
\affiliation{Newham Collegiate Sixth Form Centre, 326 Barking Rd, London, E6 2BB, UK}
\author{Isaac Hewins}
\affiliation{Newham Collegiate Sixth Form Centre, 326 Barking Rd, London, E6 2BB, UK}
\author{Ishaq Shah}
\affiliation{Newham Collegiate Sixth Form Centre, 326 Barking Rd, London, E6 2BB, UK}
\author{Krish Baranwal}
\affiliation{Newham Collegiate Sixth Form Centre, 326 Barking Rd, London, E6 2BB, UK}
\author{Mannan Arora}
\affiliation{Newham Collegiate Sixth Form Centre, 326 Barking Rd, London, E6 2BB, UK}
\author{Mateen Asad}
\affiliation{Newham Collegiate Sixth Form Centre, 326 Barking Rd, London, E6 2BB, UK}
\author{Mubasshirah Khan}
\affiliation{Newham Collegiate Sixth Form Centre, 326 Barking Rd, London, E6 2BB, UK}
\author{Nabian Hasan}
\affiliation{Newham Collegiate Sixth Form Centre, 326 Barking Rd, London, E6 2BB, UK}
\author{Nuh Azad}
\affiliation{Newham Collegiate Sixth Form Centre, 326 Barking Rd, London, E6 2BB, UK}
\author{Salgai Fedaiee}
\affiliation{Newham Collegiate Sixth Form Centre, 326 Barking Rd, London, E6 2BB, UK}
\author{Shakeel Majeed}
\affiliation{Newham Collegiate Sixth Form Centre, 326 Barking Rd, London, E6 2BB, UK}
\author{Shayam Bhuyan}
\affiliation{Newham Collegiate Sixth Form Centre, 326 Barking Rd, London, E6 2BB, UK}
\author{Tasfia Tarannum}
\affiliation{Newham Collegiate Sixth Form Centre, 326 Barking Rd, London, E6 2BB, UK}
\author{Yahya Ali}
\affiliation{Newham Collegiate Sixth Form Centre, 326 Barking Rd, London, E6 2BB, UK}
\author{Dan E.\ Browne}
\affiliation{Department of Physics and Astronomy, UCL, London WC1E 6BT, UK}
\author{P.\ A.\ Warburton}
\affiliation{London Centre for Nanotechnology, UCL, London WC1H 0AH, UK}
\affiliation{Department of Electronic \& Electrical Engineering, UCL, London WC1E 7JE, UK}
\maketitle

\begin{abstract}
  By exploiting the link between time-independent Hamiltonians and thermalisation, heuristic predictions on the performance of continuous-time quantum walks for MAX-CUT are made. The resulting predictions depend on the number of triangles in the underlying MAX-CUT graph. We extend these results to the time-dependent setting with multi-stage quantum walks and Floquet systems. The approach followed here provides a novel way of understanding the role of unitary dynamics in tackling combinatorial optimisation problems with continuous-time quantum algorithms. 
\end{abstract}

\section{Introduction}
Continuous-time quantum walks (CTQWs) were first proposed for quantum computation by Farhi et al.\ \cite{Far98}. Since then CTQWs have been applied to numerous other problems \cite{Chi09, Wan20,wan19,Wan22, Chi03, Iza17, Lo16}, including unstructured search, and found to recover the same scaling as Grover's algorithm \cite{Chi04}.

More recently CTQWs have been applied to a wider range of combinatorial optimisation problems, such as Sherrington-Kirkpatrick spin glasses and the random energy model \cite{Cal19}. For both of these problems CTQWs were found to have  a better scaling than Grover's algorithm. CTQWs have also been applied to MAX-2-SAT \cite{Mir22,Cal21}. The approach is relatively simple: a time-independent Hamiltonian is applied and the resulting distribution is sampled to try and find the solution or approximate solutions to the optimisation problem. Due to the exponentially growing Hilbert-space, previous work on combinatorial optimisation problems with CTQW has largely been limited to tens of qubits \cite{Cal19,Mir22,Cal21}, with the exception of unstructured search. However, it has been shown CTQWs can always do better than random guessing \cite{Cal21a}.

In this paper we explore the performance of CTQWs on MAX-CUT, though we expect many of the arguments to hold for a wide range of combinatorial optimisation problems. In Sec.\ \ref{sec:QWMC} we introduce CTQWs for MAX-CUT in more depth. By exploiting the effective low dimensionality of the evolution for short run-times we explore the performance of CTQWs in this limit, independent of problem size in Sec.\ \ref{sec:stl}.

Given the intractability of the Schr\"odinger equation for large problem sizes and long times, in Sec \ref{sec:QWTL} we turn to the explanatory power of the Eigenstate Thermalisation Hypothesis (ETH) to provide some insight. First proposed by Deutsch \cite{Deu91} and Srednicki \cite{Sre94}, the ETH conjectures that an isolated system will reach thermal equilibrium. See \cite{Deutsch_2018} for a review. The ETH is a hypothesis not expected to hold for every system. Indeed it is known not to hold for integrable systems \cite{Deu91, Rig09, Ess16, Bre20, Noh21} nor systems that exhibit many-body localisation \cite{Hus13,Nan15, Alt18}. The ETH has been observed both numerically \cite{Rig08,Bir10,San10,Ste13,Kim14,Ste14,Fra15,Kho15,Mon17,Yos18, Jan19} and experimentally \cite{Tro12,Clo16,Kau16,Kuc18,Tan18} for a variety of systems. In Sec.\ \ref{sec:MSQW} we turn the tools developed for a single-stage CTQW to multi-stage quantum walks (MSQWs) \cite{Cal21a}. We also briefly discuss some implications of the ETH for gate-based implementation of CTQWs in Sec.\ \ref{sec:QWF}.

The key take-home of this paper is that by assuming thermalisation we can analytically reason about the behaviour of CTQWs, using a phenomenologically motivated model discussed in Sec.\ \ref{sec:dos}. CTQWs for optimisation being a relatively new and promising method for tackling combinatorial optimisation problems \cite{Cal19,Mir22,Cal21,Cal21a, Chi04}, this approach provides physical insight where exact diagonalisation is prohibitively difficult.

We emphasise all the discussion in this paper relates to CTQWs in the closed-system setting. Throughout this paper we adopt the convention $\hbar=1$ and $k_B=1$. The Pauli matrices are denoted by $X$, $Y$ and $Z$. We also make use of the python packages QuTip \cite{QuT1,QuT2} and NetworkX \cite{Ari08}.

\section{Introducing the Eigenstate Thermalisation Hypothesis}
\label{sec:ETH}

The aim of this section is to provide a very brief introduction to the ETH, focusing on the results pertinent to this paper. This section draws heavily from \cite{Deutsch_2018}. For a more comprehensive review the reader is referred to the original work. This section is split into two parts: the first will discuss what is meant by a steady state, while the second part will introduce the consequences of the ETH.

\subsection{The steady-state}

Typically, classical random walks tend to steady states \cite{Xia20}. Due to the oscillatory behaviour of the Schr\"odinger equation we generally do not expect this to be true for the state-vector of a CTQW. Instead of looking at the state-vector we might look at the expectation value of an observable $\mathcal{O}$, with initial state $\ket{\psi_i}$ and Hamiltonian $H$:

\begin{align}
    \label{eq:obs_av}
    &\langle \mathcal {O}(t)\rangle  =  \bra{\psi_i}e^{iHt}\mathcal{O} e^{-iHt}\ket{\psi_i} \nonumber \\
    &=  \sum_{m,n}e^{i\left(E_m-E_n\right)t}\bra{\psi_i}\ket{E_m}\bra{E_m}\mathcal{O}\ket{E_n}\bra{E_n}\ket{\psi_i} \nonumber \\
    &= \sum_{E_m=E_n} \bra{\psi_i}\ket{E_m}\bra{E_m}\mathcal{O}\ket{E_n}\bra{E_n}\ket{\psi_i}+ \nonumber\\
    &\sum_{E_m \neq E_n}e^{i\left(E_m-E_n\right)t}\bra{\psi_i}\ket{E_m}\bra{E_m}\mathcal{O}\ket{E_n}\bra{E_n}\ket{\psi_i},
\end{align}
where $\ket{E_k}$ denotes an eigenstate of $H$. It follows that if the expectation of $\mathcal{O}$ is going to tend to a certain value, it will tend to its long time-averaged value:

\begin{align}
    \label{eq:obar}
    \langle \bar{\mathcal {O}}\rangle & :=  \lim_{T\rightarrow \infty}\frac{1}{T} \int_0^T \bra{\psi_i}e^{iHt}\mathcal{O} e^{-iHt}\ket{\psi_i} \nonumber \\
    &=\sum_{E_m=E_n} \bra{\psi_i}\ket{E_m}\bra{E_m}\mathcal{O}\ket{E_n}\bra{E_n}\ket{\psi_i}.
\end{align}
From Eq.\ \ref{eq:obs_av} it is clear that for the system to reach Eq. \ref{eq:obar}  there must be sufficient dephasing on sufficiently short time scales, such that the second sum in Eq.\ \ref{eq:obs_av} becomes negligible compared to the first sum. In practice for any finite system we expect there to be fluctuations around the steady state value of $\mathcal {O}$ (i.e., around $\langle \bar{\mathcal {O}}\rangle$). In short, a system is in its steady state if $\langle \mathcal {O}(t)\rangle = \langle \bar{\mathcal {O}}\rangle$ for all $t$ after a certain value, up to negligible fluctuations in $\langle \mathcal {O}(t)\rangle$. For discussion about the time taken to reach equilibrium, please see \cite{Wilming_2018}.

Though Eq.\ \ref{eq:obar} is useful, it requires complete knowledge of the spectrum of $H$, which quickly becomes intractable for large problem sizes. It is also perhaps not clear from this expression which observables correspond to steady states. The next section begins to address some of these questions.

\subsection{The Eigenstate Thermalisation Hypothesis}

The ETH is an attempt to explain how thermalisation occurs in a closed-system. The details can be found in \cite{Deutsch_2018}. A consequence is that for a large class of observables the steady state value is equal to averaging over the microcanonical ensemble in the thermodynamic limit. The fluctuations from this value are exponentially suppressed with the degrees of freedom in the system \cite{Deutsch_2018}. In certain cases a single eigenstate is sufficient to carry out the microcanonical averaging \cite{Sre94}. 

As mentioned prior, not all Hamiltonians exhibit ETH and not all observables thermalise \cite{Gar18}. Typically, the Hamiltonian is assumed to be highly non-degenerate \cite{Sre94,Deu91,Rig08} and to provide a locally conserved quantity \cite{Ess16}. The observables that exhibit thermalisation are typically considered to be local or few-body operators \cite{Sre94,Rig08}. For example, consider the highly non-local operator for a non-trivial system \cite{Ess16}:
\begin{equation}
    \mathcal{O}_{jk}=-i \left(\ket{E_j}\bra{E_k}-\ket{E_k}\bra{E_j}\right),
\end{equation}
where $E_j\neq E_k$. This will clearly not approach a steady state. 

In summary, the ETH tells us that for large systems the steady state, exhibited by local observables, can be well approximated by the microcanonical ensemble. Though there are known exceptions, this behaviour is thought to be very common \cite{Nan15,Rei15}. Note that it is possible for a system to approach a steady state that is not a thermal state \cite{Ess16}.

For ease of computation, throughout this paper we make use of the canonical ensemble, which is equivalent to the microcanonical ensemble in the thermodynamic limit \cite{Rig08,Kim14}. That is to say, throughout this paper we assume the pure quantum state is locally well approximated by a thermal Gibbs state:
\begin{equation}
    \rho_{\beta}(H)=\frac{e^{-\beta H}}{\Tr e^{-\beta H}},
\end{equation}
with temperature $1/\beta$. 

A CTQW is a time-independent Hamiltonian and in most settings unlikely to be integrable. We might therefore expect it to exhibit thermalisation for local observables. We provide numerical evidence for this in Sec.\ \ref{sec:therm_check}. By exploiting the observation that the system is well approximated by a thermal state we make predictions on the performance of CTQWs.

\section{Quantum walks for MAX-CUT}
\label{sec:QWMC}

\subsection{Set-up of the CTQW}
In this section we introduce MAX-CUT for CTQWs. Given a graph $G=(V,E)$, a cut separates the nodes into two disjoint sets. The weight of the cut is given by the number of edges which connect the two sets. The aim of MAX-CUT is to find the maximum cut.

This can be encoded as finding the ground state of an Ising Hamiltonian,
\begin{equation}
    H_p=\sum_{(i,j)\in E} Z_iZ_j.
\end{equation}

In this paper we focus on graphs with fixed degrees and binomial graphs. To generate the binomial graphs each edge is selected with a certain probability, throughout this paper this probability is taken to be $2/3$. Binomial graphs are also known as Erd\H os R\'enyi random graphs.

As the corresponding driver Hamiltonian we take the transverse-field Hamiltonian:
\begin{equation}
    H_d=-\sum_{i=1}^n X_i,
\end{equation}
where $n=\abs{V}$. This corresponds to a quantum walk on the Boolean hypercube.

The CTQW Hamiltonian is given by:
\begin{equation}
    H_{QW}=H_d+\gamma H_p,
\end{equation}
where $\gamma>0$ is a parameter that needs to be set\footnote{Normally $\gamma$ is appended to $H_d$ not $H_p$. Here we take the opposite approach so changing $\gamma$ does not change $\langle H \rangle$.}. The system is prepared in the ground-state of $H_d$, here denoted by $\ket{+}$, and evolved under $H_{QW}$. The resulting state, after a time $t$, is
\begin{equation}
    \label{eq:CTQW}
    \ket{\psi(t)}=e^{-iH_{QW}t}\ket{+}.
\end{equation}

To assess the performance of Eq.\ \ref{eq:CTQW} we look at $\langle H_p \rangle$. If $\langle H_p \rangle=E_0^{(p)}$, where $E_0^{(p)}$ is the ground state energy of $H_p$, then the CTQW has found the ground-state. Random guessing would correspond to $\langle H_p \rangle=0$. The value of $\langle H_p \rangle$ can be related to the length of cut by:
\begin{multline}
     \langle H_p \rangle=-2\times\left(\text{Average cut value from the CTQW}\right.  \\
     - \left. \text{Average cut value from random guessing}\right).
\end{multline}
In short, the more negative $\langle H_p \rangle$ is, the better the performance of the CTQW. 

Energy is conserved during the evolution. Since the evolution starts in the ground-state of $\langle H_d \rangle$, $\langle H_d \rangle$ cannot decrease initially. Assuming non-trivial dynamics, this means that $\langle H_p \rangle$ must decrease to conserve energy. Hence, why Eq.\ \ref{eq:CTQW} will do better than random guessing \cite{Cal21a}. 

A second consequence of the conservation of energy is $\langle H_{QW} \rangle=-n=E_0^{(d)}$, where $E_0^{(d)}$ is the ground-state energy of $H_d$. This means that the evolution cannot completely occupy the ground-state of the CTQW Hamiltonian. This is in stark contrast to a number of other popular approaches to tackling optimisation problems with NISQ algorithms \cite{Alb18,Hau20,Zho20}.

Finally, we note the Hamiltonian in Eq.\ \ref{eq:CTQW} commutes with the spin-flip operator:
\begin{equation}
    G=\prod_{k=1}^n X_k.
\end{equation}

Since, $\ket{+}$ is an eigenstate of $G$ with eigenvalue plus one, the evolution is restricted to the plus one eigenspace of $G$. This can be utilised in simulation to reduce the size of the Hilbert space by half. It also means that the CTQWs cannot reach a separable computational basis state, resulting in the ideal final state having non-zero entanglement. We also touch upon this in Appendix \ref{app:sfs} in the context of the ETH.  In the next section we discuss how to optimise $\gamma$ to give the best possible performance.

\subsection{Optimising the free parameter}
\label{sec:gamma}

The CTQW contains a free-parameter $\gamma$. Heuristically this controls the amount of dynamics present in the evolution. If $\gamma$ is too large, we have approximate evolution under $H_p$, leaving the measurement statistics unchanged. Correspondingly $\langle H_p \rangle \approx 0$. If $\gamma$ is too small we have approximate evolution under $H_d$, resulting in trivial dynamics and $\langle H_p \rangle \approx 0$. Clearly for a CTQW to be successful $\gamma$ must be chosen somewhere between these two limits.

In this section we attempt to optimise the free parameter $\gamma$ to give the best time-averaged value of $\langle H_p \rangle$. More explicitly the time-averaged value is given by:
\begin{equation}
    \langle \bar{H_p} \rangle =\lim_{T \rightarrow \infty} \frac{1}{T} \int_0^T \langle H_p(t) \rangle \, \dd t.
\end{equation}
Numerically, to evaluate $\langle \bar{H_p} \rangle$ we calculate: 
\begin{equation}
    \label{eq:hpbar}
    \langle \bar{H_p} \rangle =\sum_{E_m=E_n} \bra{+}\ket{E_n}\bra{E_m}\ket{+} \bra{E_n}H_p\ket{E_m},
\end{equation}
where $\ket{E_k}$ is an eigenstate of $H_{QW}$ with corresponding eigenvalue $E_k$. The intuition behind these equations can be found in Sec.\ \ref{sec:ETH}.

\begin{figure}
    \centering
    \includegraphics[width=0.48\textwidth]{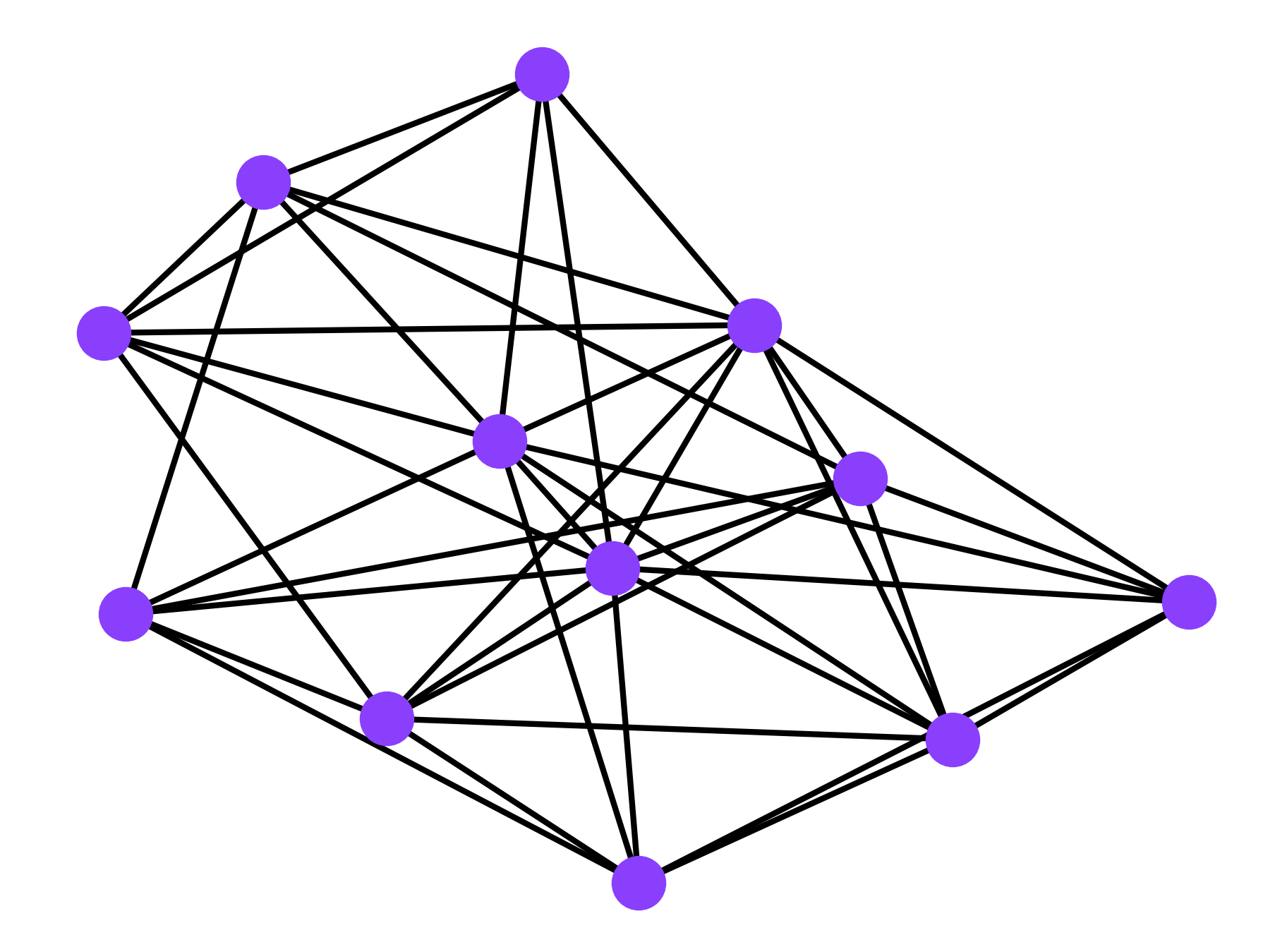}
    \caption{The single 12-qubit binomial graph instance discussed in Sec.\ \ref{sec:gamma}.}
    \label{fig:graph_first}
\end{figure}

In the rest of this section we find by brute force optimisation the optimal choice of $\gamma$ for different graph choices. We also make comparisons to a reasonable heuristic choice of $\gamma$. To begin with we focus on a specific example of a 12-qubit binomial graph shown in Fig.\ \ref{fig:graph_first}, before looking at statistics gathered from multiple graph instances.

\begin{figure}
    \centering
    \includegraphics[width=0.48\textwidth]{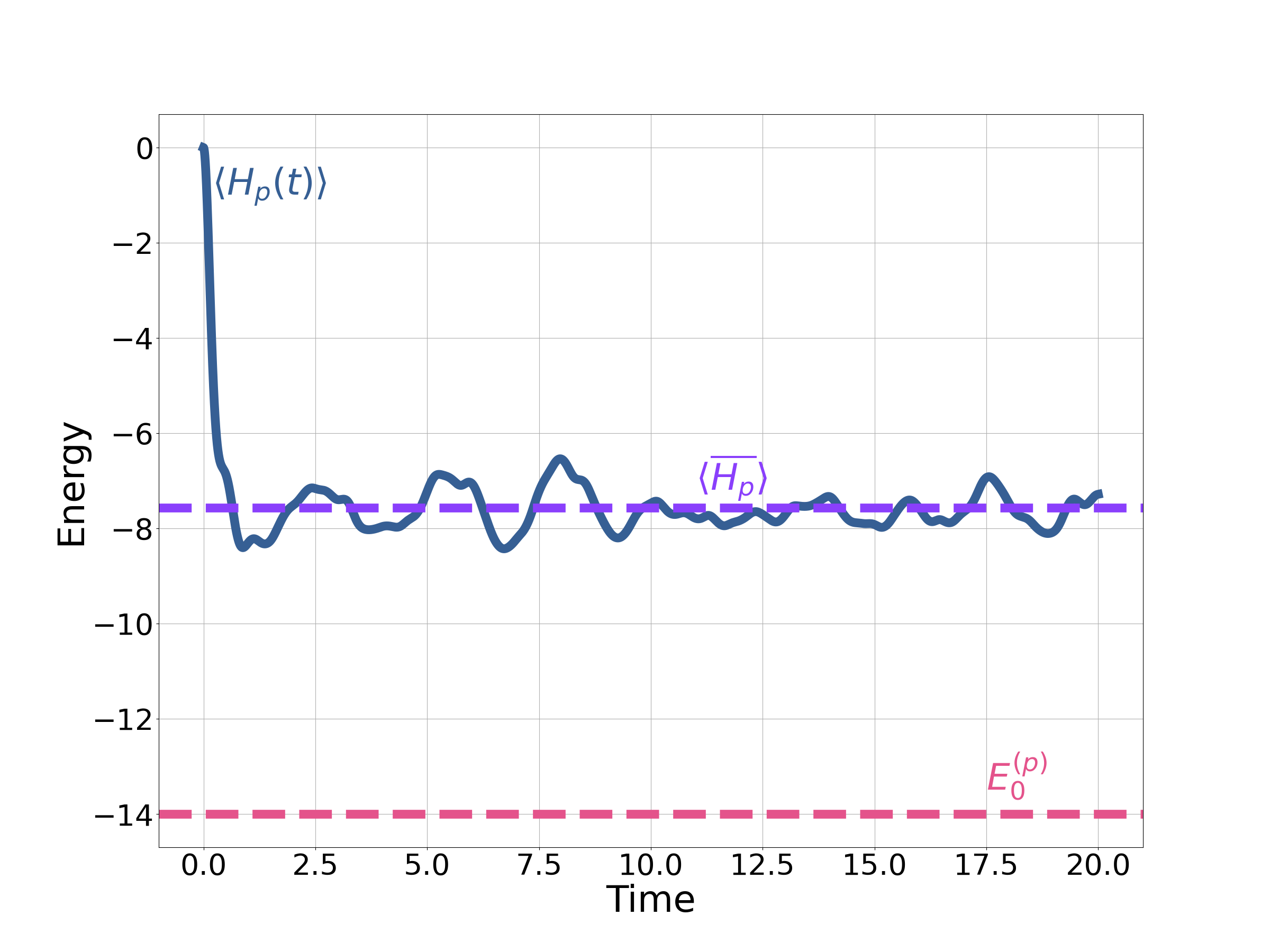}
    \caption{A time domain plot of $\langle H_p \rangle$ (the blue line) for a randomly generated 12-qubit instance shown in Fig.\ \ref{fig:graph_first}. The dashed pink line shows the minimum energy of $H_p$. The dashed purple line is the time-averaged value, according to Eq.\ $\ref{eq:hpbar}$.}
    \label{fig:cinst}
\end{figure}

\begin{figure}
    \centering
    \includegraphics[width=0.48\textwidth]{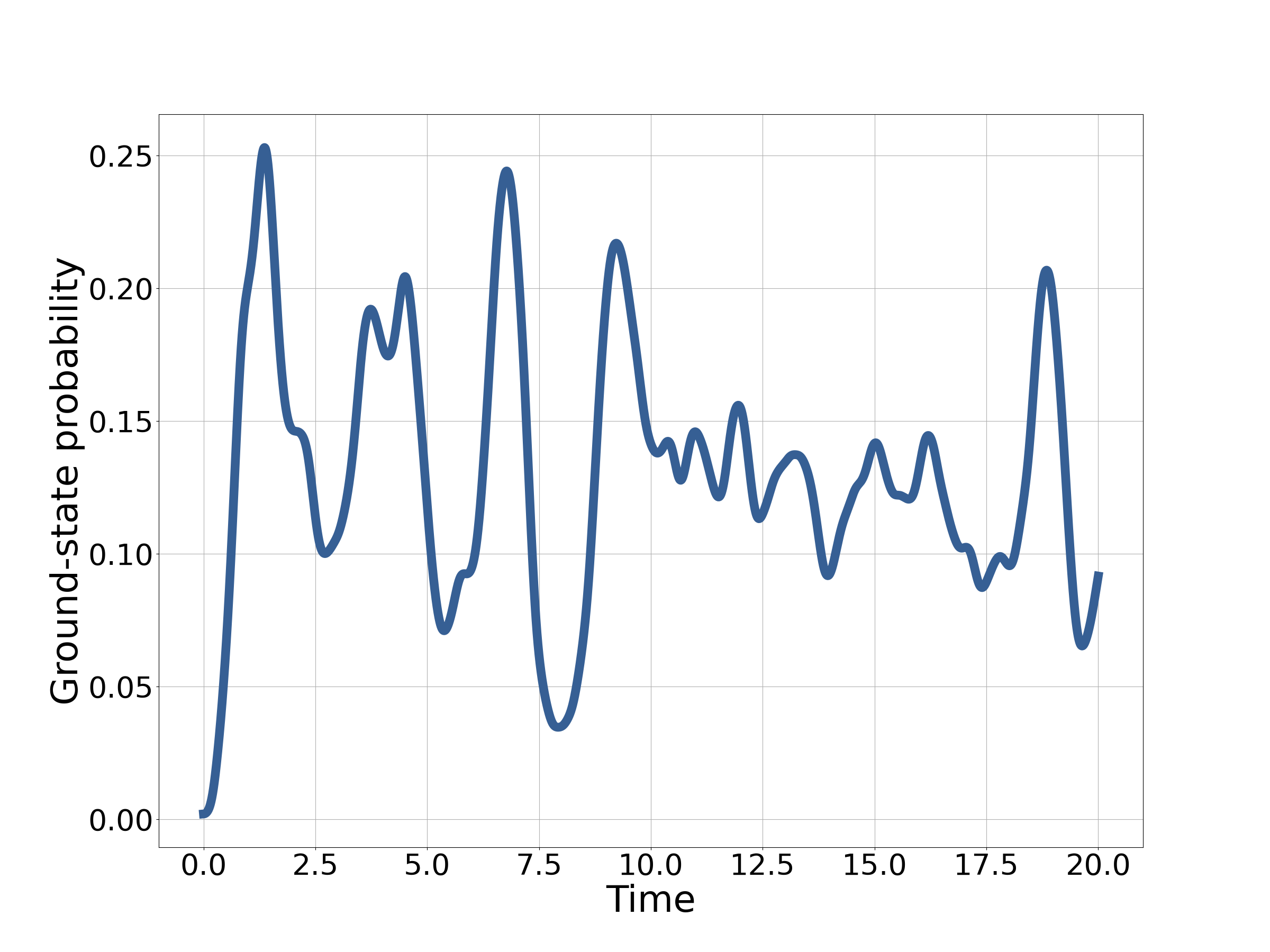}
    \caption{A time domain plot of the ground-state probability for the randomly generated 12-qubit instance shown in Fig.\ \ref{fig:graph_first}.}
    \label{fig:gspinst}
\end{figure}

Starting with a specific choice of $\gamma$ for the graph instance shown in Fig.\ \ref{fig:graph_first}, Fig.\ \ref{fig:cinst} shows a CTQW  with $\gamma=1$. The blue line in Fig.\ \ref{fig:cinst} shows the average value of $H_p$ from the Schr\"odinger equation, or more explicitly:
\begin{equation}
    \langle H_p(t) \rangle=\bra{+}e^{i H_{QW} t} H_{p} e^{-i H_{QW} t} \ket{+}.
\end{equation}
The dashed pink line shows the ground state energy of $H_p$ (i.e. $E_0^{(p)}$). For the majority of the evolution shown $\langle H_p (t) \rangle$ is fluctuating around the steady-state value, $\langle \bar{H_p} \rangle$ (the dashed purple line). We can compare this to the ground state probability, shown in Fig.\ \ref{fig:gspinst}. The ground state probability shows significant oscillations, illustrating why we avoid optimising over this quantity.

\begin{figure}
    \centering
    \includegraphics[width=0.48\textwidth]{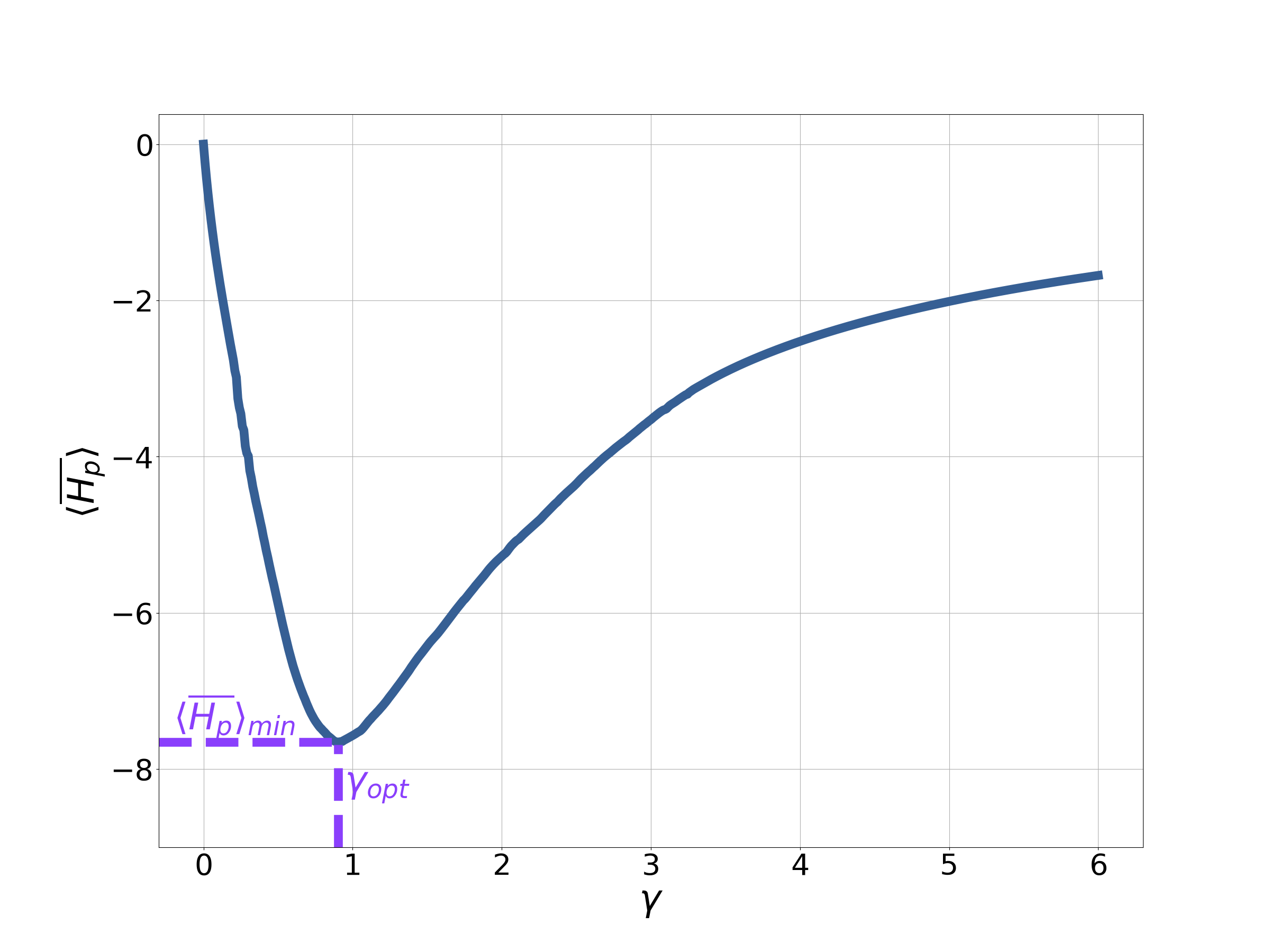}
    \caption{The time-averaged value $\langle H_p \rangle$ as $\gamma$ is changed for the 12 qubit graph shown in Fig.\ \ref{fig:graph_first}. The dashed purple line shows the location of $\gamma_{opt}$ and $\langle \bar{H_p} \rangle_{min}$.}
    \label{fig:gsweep_inst}
\end{figure}

Fig.\ \ref{fig:gsweep_inst} shows how $\langle \bar{H_p} \rangle$ varies with $\gamma$ for the same problem instance as Fig.\ \ref{fig:cinst}. The optimal $\gamma$ (i.e. $\gamma_{opt}$) occurs at the minimum value of $\langle \bar{H_p} \rangle$ (i.e., $\langle \bar{H_p} \rangle_{min}$). For the problem instance considered in Fig.\ \ref{fig:gsweep_inst}, $\gamma_{opt}\approx 0.90$ and $\langle \bar{H_p} \rangle_{min} \approx-7.65$.

In previous works, the free parameter in a CTQW has been heuristically chosen in an attempt to maximise dynamics. One approach to do this has been to match the energy scales of the driver and problem Hamiltonian \cite{Cal19,Mir22}. For a MAX-CUT problem, this could be interpreted as $\Tr\gamma^2 H_p^2=\Tr H_d^2$ such that:
\begin{equation}
    \label{eq:heur_gam}
    \gamma_{heur}^2\kappa_2=n,
\end{equation}
where $\kappa_2=\abs{E}$ is the number of edges in the graph.

Focusing first on the performance of the binomial graph shown in Fig.\ \ref{fig:gsweep_inst}, evaluating Eq.\ \ref{eq:heur_gam} gives $\gamma_{heur}=0.5$,  corresponding to $\langle \bar{H_p} \rangle_{min} \approx -5.88$. This is a significant reduction in the performance of the CTQW.

\begin{figure}
    \centering
    \includegraphics[width=0.48\textwidth]{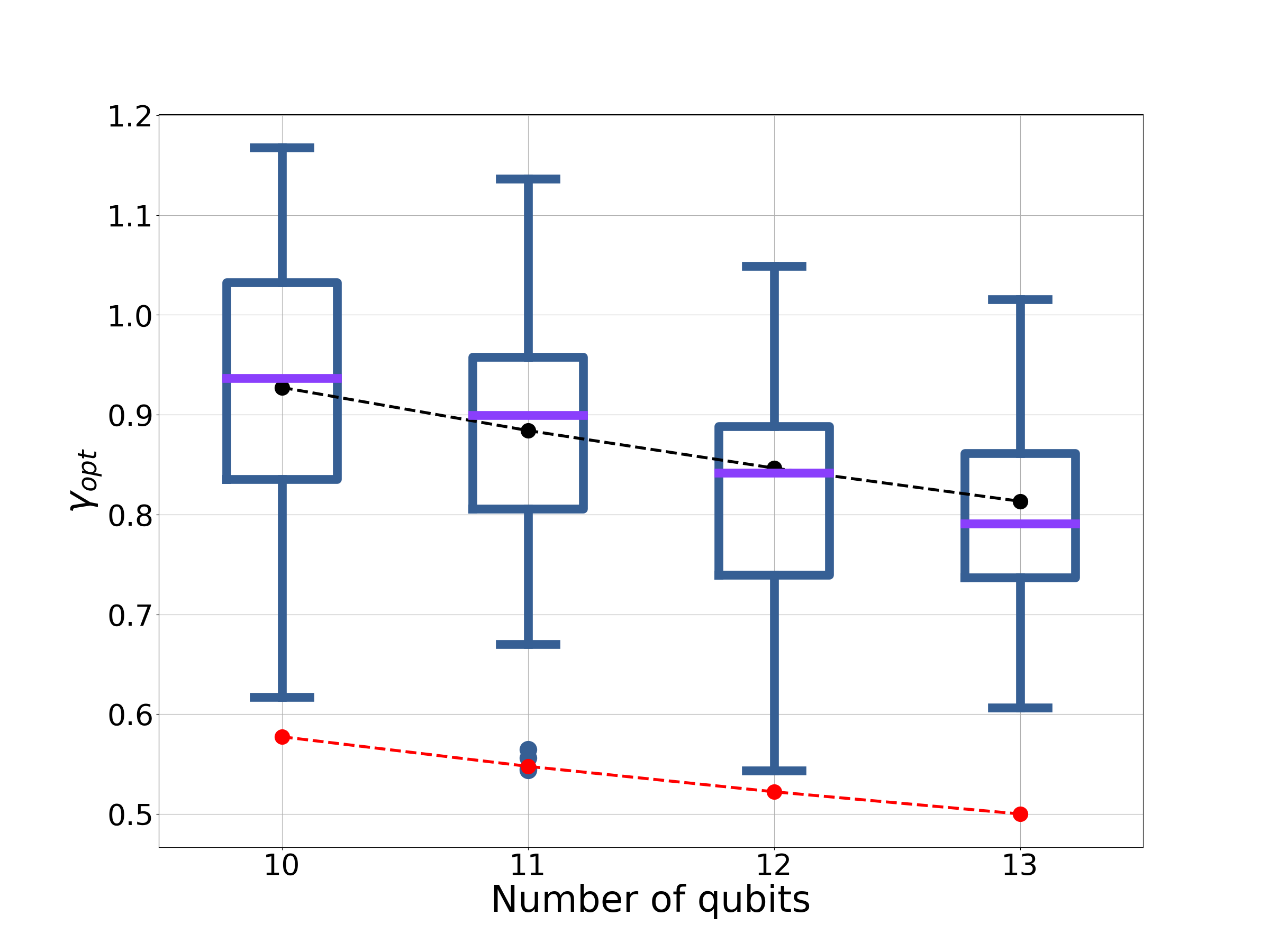}
    \caption{The optimal value of $\gamma$ for one hundred problem instances for each problem size for binomial graphs. The  red dashed line reflects the heuristic from Eq.\ \ref{eq:heur_gam}, explicitly $\gamma=\sqrt{3/(n-1)}$. The dashed black line corresponds to fitting the curve $\gamma_{opt}=an^{-1/2}$ to the medians of the data, yielding $a\approx 2.93$.}
    \label{fig:gamma_rand}
\end{figure}

Fig.\ \ref{fig:gamma_rand} shows the optimal $\gamma$ for one hundred randomly generated problem instances for binomial graphs with problem sizes ranging between ten and thirteen qubits. The optimal $\gamma$ is seen to decrease with the problem size. The range of optimal values of $\gamma$ is on the order of 0.1.  The dashed red line represents the heuristic in Eq.\ \ref{eq:heur_gam} assuming the number of edges is $n(n-1)/3$. Clearly Eq.\ \ref{eq:heur_gam} is significantly underestimating the optimal $\gamma$. 

As mentioned in the introduction of this section, it is expected that the optimal $\gamma$ will balance $H_p$ and $H_d$, hence why it is reasonable to assume  $\gamma \propto n^{-1/2}$. This is the same functional dependence on $n$ as Eq.\ \ref{eq:heur_gam}. The dashed black line in Fig.\ \ref{fig:gamma_rand} shows $\gamma \propto n^{-1/2}$ fitted to the available data. The curve is not inconsistent with the data.

\begin{figure}
    \centering
    \includegraphics[width=0.48\textwidth]{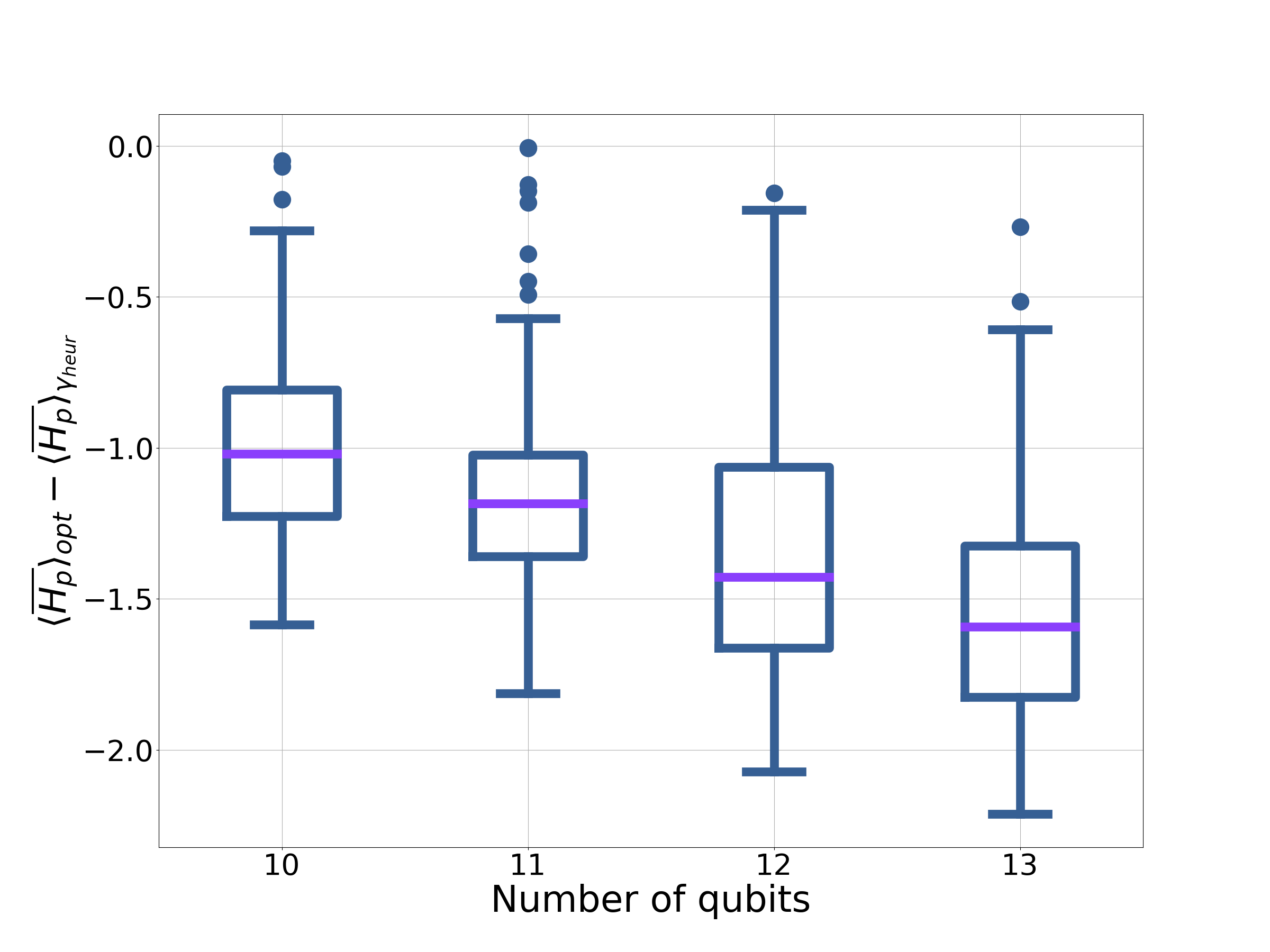}
    \caption{The difference in $\langle \bar{H_p} \rangle$ for $\gamma_{opt}$ and $\gamma_{heur}$ for the same binomial graphs as Fig.\ \ref{fig:gamma_rand}.}
    \label{fig:hpdiffRandEst}
\end{figure}

Fig.\ \ref{fig:hpdiffRandEst} shows how $\langle \bar{H_p} \rangle$ changes between $\gamma_{opt}$ and $\gamma_{heur}$ for the same problem instances as Fig.\ \ref{fig:gamma_rand}. The difference in performance is on the order of one and appears to increasing with the problem size. There is clearly scope for improvement on the heuristic choice of $\gamma$. 

\begin{figure}
    \centering
    \includegraphics[width=0.48\textwidth]{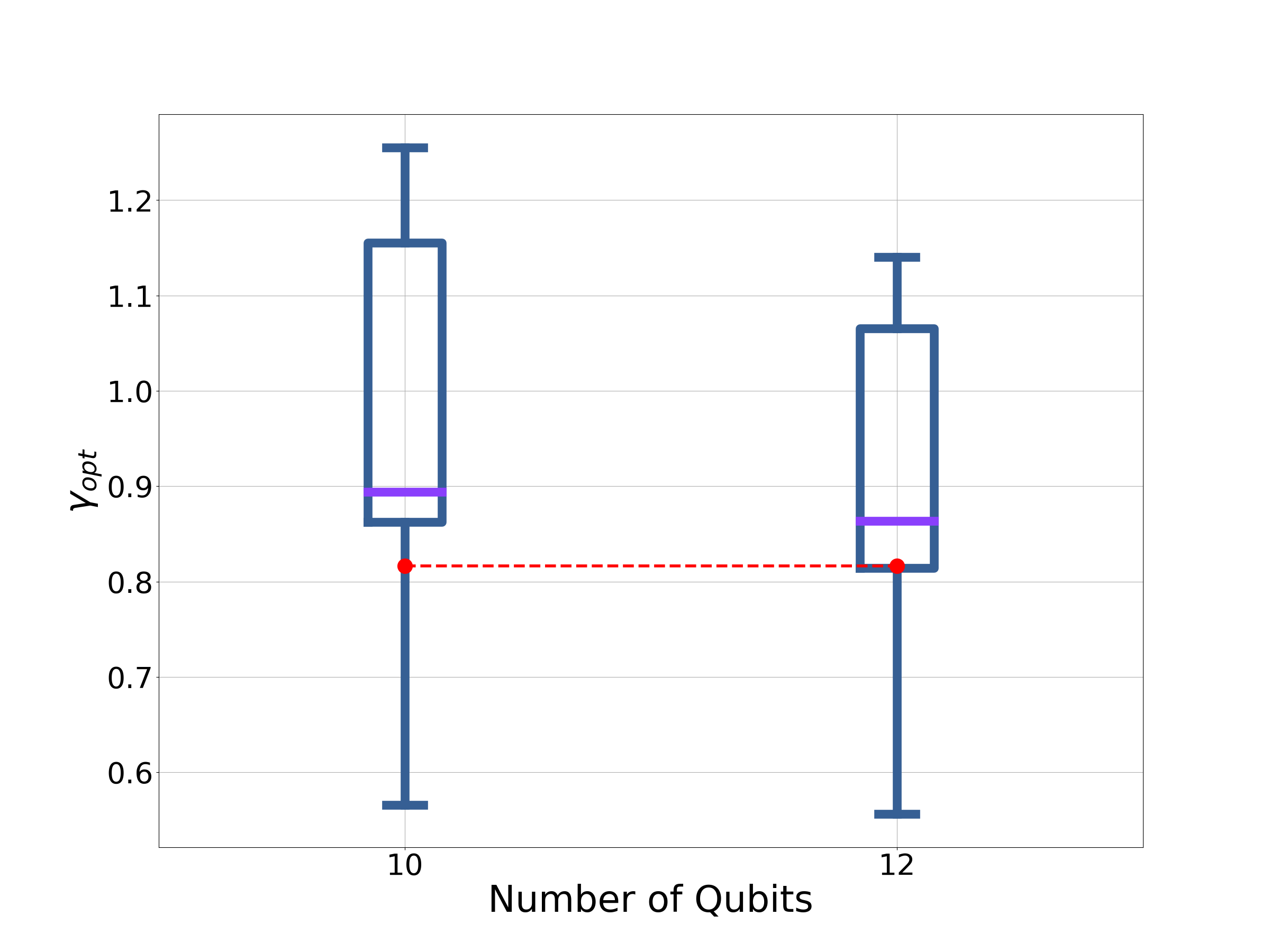}
    \caption{The optimal value of $\gamma$ for three-regular graphs. In ascending problem size, there are: 16, 45 problem instances. The red dashed line is the heuristic from Eq.\ \ref{eq:heur_gam}, explicitly $\gamma=\sqrt{2/3}$.}
    \label{fig:gamma_reg}
\end{figure}

\begin{figure}
    \centering
    \includegraphics[width=0.48\textwidth]{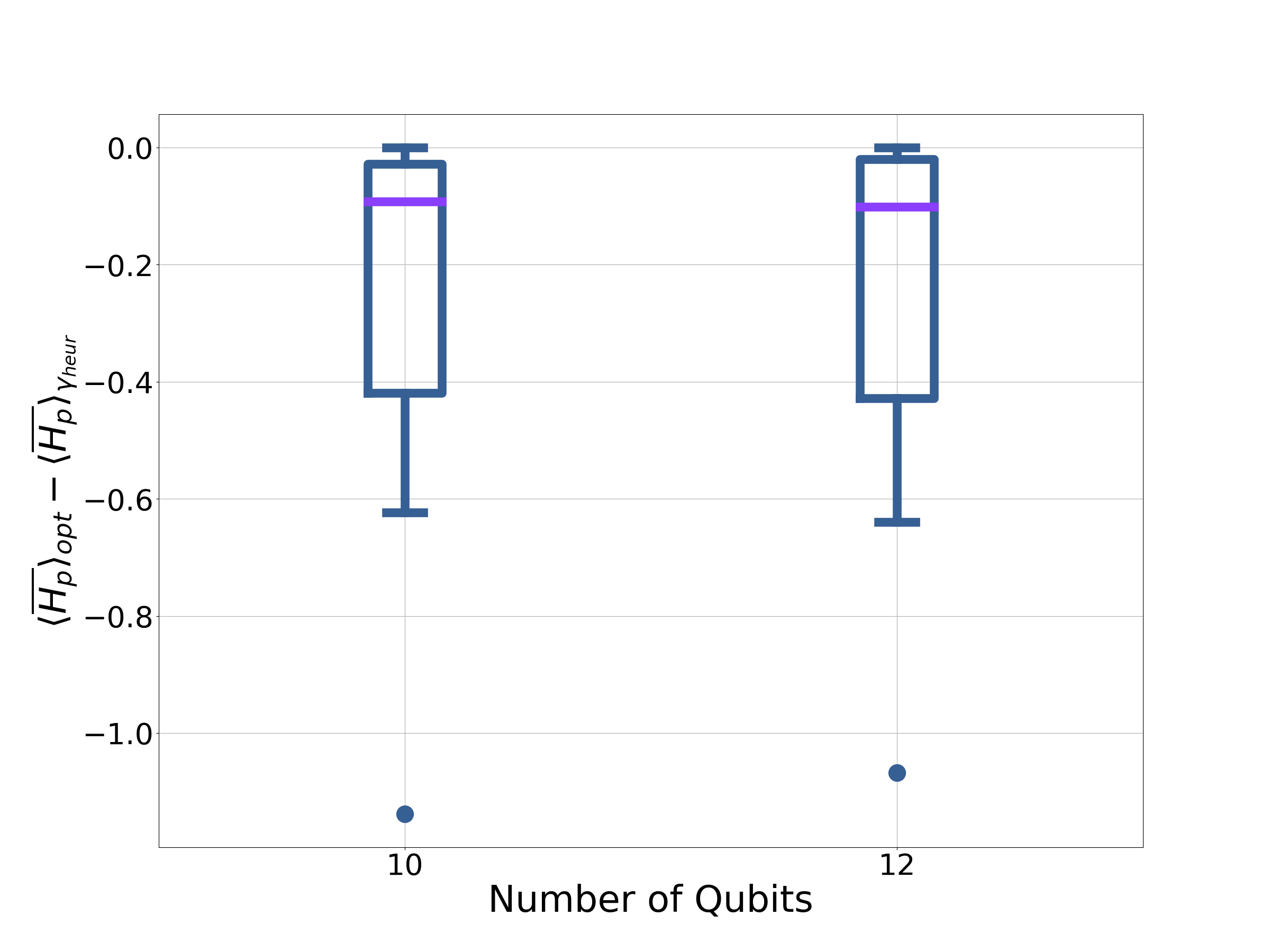}
    \caption{The difference in $\langle \bar{H_p} \rangle$ for $\gamma_{opt}$ and $\gamma_{heur}$ for the same three-regular graphs as Fig.\ \ref{fig:gamma_reg}.}
    \label{fig:hpdiffRegEst}
\end{figure}

We now turn to three-regular graphs, Fig.\ \ref{fig:gamma_reg} shows $\gamma_{opt}$ for different problem sizes. For regular graphs the number of edges for a given problem size is fixed and scales with $n$. Hence, $\gamma_{heur}$ evaluates to be the same for all instances of a $d$-regular graph. This is the red dashed line in the figure. For three-regular graphs $\gamma_{heur}$ also appears to be generally underestimating $\gamma_{opt}$ but less so than in the binomial case. This is reflected in the change in performance, as shown in Fig.\ \ref{fig:hpdiffRegEst}.

In this section we have focused on finding the optimal $\gamma$. The closer $\gamma$ is to the optimal value, the better the approximation ratio is. We have also explored the use of a heuristic based around the mantra of maximising dynamics. The heuristic choice of $\gamma$ has elucidated how the optimal choice of $\gamma$ might scale with $n$. This is particularly pertinent to the discussion in Sec.\ \ref{sec:dos}. In Sec.\ \ref{sec:hpPred} we show how the heuristic choice of $\gamma$ (i.e. Eq. \ref{eq:heur_gam}) can be recovered assuming thermalisation and how it might be improved upon.

An alternative, more involved, method  for finding heuristic values of $\gamma$ also based on the notion of maximising dynamics can be found in  \cite{Cal21a}. 

For the majority of this paper we are interested in the steady-state behaviour of the CTQW. However, there may be some advantage to running the CTQW on a very short time scale, with the aim of having a short time to solution. In the next section, we analytically investigate the very short time behaviour of CTQWs for MAX-CUT.

\section{The very-short-time limit}
\label{sec:stl}

A CTQW consists of a relatively simple, time-independent Hamiltonian. Following \cite{Lab17,Gna22} in this section we evaluate the torsion of the quantum evolution. The torsion is a local measure of how much the evolution deviates from a two-dimensional space. By working in this sub-space we can make short-time predictions for $\langle H_p \rangle$.

The torsion depends only on expectations of powers of $H_{QW}$, hence are constants over the duration of the evolution for a time-independent Hamiltonian. For the details of the calculation see Appendix \ref{app:tor}. The torsion of the evolution is given by:

\begin{equation}
    \mathcal{T}=2\gamma^4\left[\kappa_2\left(\kappa_2-1\right)-\frac{18 \kappa_3^2}{\kappa_2} + 12 \kappa_4 \right],
\end{equation}
where $\kappa_2$ is the number of edges in the graph, $\kappa_3$ the numbers of triangles, and $\kappa_4$ the number of squares.

We can see that the torsion is reduced by the number of triangles in the graph. This suggests that the CTQW explores the solution space slower for graphs with a large number of triangles. The error from leaving a two dimensional space in a time $\Delta t$ can be approximated by \cite{Lab17}:
\begin{equation}
    \varepsilon_{2D}=\mathcal{T}\Delta t^4.
\end{equation}
So for times much less than $\mathcal{T}^{-\frac{1}{4}}$ we can approximate the evolution of CTQWs by a two dimensional system. The details can be found in Appendix \ref{app:tor}. Under these assumptions, with $\gamma>0$,  the value of $\langle H_p \rangle$ can be calculated as:
\begin{equation}
    \label{eq:cstl}
    \langle H_p \rangle = -\frac{4 \kappa_2 \gamma}{\omega^2}\sin^2 \omega t,
\end{equation}

where
\begin{equation}
    \omega^2=\gamma^2 \kappa_2 +\frac{\left(2 \kappa_2+3\gamma \kappa_3\right)^2}{\kappa_2^2}.
\end{equation}

As we can see, for very short times, $\langle H_p \rangle$ depends very little on the properties of the underlying MAX-CUT graph, depending primarily on the number of edges in the graph. The frequency term $\omega$ does depend on the number of triangles but for large problem sizes it is reasonable to expect $\omega^2  \approx \gamma^2 \kappa_2$. Physically, it is reasonable, that at short times the CTQW only sees triangles and edges. Longer evolution would result in the approach seeing larger loops in the graph. 

From a computational point of view this suggests that any useful short-time sampling must not be on a timescale much smaller than $\mathcal{T}^{-\frac{1}{4}}$ since you would like the CTQW to see the whole graph. A rough estimate for a CTQW to see the whole structure of the graph might be $l \mathcal{T}^{-\frac{1}{4}}$, where $l$ is the length of the largest loop in the problem graph. This is bounded from above by $n\mathcal{T}^{-\frac{1}{4}}$. 
\begin{figure}
    \centering
    \includegraphics[width=0.48\textwidth]{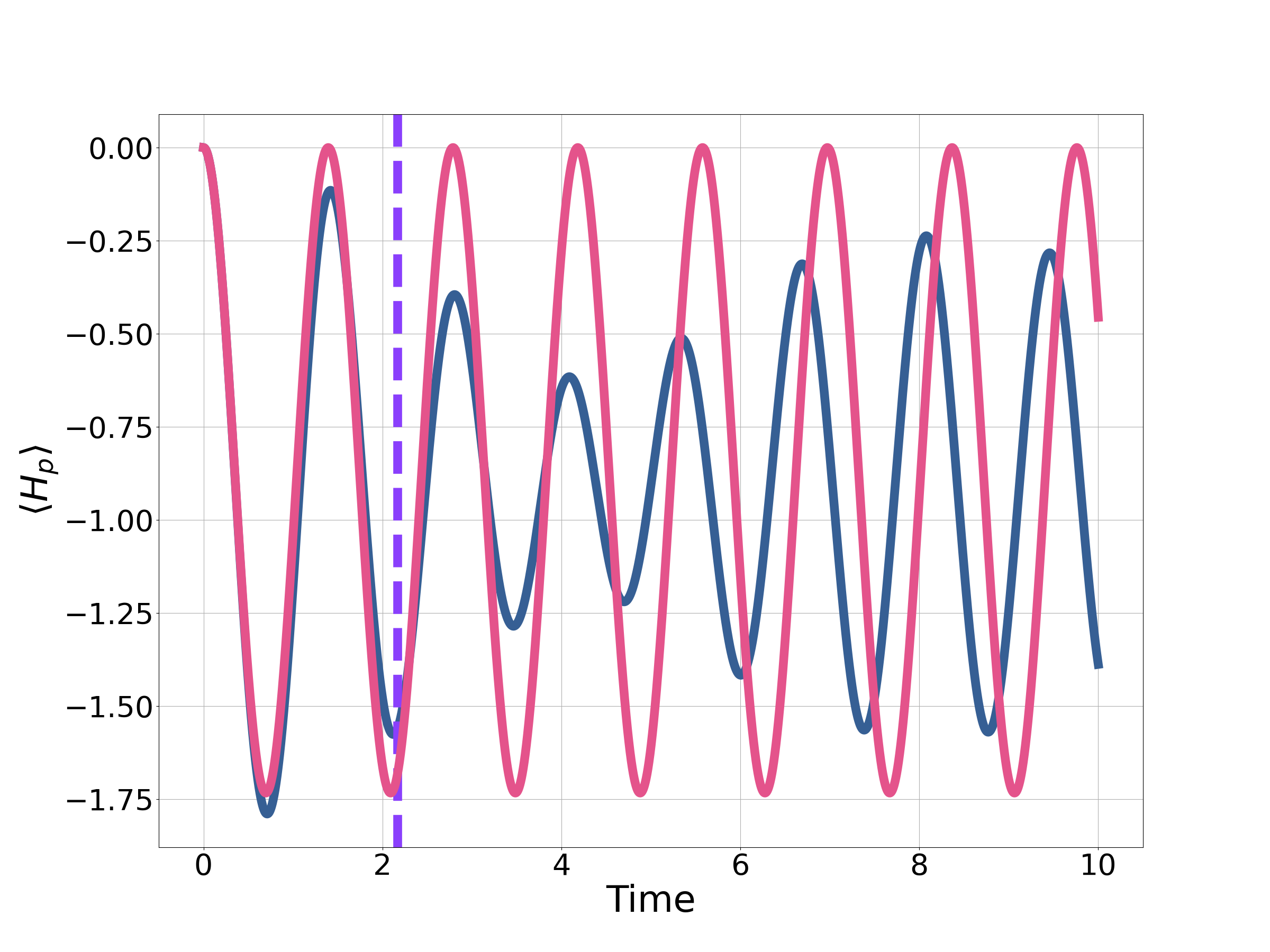}
    \caption{The performance of a CTQW on a 12 qubit binomial graph with $\gamma=0.05$. The blue line shows the result of direct integration of the Schr\"odinger equation. The pink line shows the result of the two-dimensional approximation (i.e, Eq.\ \ref{eq:cstl}) The dashed purple line shows the location of $\mathcal{T}^{-\frac{1}{4}}$.}
    \label{fig:stime_g_small}
\end{figure}

In Fig.\ \ref{fig:stime_g_small} we compare Eq.\ \ref{eq:cstl} to numerical simulation with a  12-qubit binomial graph with $\gamma=0.05$. The dashed purple line shows the location of  $\mathcal{T}^{-\frac{1}{4}}$, For short times there is good quantitative agreement between the numerical simulation  (the blue line) and the two-level prediction (the pink line). At longer times there is  reasonable qualitative agreement, with the two-dimensional approximation capturing the oscillatory nature.

\begin{figure}
    \centering
    \includegraphics[width=0.48\textwidth]{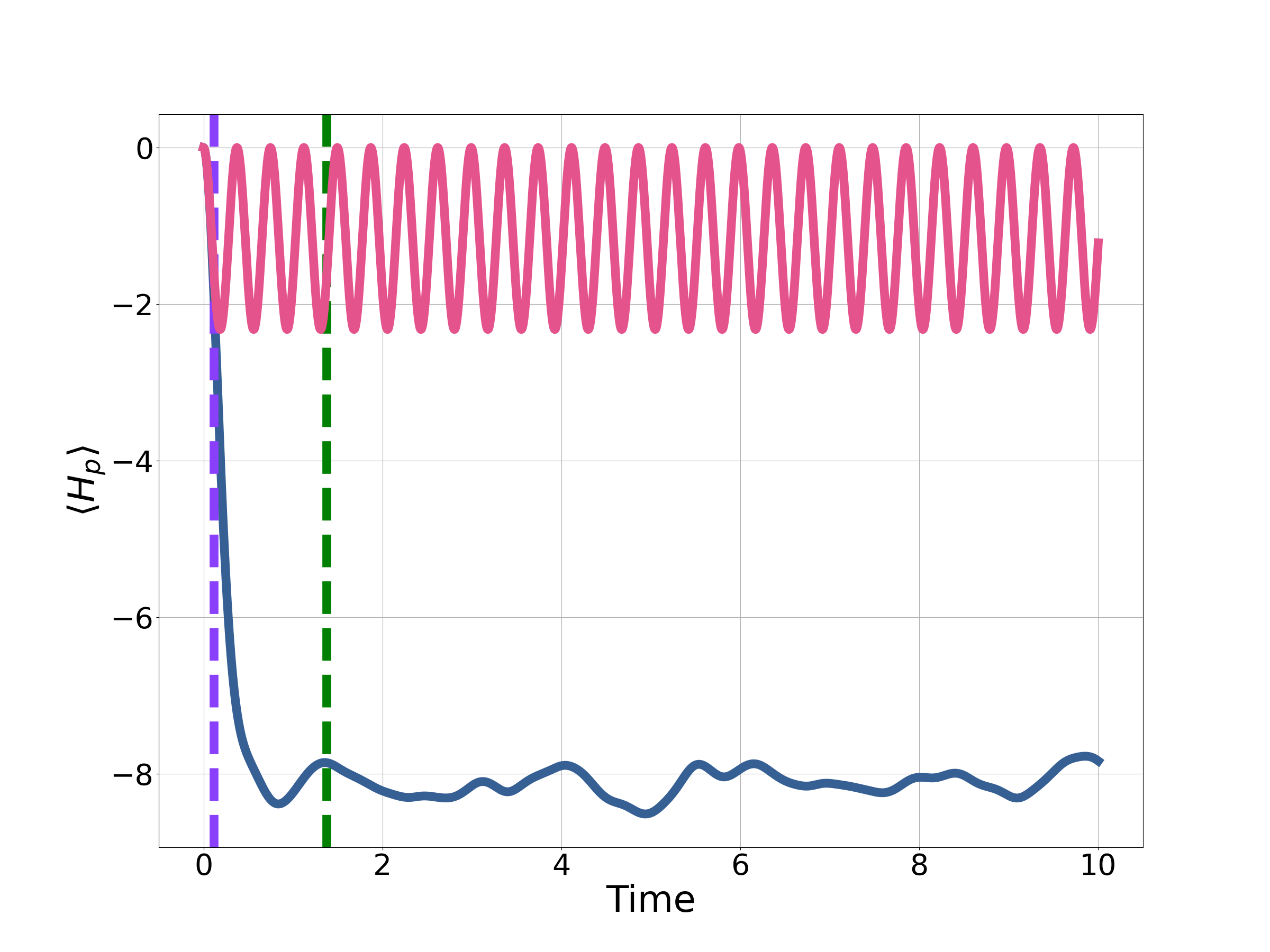}
    \caption{The performance of a CTQW on a 12 qubit binomial graph with $\gamma=1$. The blue line shows the result of direct integration of the Schr\"odinger equation. The pink line shows the result of the two-dimensional approximation (i.e, Eq.\ \ref{eq:cstl}) The dashed purple line shows the location of $\mathcal{T}^{-\frac{1}{4}}$. The dashed green line shows the location of $n\mathcal{T}^{-\frac{1}{4}}$, with $n=12$.}
    \label{fig:stime_g_big}
\end{figure}

Examining a different 12-qubit binomial graph with $\gamma=1$ gives Fig.\ \ref{fig:stime_g_big}. Again, we can see good agreement between the numerical simulation  (the blue line) and the two-level prediction (the pink line) for short times. The dashed purple line corresponds to $\mathcal{T}^{-\frac{1}{4}}$ with the dashed green line corresponding to $n\mathcal{T}^{-\frac{1}{4}}$. Here the two-level approximation provides poor qualitative insight into the CTQW outside of the short time limit, with the CTQW  approaching an approximate steady state. Noticeably, by  $n\mathcal{T}^{-\frac{1}{4}}$, the system has settled into the steady-state.

In this section we have explored the short-time limit of CTQW for MAX-CUT. The argument has relied on a geometric property of the quantum evolution (i.e., the torsion). By integrating the Schr\"odinger equation in the reduced two-dimensional sub-space we have made a prediction on the performance of CTQWs at short times, without limiting the analysis to small problem sizes.  For longer times, Eq.\ \ref{eq:cstl} breaks down. In general we are unable to integrate the Schr\"odinger equation for arbitrarily large problem sizes. For long times we turn to the Eigenstate Thermalisation Hypothesis (ETH).

\section{Continuous-time Quantum walks in the Thermodynamic limit}
\label{sec:QWTL}

As described in Sec.\ \ref{sec:ETH} a closed quantum system under evolution by a constant Hamiltonian can result in a state that is locally indistinguishable from a thermal state. In terms of CTQWs we are primarily interested in $\langle \bar{H_p} \rangle$, a sum over local observables. Thus if a CTQW close to the optimal $\gamma$ exhibits thermalisation, then $\langle \bar{H_p} \rangle$ should correspond to that of a thermal state. In Sec.\ \ref{sec:therm_check} we  provide numerical evidence that CTQWs do exhibit thermalisation even for small problem sizes. The aim of the rest of the section is to use this insight to estimate the optimal choice of $\gamma$, and ideally the associated value of $\langle \bar{H_p} \rangle$. To do this, in Sec.\ \ref{sec:dos} a model for the density of states (DOS) is provided. The DOS is then used to predict the temperatures and entropy of entanglement associated with CTQWs in Sec.\ \ref{sec:dosPred}. In Sec.\ \ref{sec:hpPred} we use the model to estimate $\langle \bar{H_p} \rangle$  and to provide estimates for the optimal choice of $\gamma$.

\subsection{Continuous-time quantum walks are well modelled by thermal states}
\label{sec:therm_check}
Here we demonstrate, for a few examples, that CTQWs are well modelled by thermal states, taking $\langle \bar{H_p} \rangle$ as a measure of success of the CTQW. If we know the temperature associated with the CTQW, this provides us with a second route to find the same value of $\langle \bar{H_p} \rangle$ by preparing a thermal state, thus side-steping the need for complete unitary dynamics.

Since the Hamiltonian is constant during the evolution under a CTQW, energy is conserved in a closed-system. Hence the following must hold true for the thermal state \cite{Ess16}:
\begin{equation}
    \label{eq:betaFix}
    \Tr \left[H_{QW} \rho_{\beta}\left(H_{QW}\right)\right]=-n,
\end{equation}
where $\rho_{\beta}(H_{QW})$ denotes a thermal state with Hamiltonian $H_{QW}$ and inverse temperature $\beta$. This equation fixes $\beta$, therefore there are no free parameters to fit. 

\begin{figure}[H]
     \centering
     \begin{subfigure}[l]{0.48\textwidth}
         \centering
         \includegraphics[width=\textwidth]{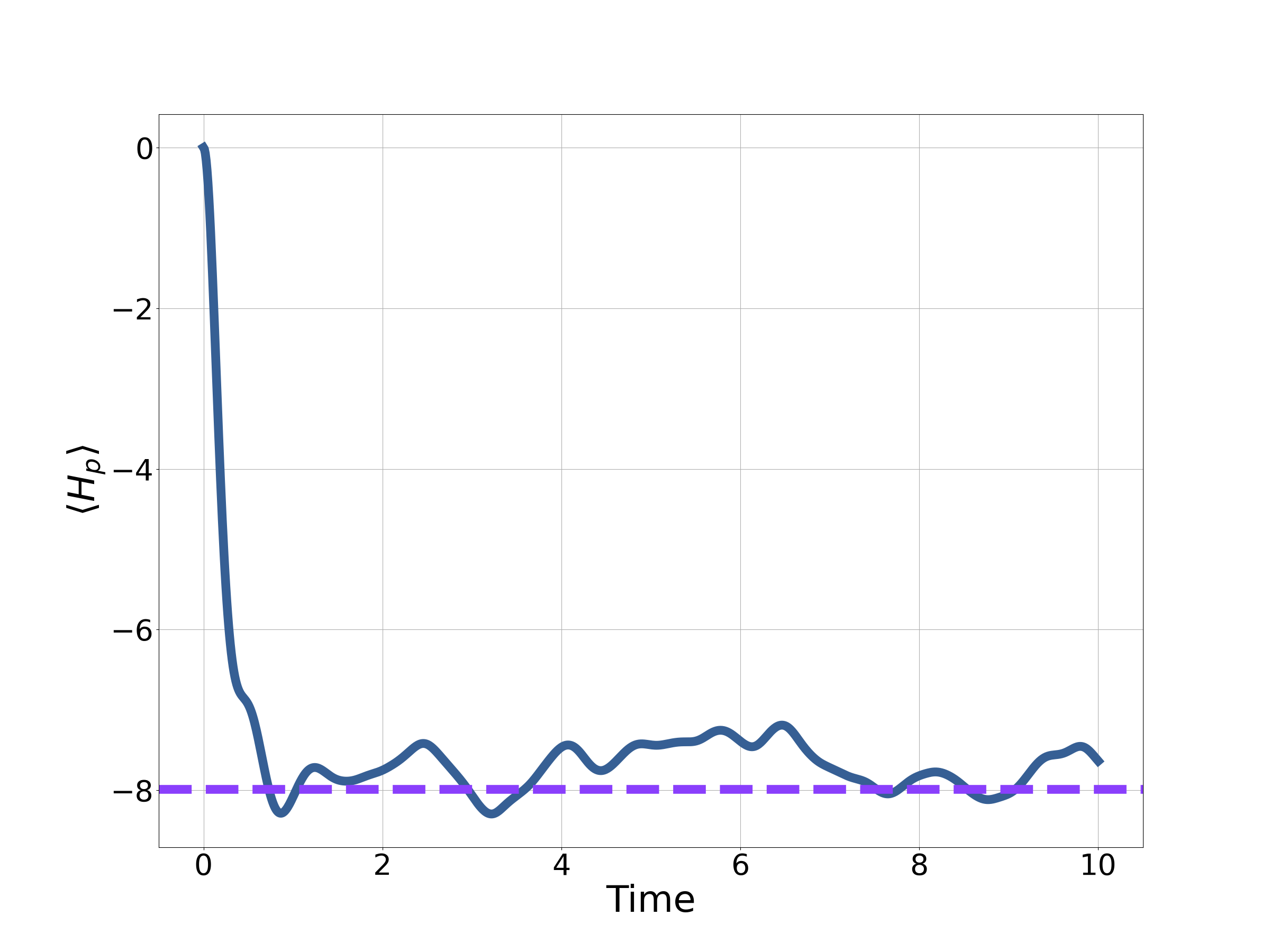}
         \caption{A 12 qubit binomial graph.}
         \label{fig:singRandReg}
     \end{subfigure}
     \\
     \begin{subfigure}[l]{0.48\textwidth}
         \centering
         \includegraphics[width=\textwidth]{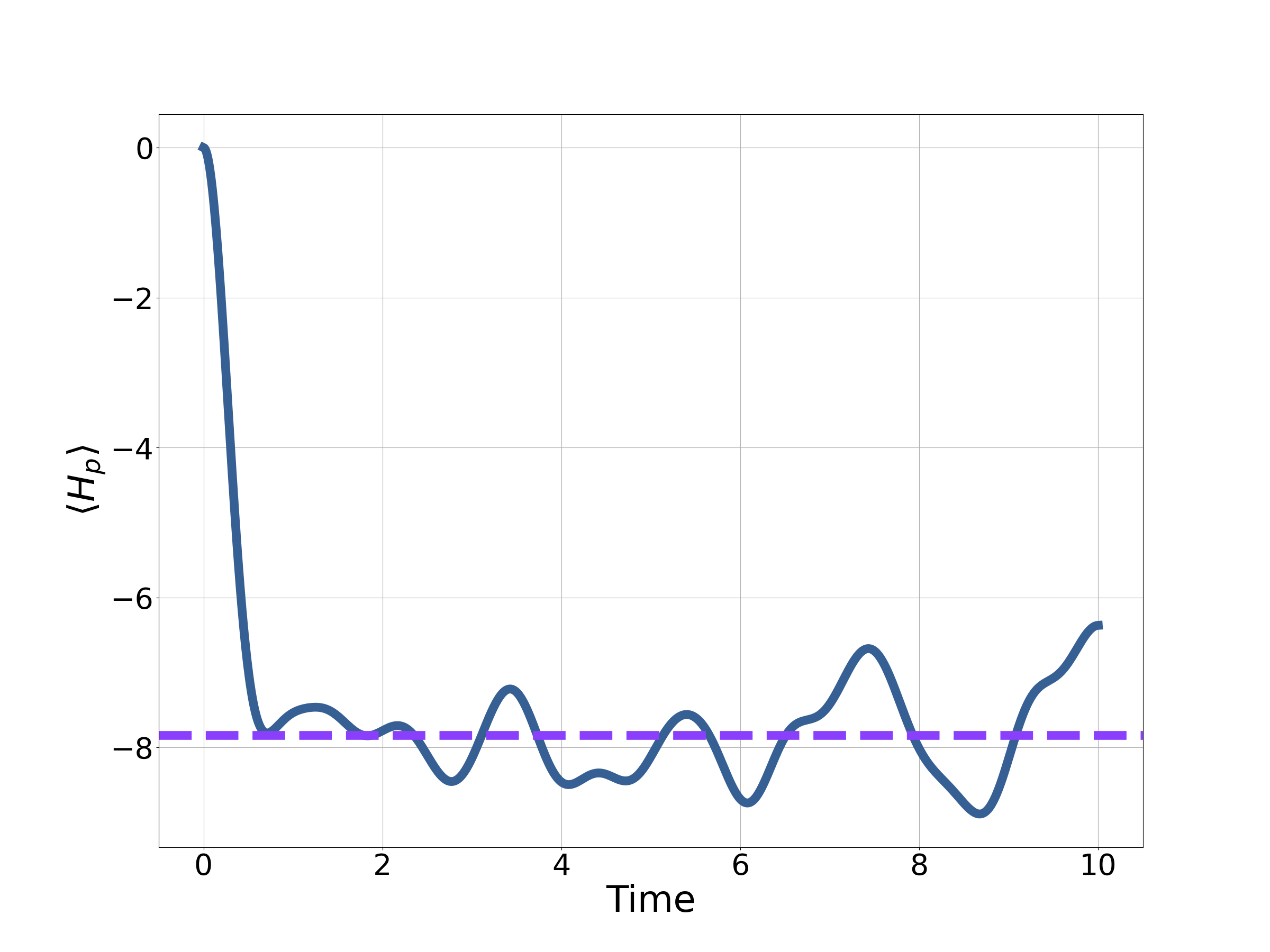}
         \caption{A 12 qubit three-regular graph. }
         \label{fig:singThreeReg}
     \end{subfigure}
        \caption{The performance of a CTQW with $\gamma$ optimised to give the best value of $\langle \bar{H_p}\rangle$. The dashed purple line shows the thermal state prediction. The temperature is fixed using Eq.\ \ref{eq:betaFix}.}
        \label{fig:fewThermInst}
\end{figure}

Fig.\ \ref{fig:fewThermInst} shows $\langle H_p \rangle$ the Schr\"odinger evolution for a binomial graph and a three-regular graph and the corresponding prediction for $\langle H_p \rangle$ from the thermal state. For both problem instances it appears that $\langle H_p \rangle$ is fluctuating around a steady state value. Importantly, despite being far from the thermodynamic limit at only 12-qubits, the thermal state prediction is capturing the steady state behaviour well. 

\begin{figure}[H]
     \centering
     \begin{subfigure}[l]{0.48\textwidth}
         \centering
         \includegraphics[width=\textwidth]{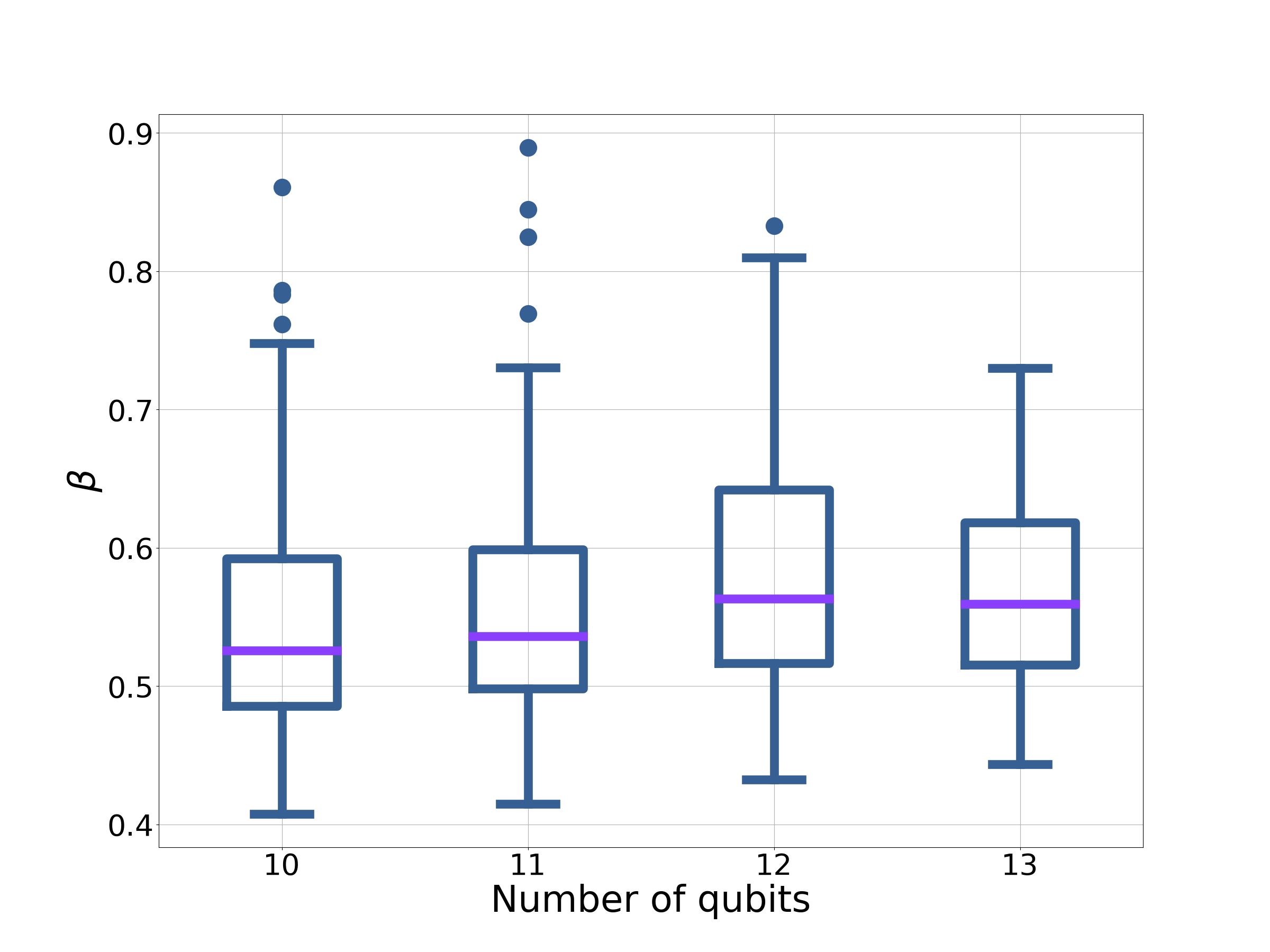}
         \caption{Binomial graphs.}
         \label{fig:manyTampsRand}
     \end{subfigure}
     \\
     \begin{subfigure}[l]{0.48\textwidth}
         \centering
         \includegraphics[width=\textwidth]{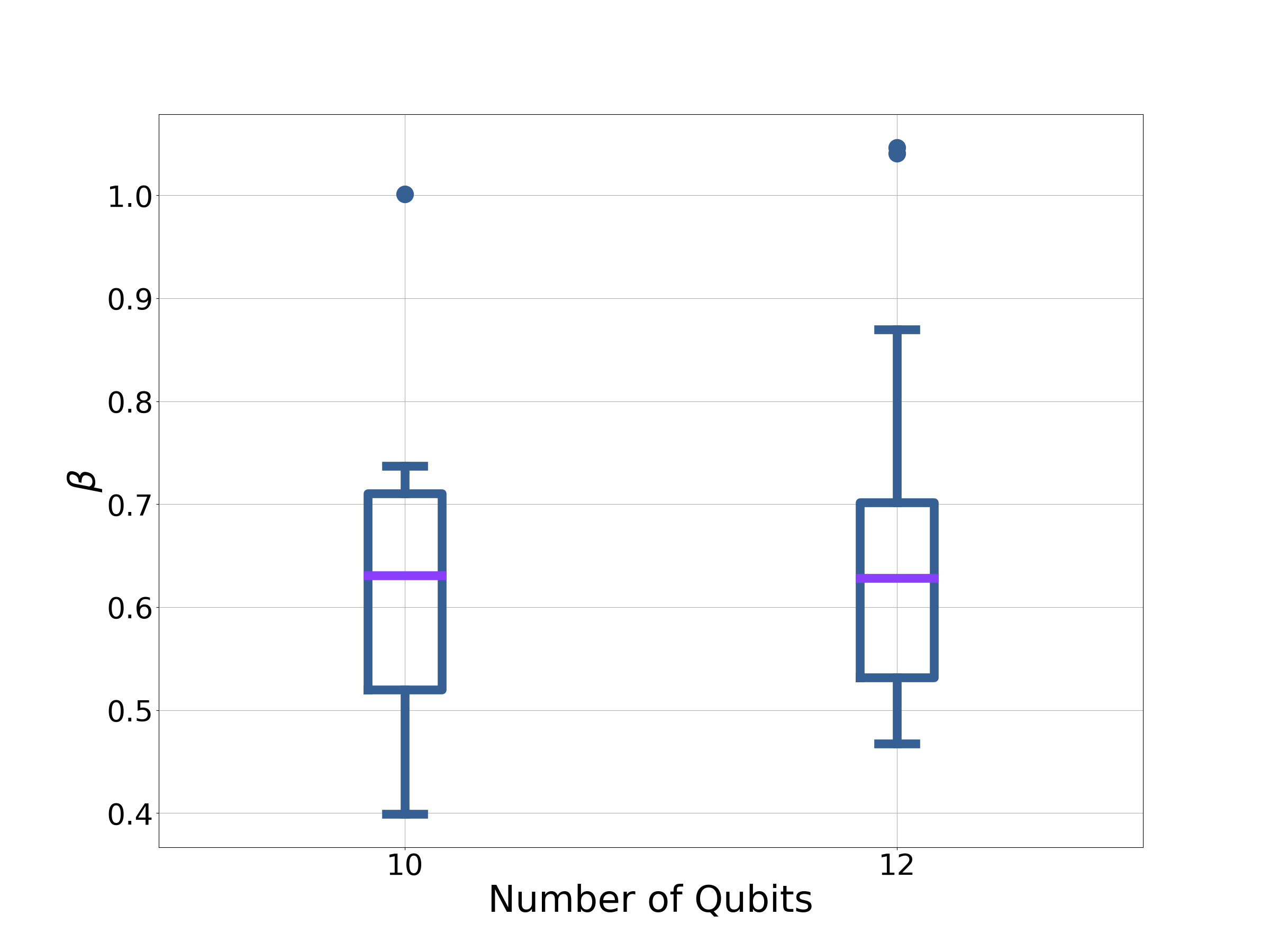}
         \caption{Three-regular graphs.}
         \label{fig:manyTempsReg}
     \end{subfigure}
        \caption{The inverse temperature associated with CTQWs (i.e. the solution to Eq.\ \ref{eq:betaFix}). For each problem instance $\gamma$ is equal to $\gamma_{opt}$, shown in Fig.\ \ref{fig:gamma_rand} and Fig.\ \ref{fig:gamma_reg}. }
        \label{fig:manyTemps}
\end{figure}

Fig. \ref{fig:manyTemps} shows the solution to Eq.\ \ref{eq:betaFix} for the optimal choice of $\gamma$ for multiple instances of binomial and three-regular graphs. For both cases the inverse temperatures are quite small, with $\beta<1$ for almost all instances. This means CTQWs correspond to Gibbs states with high temperatures.

\begin{figure}
     \centering
     \begin{subfigure}[l]{0.48\textwidth}
         \centering
         \includegraphics[width=\textwidth]{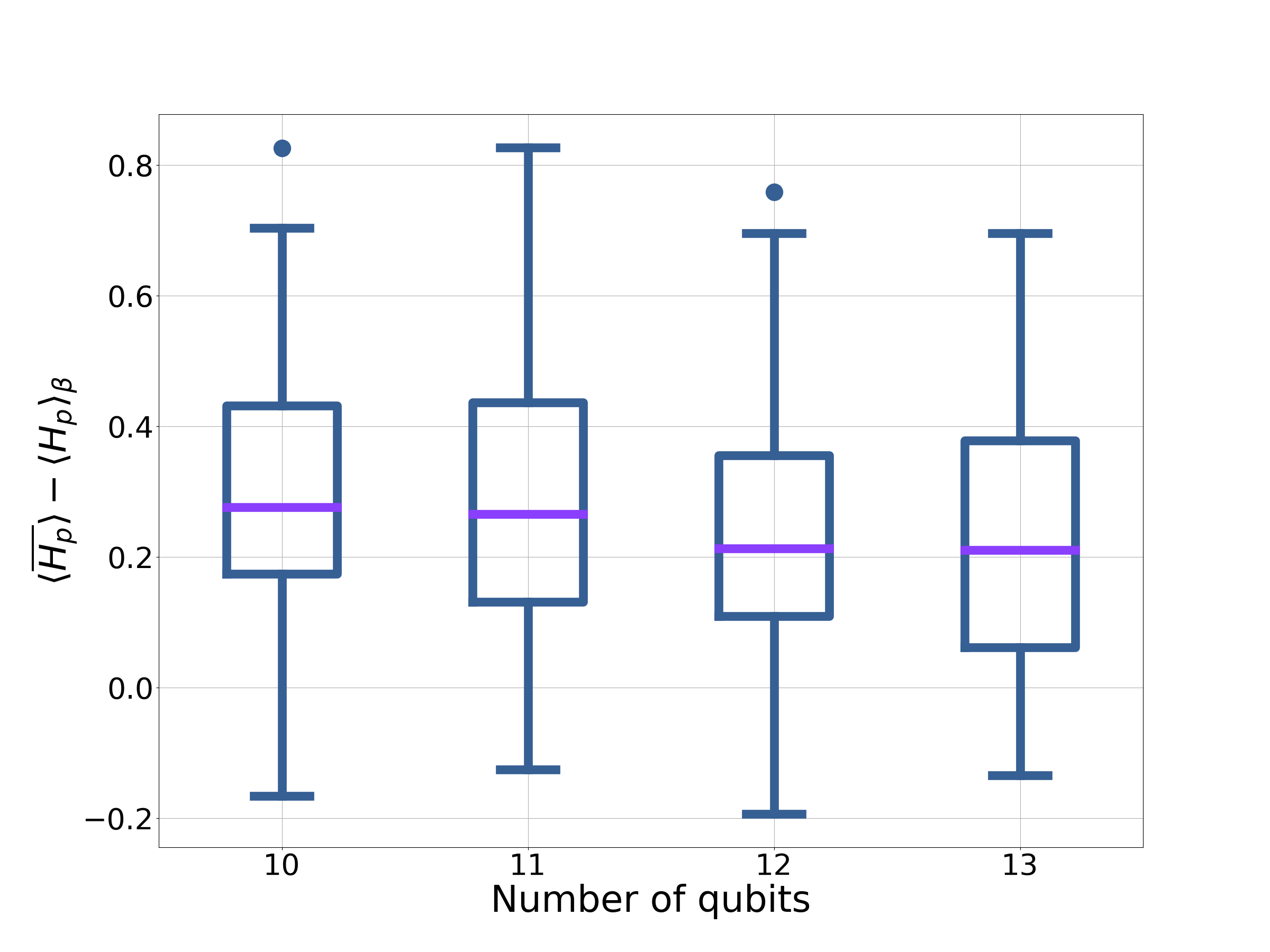}
         \caption{Binomial graphs.}
         \label{fig:manyThermRand}
     \end{subfigure}
     \\
     \begin{subfigure}[l]{0.48\textwidth}
         \centering
         \includegraphics[width=\textwidth]{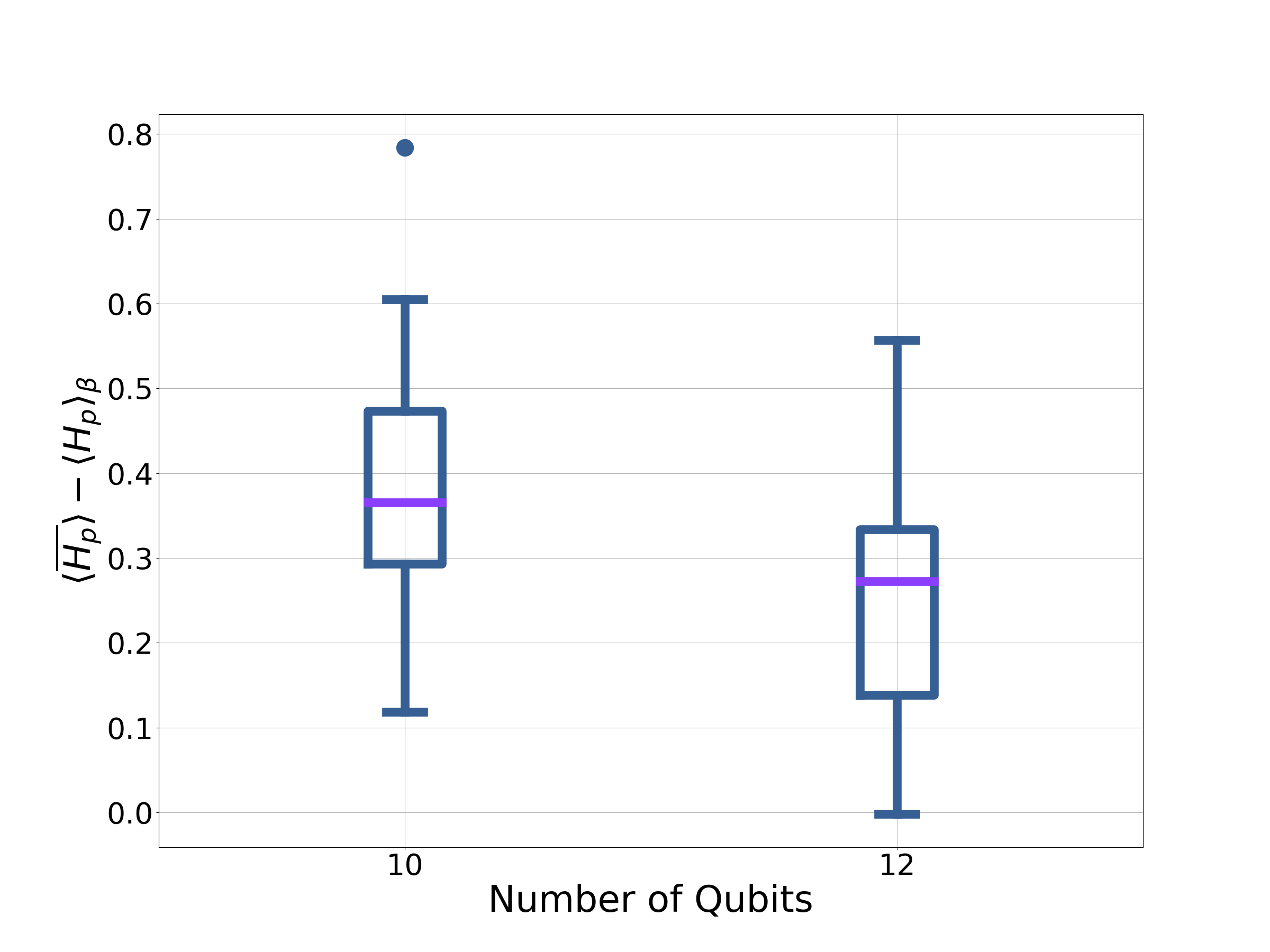}
         \caption{Three-regular graphs.}
         \label{fig:manyThermReg}
     \end{subfigure}
        \caption{The corresponding thermal prediction for the values of $\beta$ shown in Fig.\ \ref{fig:manyTemps}}
        \label{fig:manyThermInst}
\end{figure}

As mentioned in Sec.\ \ref{sec:ETH}, we expect the ETH to hold true in the thermodynamic limit. Here we are working with of order ten qubits, hence we expect there to be finite-size effects meaning there will be some error between the Schr\"odinger equation and the thermal prediction. This is captured in Fig.\ \ref{fig:manyThermInst}. Generally, the thermal prediction, $\langle H_p \rangle _{\beta}$ is overestimating the performance of the CTQW. As $n$ is increased we would expect this error to decrease. However, even for these very small systems the error is relatively small, with no significant outliers in this data set.

\clearpage

\subsection{Modelling the density of states}
\label{sec:dos}
In the previous section we numerically demonstrated that even for relatively small systems, $\langle \bar{H_p} \rangle$ is well predicted by a thermal state. Numerically solving Eq.\ \ref{eq:betaFix} is however difficult, requiring finding the exponential of a matrix that increases exponentially with the problem size. In Sec.\ \ref{sec:dosPred} we will attempt to estimate the temperature associated with a CTQW. Before this is possible we need to model the density of states (DOS) for $H_{QW}$.

To make progress we assume that the DOS is well modelled by a continuous distribution. We justify this assumption by arguing that the eigenstates of $H_{QW}$ are likely to become exponentially close as the system size grows. This follows from the largest eigenvalue of $H_{QW}$ being polynomial in $n$, and there being exponentially many states. Assuming no (or little) degeneracy, the difference between energy levels must shrink exponentially with the problem size. We also assume that the DOS is well modelled by a uni-modal distribution. This assumption will break down if $\gamma$ is too large or too small. However, as seen in Sec.\ \ref{sec:gamma} we expect useful values of $\gamma$ to correspond to somewhere between these two limits.

Intuitively, for large systems, we would expect the energy of a randomly chosen eigenstate to be close to the average energy. The average energy, $\mu$, of $H_{QW}$ is given by:
\begin{align*}
    \mu=\frac{1}{2^{n}}\sum_k E_k&=\frac{1}{2^{n}}\Tr H_{QW}\\
    &=0.
\end{align*}
Now consider the variance, $\sigma^2$:
\begin{align*}
    \sigma^2=\frac{1}{2^{n}}\sum_k E_k^2&=\frac{1}{2^{n}}\Tr H_{QW}^2\\
    &=n+\gamma^2 \kappa_2.
\end{align*}
Therefore, as $n$ goes to infinity, $\sigma/2^{n}$ tends to zero. So we expect a tightly peaked distribution for large $n$. The details for calculating moments of the DOS can be found in Appendix \ref{sec:mom}.

\begin{figure}
     \centering
     \begin{subfigure}[b]{0.48\textwidth}
         \centering
         \includegraphics[width=\textwidth]{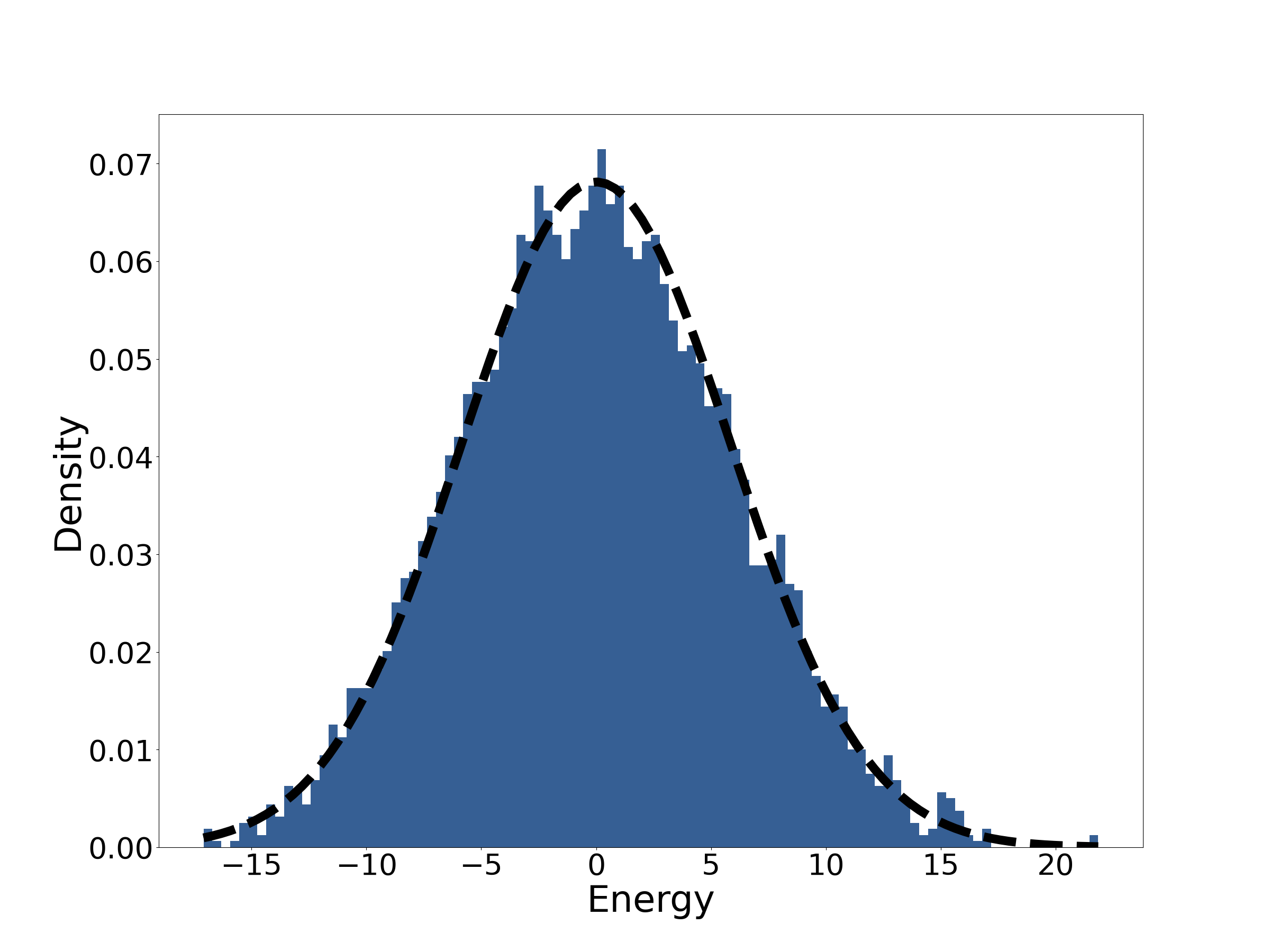}
         \caption{A 12-qubit three-regular graph with $\gamma=1.11$.}
         \label{fig:normDOSReg}
     \end{subfigure}
     \vfill
     \begin{subfigure}[b]{0.48\textwidth}
         \centering
         \includegraphics[width=\textwidth]{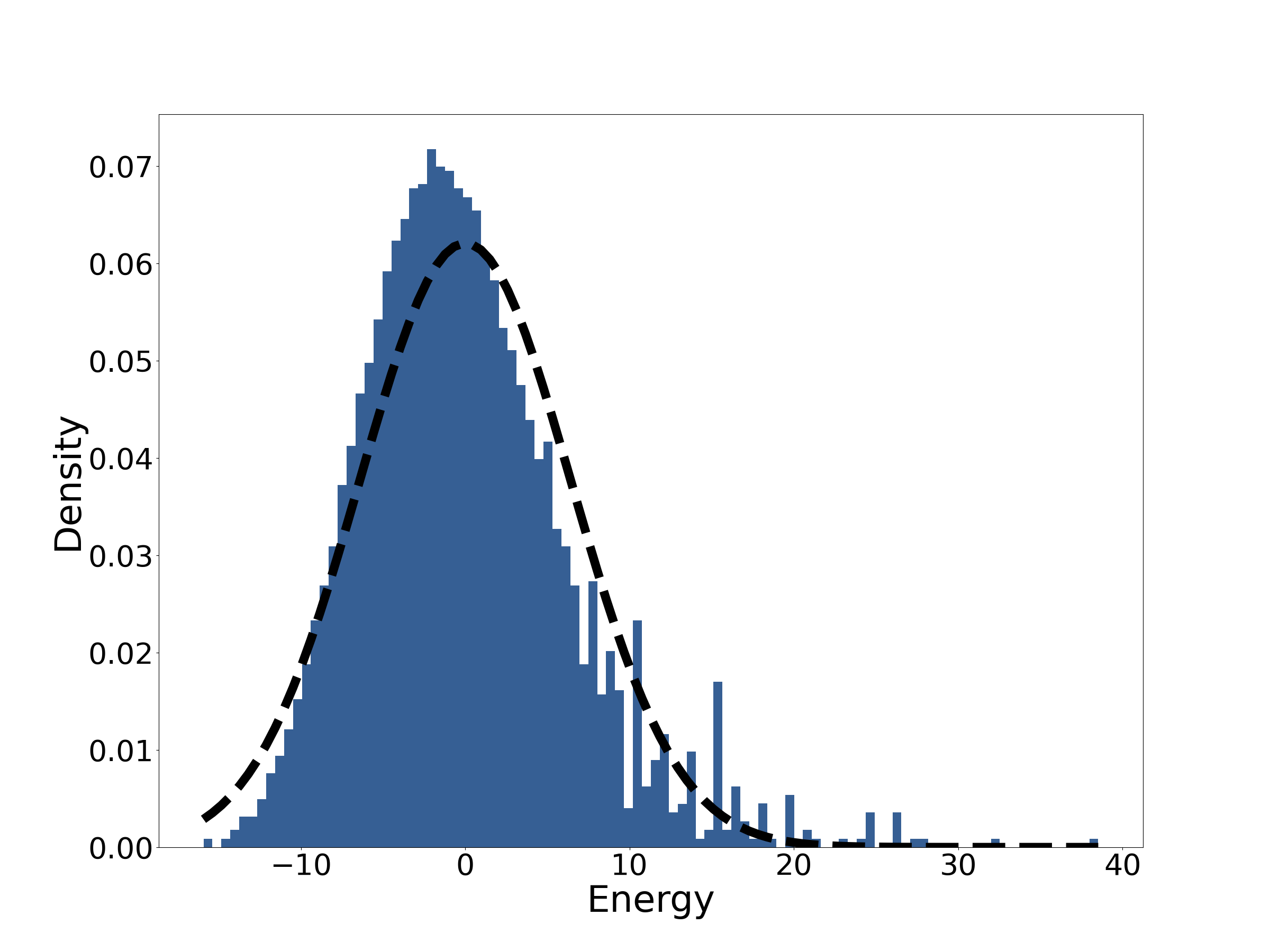}
         \caption{A 12-qubit binomial graph with $\gamma=0.78$. }
         \label{fig:normDOSRand}
     \end{subfigure}
        \caption{Histogram for the DOS for two graphs with optimised $\gamma$. The energies have been binned into 100 bins. The DOS has been normalised such that the total density is equal to one. The dashed black line shows the fitted Gaussian distribution.}
        \label{fig:normDOS}
\end{figure}

The DOS for thermalising systems has previously been modelled with a Gaussian distribution \cite{DAl16}. Since the CTQW exhibits thermalisation we adopt this approach. Fig.\ \ref{fig:normDOS} shows a Gaussian fit to the DOS for a 12-qubit three-regular graph (Fig.\ \ref{fig:normDOSReg}) and a binomial graph (Fig.\ \ref{fig:normDOSRand}) with $\gamma$ optimised for each problem. Visually this appears to be an acceptable approximation for the regular graph, less so for the binomial graph. Given the simplicity of this approximation we utilise it throughout the paper to make simple analytic predictions.

To verify the Gaussianity of the DOS, we look at two typical tests of normality: the skewness, $s$, of the distribution and the excess kurtosis $k$. The skewness is the ratio of the third moment of the distribution to the cube of the standard deviation. The kurtosis is the ratio of the fourth moment of the distribution to the variance squared. The excess kurtosis is the kurtosis minus the expected kurtosis of a Gaussian distribution, which is equal to 3. 

Starting with the skewness:

\begin{equation}
    s=\frac{6\gamma^3\kappa_3}{\left(n+\gamma^2 \kappa_2\right)^{3/2}}.
\end{equation}
Further details can be found in Appendix \ref{sec:mom}. Note that triangle-free graphs have no skewness and that for positive $\gamma$, the skewness is always non-negative.

For a $d$-regular graph, $\kappa_2=dn/2$ and $\kappa_3$ is bounded by $\mathcal{O}\left(d^2 n\right)$. So the skewness, following the heuristic in Eq.\ \ref{eq:heur_gam} (provided $\gamma$ is held constant) scales approximately as $n^{-1/2}$. Therefore it will tend to zero as the problem size is increased. 

Conversely, for binomial graphs, provided $\gamma$ scales proportional to $1/\sqrt{n}$, then the skewness will not scale with $n$. Hence, the Gaussian approximation will not hold as well for a binomial graph.

Examining now the excess kurtosis gives:
\begin{equation}
    k=-2\frac{n+4\gamma^2 \kappa_2+\gamma^4(\kappa_2-12\kappa_4)}{\left(n+\gamma^2 \kappa_2\right)^2}
\end{equation}
Under the same assumptions above for the scaling of the skewness, the excess kurtosis will tend to zero as $n$ tends to infinity for regular graphs. For binomial graphs the excess kurtosis is unlikely to vanish. 

For large regular graphs the Gaussian approximation looks to hold well. The same cannot be said to be true for binomial graphs. This is a consequence of regular graphs looking locally tree-like in the infinite sized limit \cite{Farhi_2012}. Hence it may be necessary to consider other models for the DOS for binomial graphs and for regular graphs away from the infinite-size limit.

Here we propose using an exponentially modified Gaussian (EMG) distribution \cite{Jea91,Kal11} to model the DOS for a CTQW. The EMG is a skewed unimodal distribution. The skew can only be positive, as is the case with the DOS for a  CTQW. In the correct limit it can recover a Gaussian. It has the following probability density function:

\begin{multline}
    p(x;m,\nu, \lambda)\, \dd x =\frac{\lambda}{2}e^{\frac{\lambda}{2}\left(2m+\lambda \nu^2-2x\right)}\\
    \times \erfc \left(\frac{m+\lambda \nu^2 -x}{\sqrt{2 \nu^2}} \right)\, \dd x,
\end{multline}
where $\erfc$ is the complementary error function. The fitting parameters $m$, $\nu$ and $\lambda$ can be ascertained from $n$, $\sigma$ and $s$. For the details the reader is referred to Appendix \ref{sec:EMG}.

\begin{figure}
     \centering
     \begin{subfigure}[b]{0.48\textwidth}
         \centering
         \includegraphics[width=\textwidth]{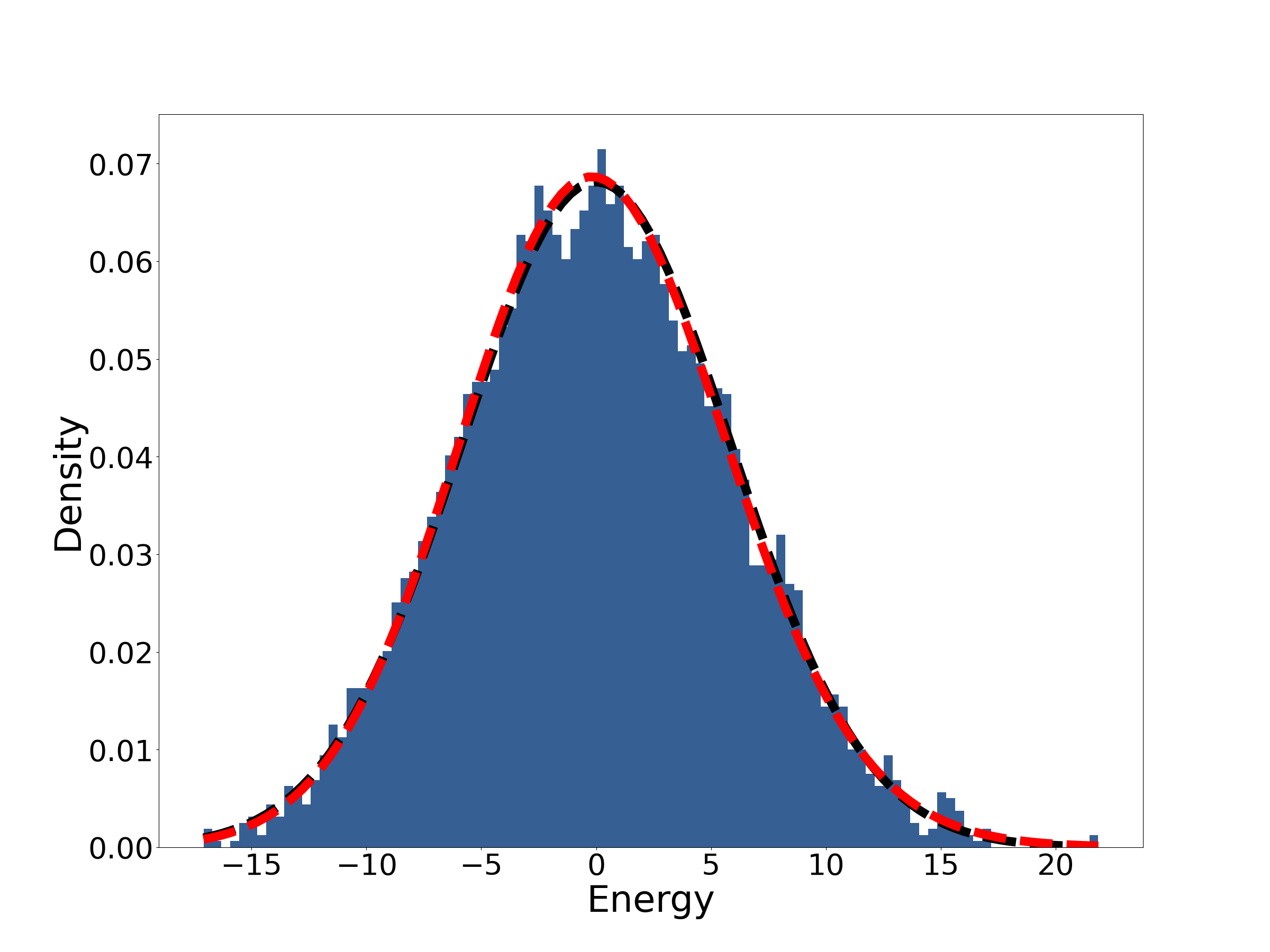}
         \caption{A 12 qubit three-regular graph with $\gamma=1.11$.}
         \label{fig:skewDOSReg}
     \end{subfigure}
     \vfill
     \begin{subfigure}[b]{0.48\textwidth}
         \centering
         \includegraphics[width=\textwidth]{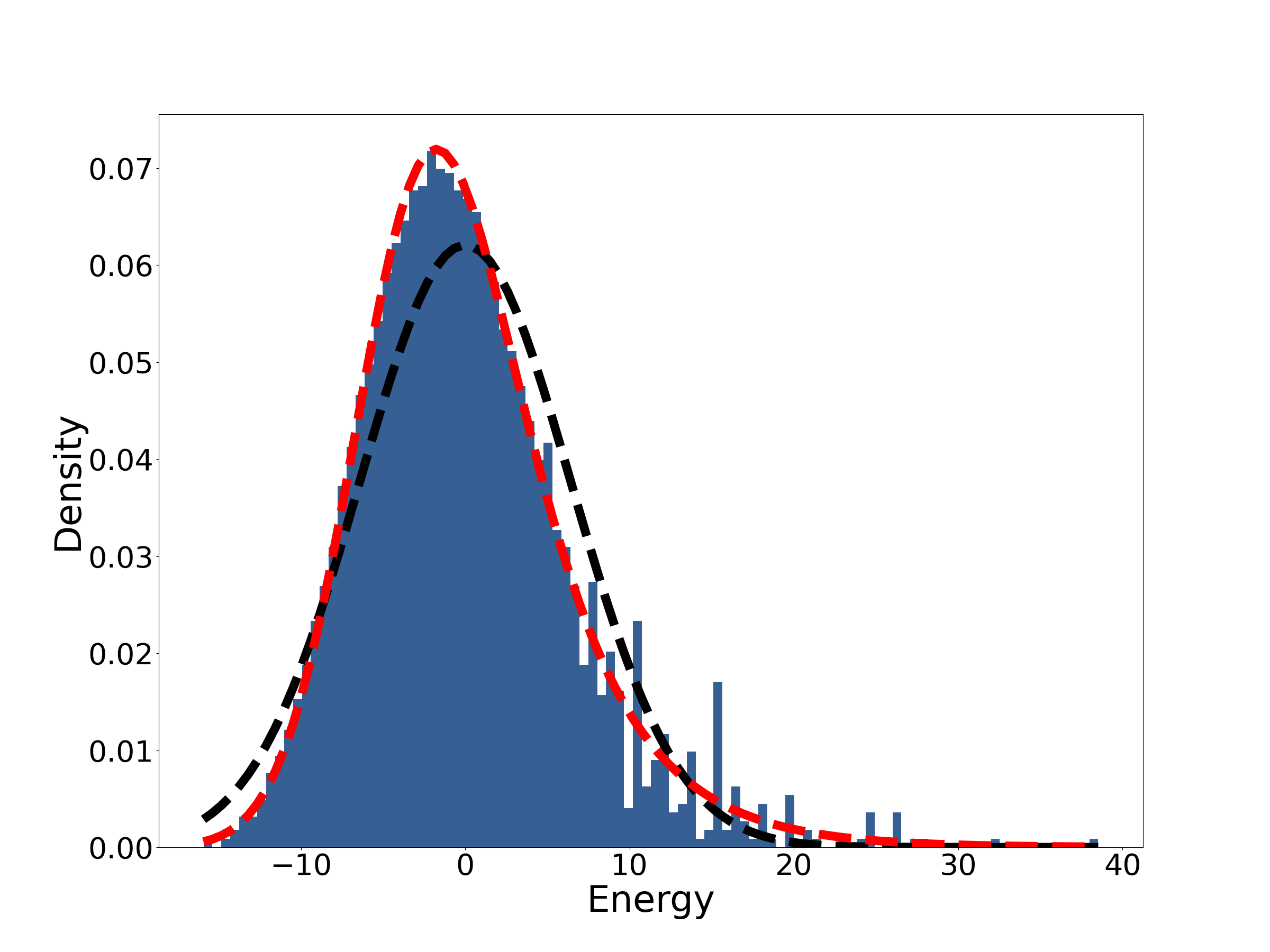}
         \caption{A 12 qubit binomial graph with $\gamma=0.78$. }
         \label{fig:skewDOSRand}
     \end{subfigure}
        \caption{Histogram for the DOS for two graphs with $\gamma$ optimised. The energies have been binned into 100 bins. The DOS has been normalised such that the total density is equal to one. The dashed black line shows the fitted Gaussian distribution. The dashed red line shows the EMG distribution.}
        \label{fig:skewDOS}
\end{figure}

The Gaussian approximation only includes information about the problem-size and therefore will make the same prediction for a large number of graphs. For instance, all regular graphs with the same degree have the same DOS under the Gaussian approximation. By incorporating the skewness into the model we are incorporating more information about the graph structure, namely the number of triangles in the problem. Visually, as shown in Fig.\ \ref{fig:skewDOS}, we can see that the EMG distribution (the dashed red line) provides a better fit than the Gaussian distribution for the binomial graph. The EMG distribution also still models the DOS for the regular graph well. 

During the rest of this paper we use the DOS to make analytic predictions about the behaviour of a CTQW in a closed-system setting. 

\subsection{Predictions from the density of states}
\label{sec:dosPred}

Having provided a model for the DOS for $H_{QW}$, in this section we make two predictions in relation to CTQWs, namely the temperature and the entropy of entanglement.

\subsubsection{Estimating the temperature}
\label{sec:beta}
\begin{figure}
     \centering
     \begin{subfigure}[b]{0.48\textwidth}
         \centering
         \includegraphics[width=\textwidth]{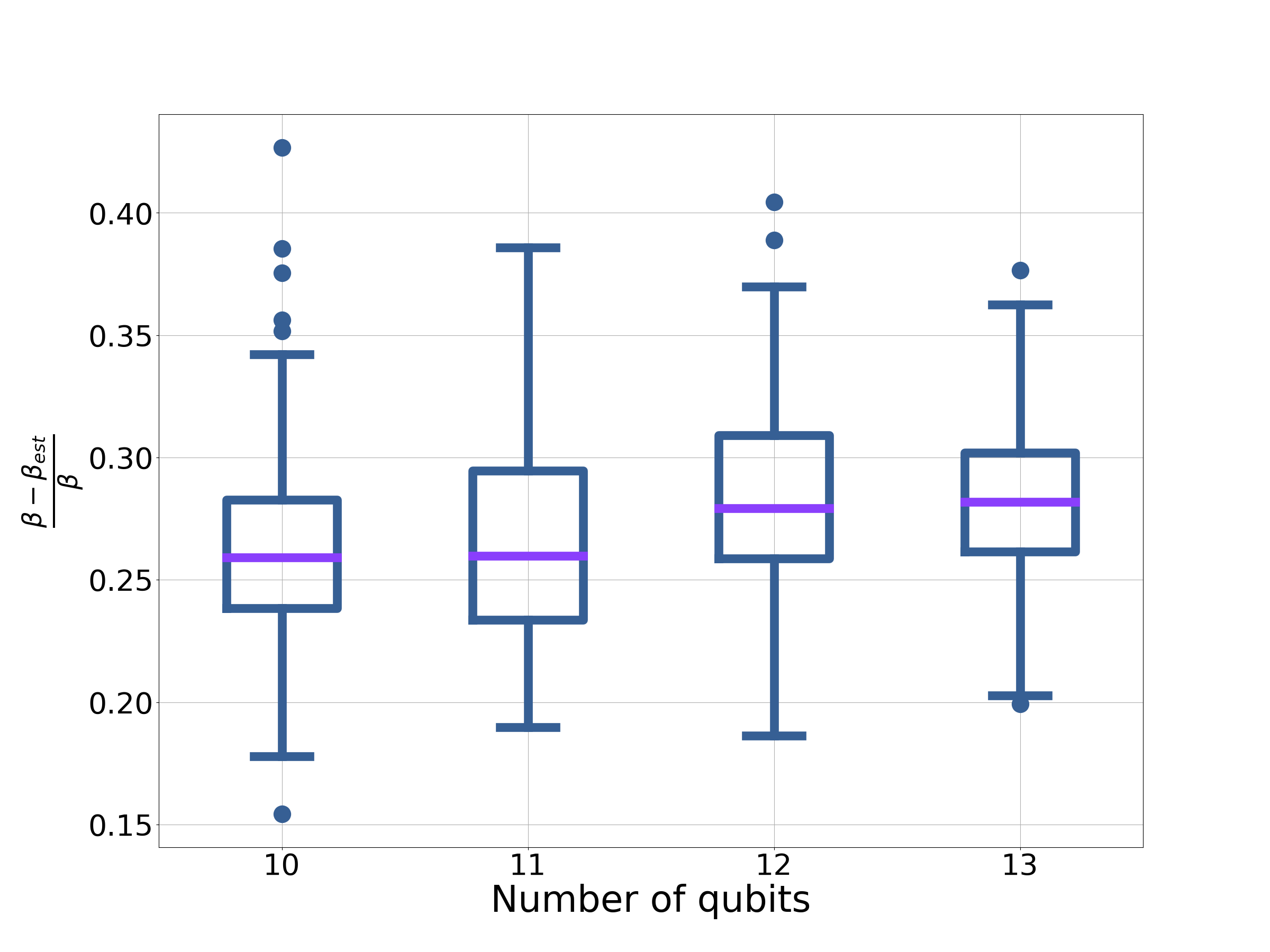}
         \caption{Binomial graphs}
         \label{fig:normTempRand}
     \end{subfigure}
     \vfill
     \begin{subfigure}[b]{0.48\textwidth}
         \centering
         \includegraphics[width=\textwidth]{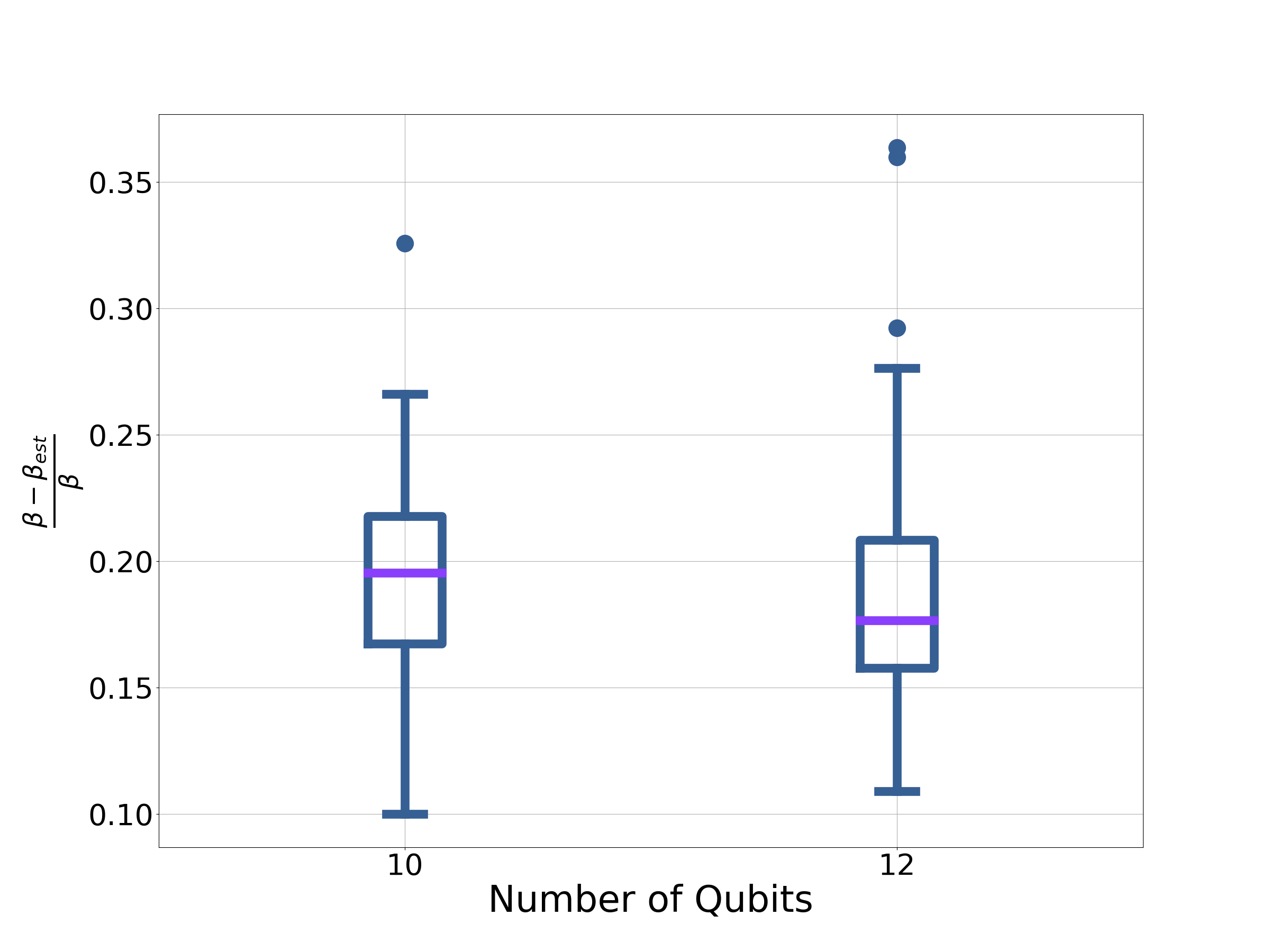}
         \caption{Three-regular graphs}
         \label{fig:normTempReg}
     \end{subfigure}
        \caption{The error in the predicted inverse temperature assuming a Gaussian DOS (i.e. Eq.\ \ref{eq:betaCrude}). For each problem instance $\gamma$ is equal to $\gamma_{opt}$, shown in Fig.\ \ref{fig:gamma_rand} and Fig.\ \ref{fig:gamma_reg}.}
        \label{fig:normTemp}
\end{figure}
Finding the temperature of the CTQW requires finding the solution to Eq. \ref{eq:betaFix}. To find approximate solutions we use the DOS models to evaluate the partition function, $\mathcal{Z}$. Eq. \ref{eq:betaFix} then becomes \cite{Blu06}:
\begin{equation}
    -n=-\frac{\partial \ln \mathcal{Z}}{\partial \beta}.
\end{equation}

 Assuming a Gaussian DOS then the estimated inverse temperature is given by:  
\begin{equation}
    \label{eq:betaCrude}
    \beta_{est}=\frac{n}{n+\gamma^2 \kappa_2}.
\end{equation}
This is perhaps what one would estimate as the temperature on dimensional grounds alone. The full details can be found in Appendix \ref{sec:norm}.

Taking the EMG to model the density of states gives:
\begin{equation}
    \label{eq:betaTri}
    \beta_{EM}=\frac{n\Delta-\sigma^2+\sqrt{\left(n\Delta+\sigma^2\right)^2-4n\Delta^3}}{2\Delta\left(\sigma^2-\Delta^2\right)},
\end{equation}
where $\sigma^2=n+\gamma^2 \kappa_2$ and $\Delta=\gamma (3\kappa_3)^{1/3}$. The details can be found in Appendix \ref{sec:EMG}.

\begin{figure}
     \centering
     \begin{subfigure}[b]{0.48\textwidth}
         \centering
         \includegraphics[width=\textwidth]{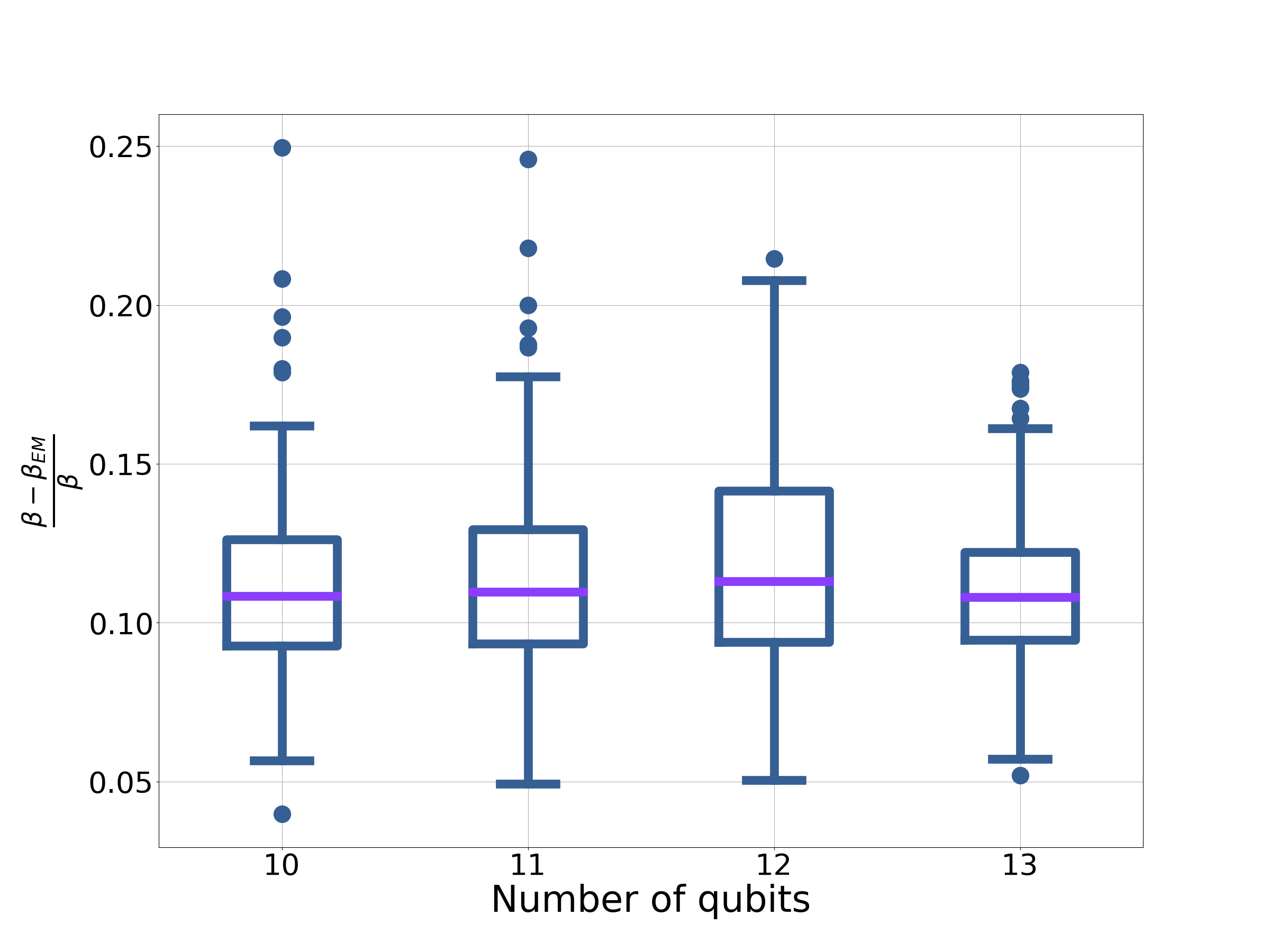}
         \caption{Binomial graphs}
         \label{fig:skewTempRand}
     \end{subfigure}
     \vfill
     \begin{subfigure}[b]{0.48\textwidth}
         \centering
         \includegraphics[width=\textwidth]{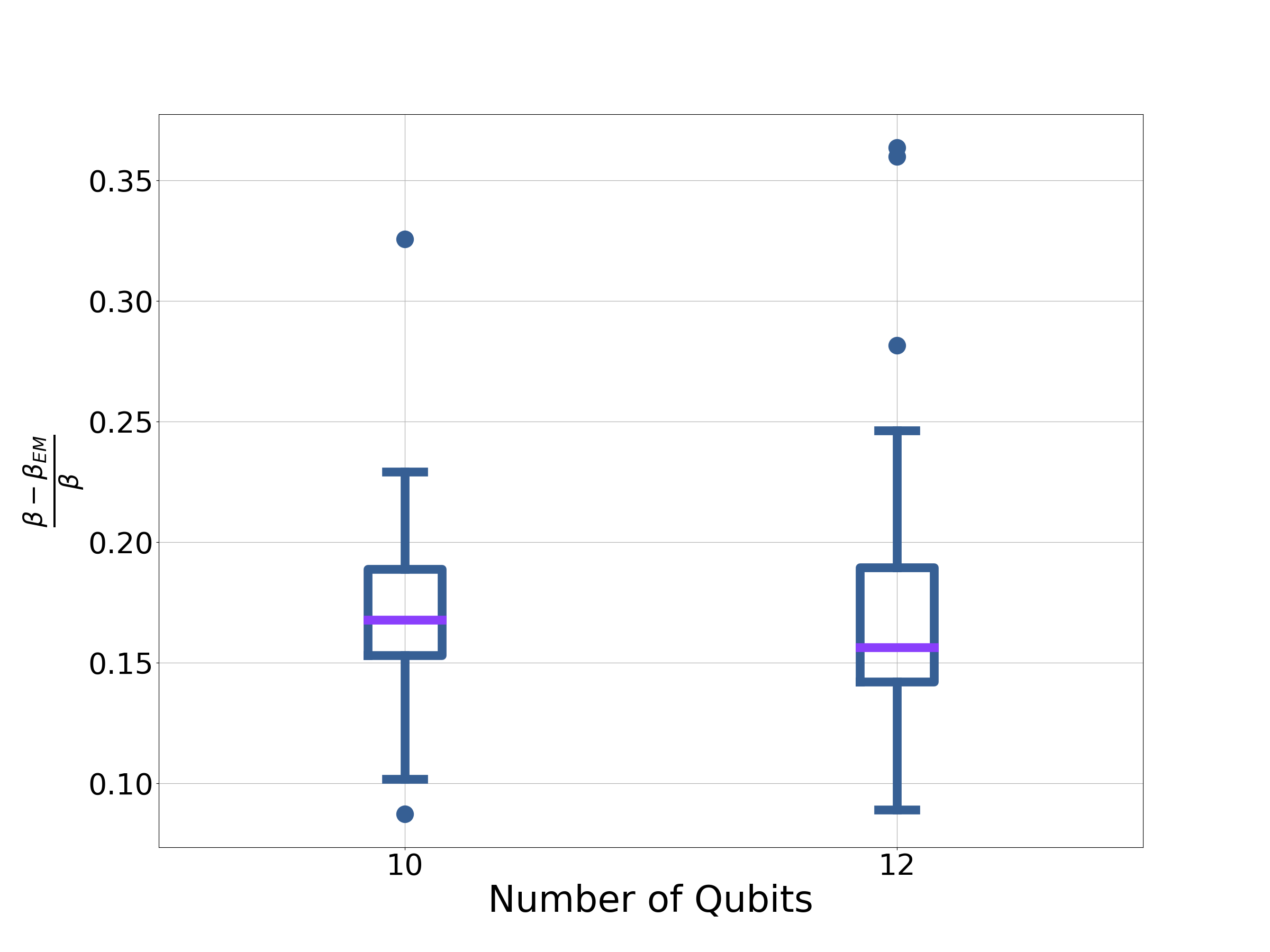}
         \caption{Regular graphs}
         \label{fig:skewTempReg}
     \end{subfigure}
        \caption{The error in the predicted inverse temperature assuming an EMG DOS (i.e. Eq.\ \ref{eq:betaTri}). For each problem instance $\gamma$ is equal to $\gamma_{opt}$, shown in Fig.\ \ref{fig:gamma_rand} and Fig.\ \ref{fig:gamma_reg}.}
        \label{fig:skewTemp}
\end{figure}

Fig.\ \ref{fig:normTemp} shows the error in temperature between $\beta_{est}$ and $\beta$ (the numerical solution to Eq.\ \ref{eq:betaFix}). The approximate solutions are consistently underestimating the inverse temperature. The error for binomial graphs is substantial, with typical errors being between $25\%$ and $30\%$ (Fig. \ref{fig:normTempRand}). For three-regular graphs, this reduces to somewhere between $15 \%$ and $25 \%$ (Fig.\ \ref{fig:normTempReg}). Using the EMG distribution (i.e. Eq.\ \ref{eq:betaTri}) provides little improvement on the estimate for the three-regular graphs (Fig.\ \ref{fig:skewTempReg}). This is unsurprising given how close to Gaussian the DOS is for these problems. For binomial graphs the improvement is more substantial with with typical errors being between $10\%$ and $15\%$. As $n$ is increased, we expect that the DOS will be better modelled by a continuous DOS, hence we would expect the error in $\beta$ to be improved.

By assigning the quantum evolution a temperature, we have mapped the challenge of understanding a dynamical problem to a static problem. So far in this section we have shown how to reasonably estimate the associated temperature. If the temperature is too high, then the associated thermal state can be efficiently classically approximated. Results by Crosson et al.\ \cite{Cro20} suggest that for values of $\beta \leq0.1$ a classical computer could simulate the associated thermal state efficiently for a three-regular graph, suggesting that for a CTQW to provide an advantage, it must operate outside this regime.

\subsubsection{Estimating the entropy}

\begin{figure}
    \centering
    \includegraphics[width=0.48\textwidth]{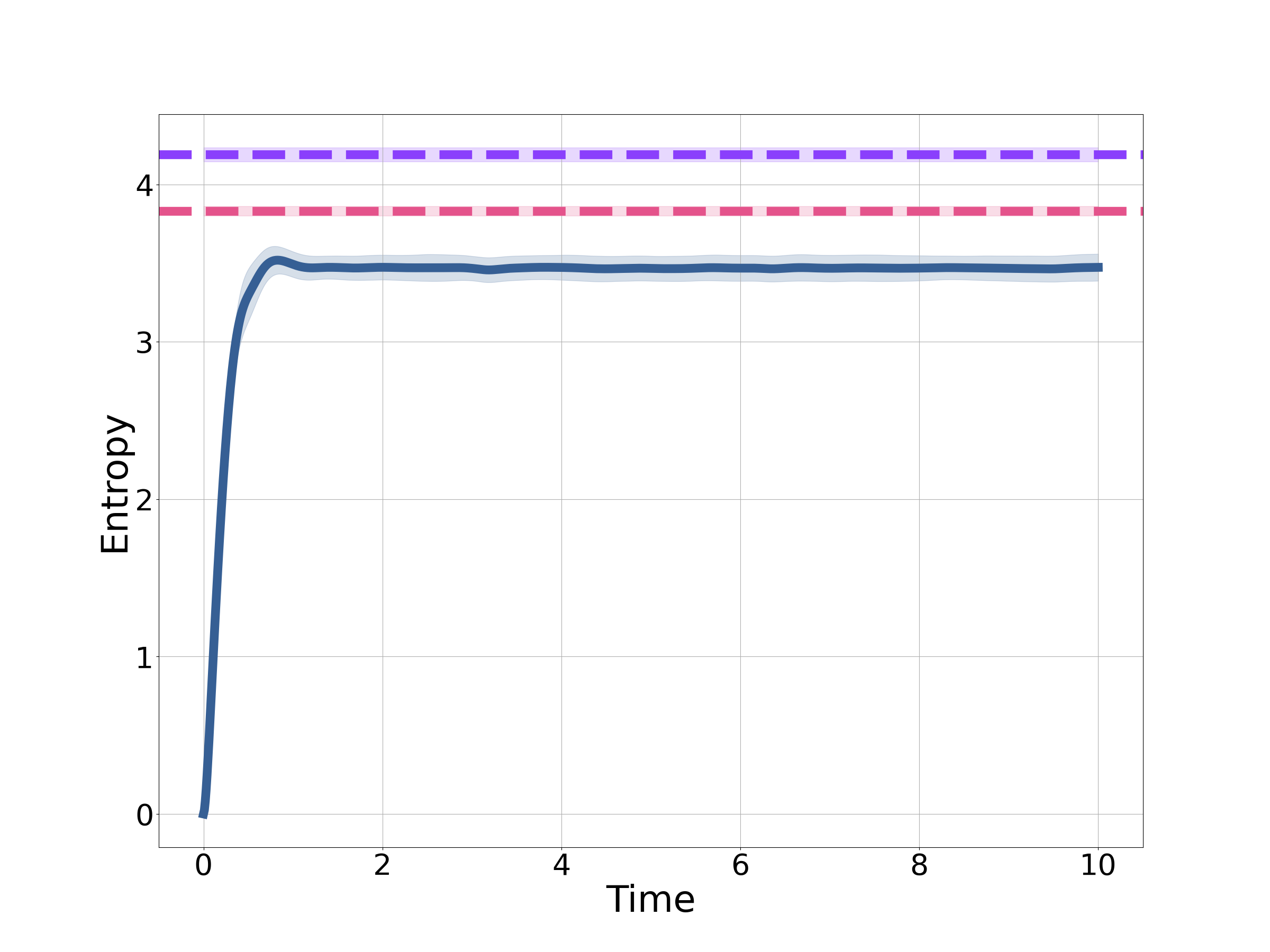}
    \caption{The entanglement entropy averaged over a hundred 14-qubit binomial instances (solid blue line). The dashed purple line shows $S^{norm}$ and the dashed pink line $S^{EM}$ The shaded regions show a single standard deviation.}
    \label{fig:ent}
\end{figure}

Often considered of high interest for a quantum algorithm is the entanglement in the system \cite{Du22}. This can be quantified by the entanglement entropy. Here we calculate the entanglement entropy by tracing out half of the qubits, which are randomly selected, and calculating the von Neumann entropy of the reduced density operator.

We compare the entanglement entropy to the thermodynamic entropy derived from the modelled DOS, given by:
\begin{equation}
    \label{eq:ent}
    S=\beta \langle H_{QW} \rangle+\ln \mathcal{Z}.
\end{equation}
Since entropy is an extensive quantity we take half of the above and compare it to the entanglement entropy. Although the ETH cannot be straightforwardly applied here as the entropy is not a local observable, the two have been shown to be linked in previous works \cite{Deu10,Deu13,San11}.

For the Gaussian distribution Eq.\ \ref{eq:ent} reduces to:
\begin{equation}
    S^{norm}=n\ln 2 -\beta_{est}n+\frac{\sigma^2 \beta_{est}^2}{2},
\end{equation}
and for the EMG:
\begin{multline}
    S^{EM}=n\ln 2+\beta_{EM} \left(\Delta -n \right)+\frac{\beta^2_{EM}}{2}\left(\sigma^2-\Delta^2\right)\\
    -\ln \left(1+\beta_{EM} \Delta \right)
\end{multline}
Fig.\ \ref{fig:ent} shows the entanglement entropy averaged over a hundred 14-qubit binomial graphs as a function of time. As is clear from the figure the entanglement entropy is approximately constant. The dashed purple line shows $S^{norm}$ and the dashed pink line $S^{EM}$. Both are overestimating the entanglement entropy. However, they appear to be reasonable estimates with $S^{EM}$ providing the better estimate. As mentioned in Sec.\ \ref{sec:QWMC}, the ideal final state will be a low-entanglement state. CTQWs lack any mechanism to dissipate entanglement, as shown by Fig.\ \ref{fig:ent}

\subsection{Estimating \texorpdfstring{$\langle \bar{H_p} \rangle$}{the time averaged expectation of the problem Hamiltonian}}
\label{sec:hpPred}
In Sec.\ \ref{sec:beta} we saw how the temperatures associated with a CTQW could be found. By assuming the system is well modelled by a Gibbs distribution, it follows that
\begin{equation}
    \label{eq:hppred}
    \langle \bar{H_p} \rangle = -\frac{1}{\beta} \frac{\partial \ln \mathcal{Z}}{\partial \gamma},
\end{equation}
holds approximately true. The derivation behind this equation can be found in Appendix \ref{app:hp_reason}. For the Gaussian approximation this gives:
\begin{equation}
    \label{eq:hpnorm}
    \langle \bar{H_p} \rangle_{est} = -\frac{\gamma n \kappa_2 }{n+\gamma^2 \kappa_2},
\end{equation}
which is optimised by 
\begin{equation}
    \gamma_{est}^{opt}=\sqrt{\frac{n}{\kappa_2}}.
\end{equation}
The details can be found in Appendix \ref{sec:norm}. Note that this is the same result as the heuristic of maximising the dynamics (i.e. Eq. \ref{eq:heur_gam}). The same approach can be taken for the EMG model. The corresponding result is too cumbersome to be written out in full.

\begin{figure}
    \centering
    \includegraphics[width=0.48\textwidth]{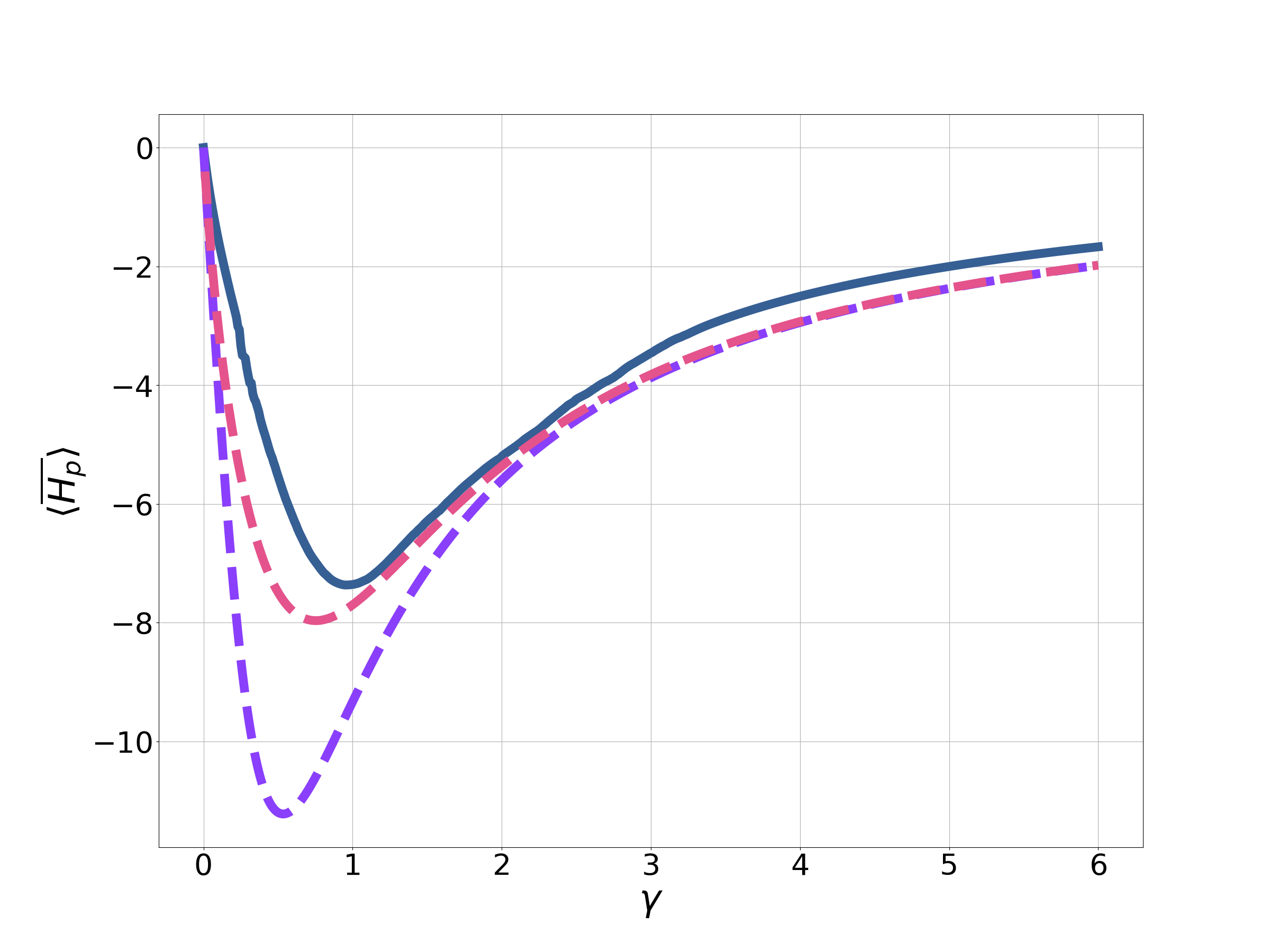}
    \caption{The solid blue line shows $\langle \bar{H_p} \rangle$ for a 12-qubit binomial graph. The dashed purple (pink) line shows the prediction from Eq.\ \ref{eq:hppred} assuming a Gaussian (EMG) DOS. }
    \label{fig:hpPred}
\end{figure}

Fig.\ \ref{fig:hpPred} compares $\langle \bar{H_p} \rangle$ to the predictions from the Gaussian and EMG predictions for a single 12-qubit binomial graph. Though both overestimate the performance of the CTQW, both provide reasonable estimates on the optimal $\gamma$. The EMG clearly provides a better model for $\langle \bar{H_p} \rangle$ than the Gaussian model. Indeed it is remarkably close to the direct numerical calculation of $\langle \bar{H_p} \rangle$. Note that although the thermal model appears to provide good estimates for $\langle \bar{H_p} \rangle$ for large and small $\gamma$, in this regime $\langle H_p(t) \rangle$ will not necessarily display steady-state behaviour. For example in Fig.\ \ref{fig:stime_g_small}, $\gamma$ is very small and there are significant oscillations present. However, this regime of $\gamma$, as discussed in Sec.\ \ref{sec:QWMC}, is unlikely to be of practical interest to CTQWs.

For the performance of $\gamma_{est}^{opt}$ the reader is referred to Sec.\ \ref{sec:QWMC}. For the EMG distribution there are two important quantities to examine:
\begin{itemize}
    \item How much does the performance of $\langle \bar{H_p} \rangle$ change between the optimal $\gamma$ and the $\gamma$ predicted by the EMG DOS?
    \item What is the difference between the prediction of $\langle \bar{H_p} \rangle$ and the true value?
\end{itemize}

\begin{figure}
     \centering
     \begin{subfigure}[b]{0.48\textwidth}
         \centering
         \includegraphics[width=\textwidth]{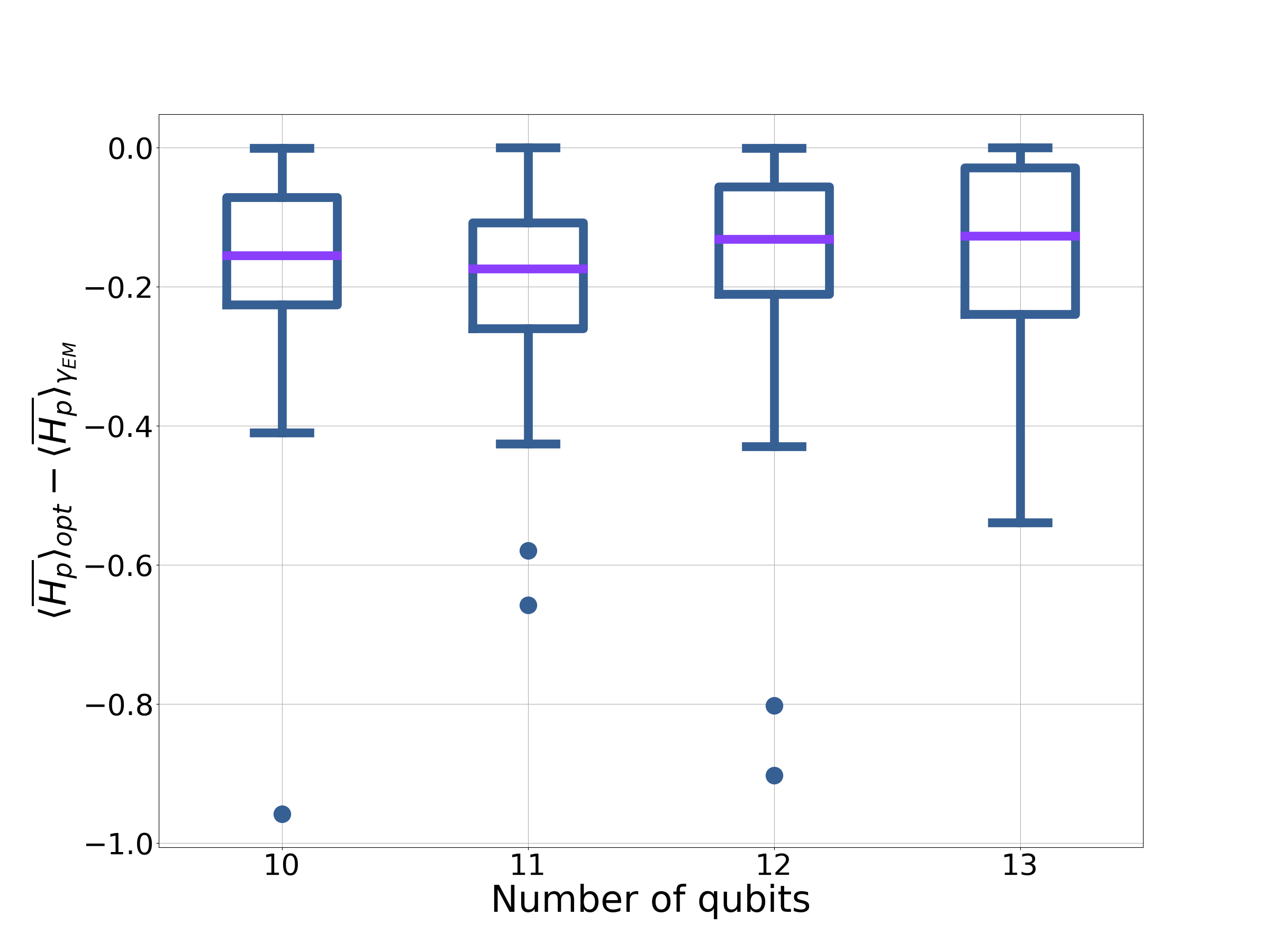}
         \caption{Binomial graphs}
         \label{fig:skewDhpRand}
     \end{subfigure}
     \vfill
     \begin{subfigure}[b]{0.48\textwidth}
         \centering
         \includegraphics[width=\textwidth]{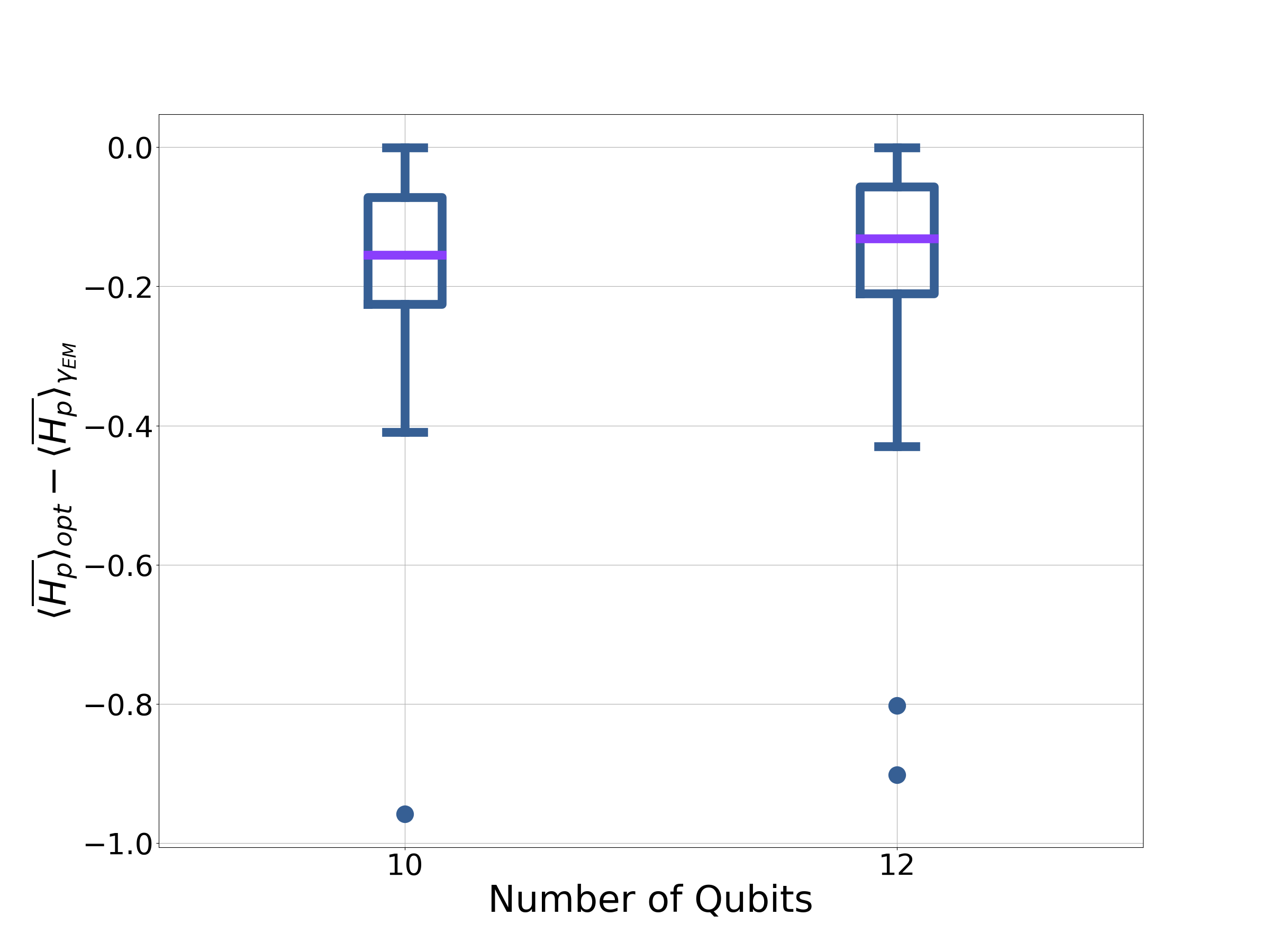}
         \caption{Three-regular graphs}
         \label{fig:skewDhpReg}
     \end{subfigure}
        \caption{The difference in $\langle \bar{H_p} \rangle$ between the optimal choice of $\gamma$ and the prediction from the EMG DOS ($\gamma_{EM}$). }
        \label{fig:skewDhp}
\end{figure}

The first of these is addressed for small problem sizes in Fig.\ \ref{fig:skewDhp}. The performance is improved over $\gamma_{heur}$ from simply balancing the drive and problem Hamiltonians, especially for binomial graphs. Hence, we believe we have found a good heuristic method for finding $\gamma$ for CTQWs applied to MAX-CUT.

\begin{figure}
    \centering
    \includegraphics[width=0.48\textwidth]{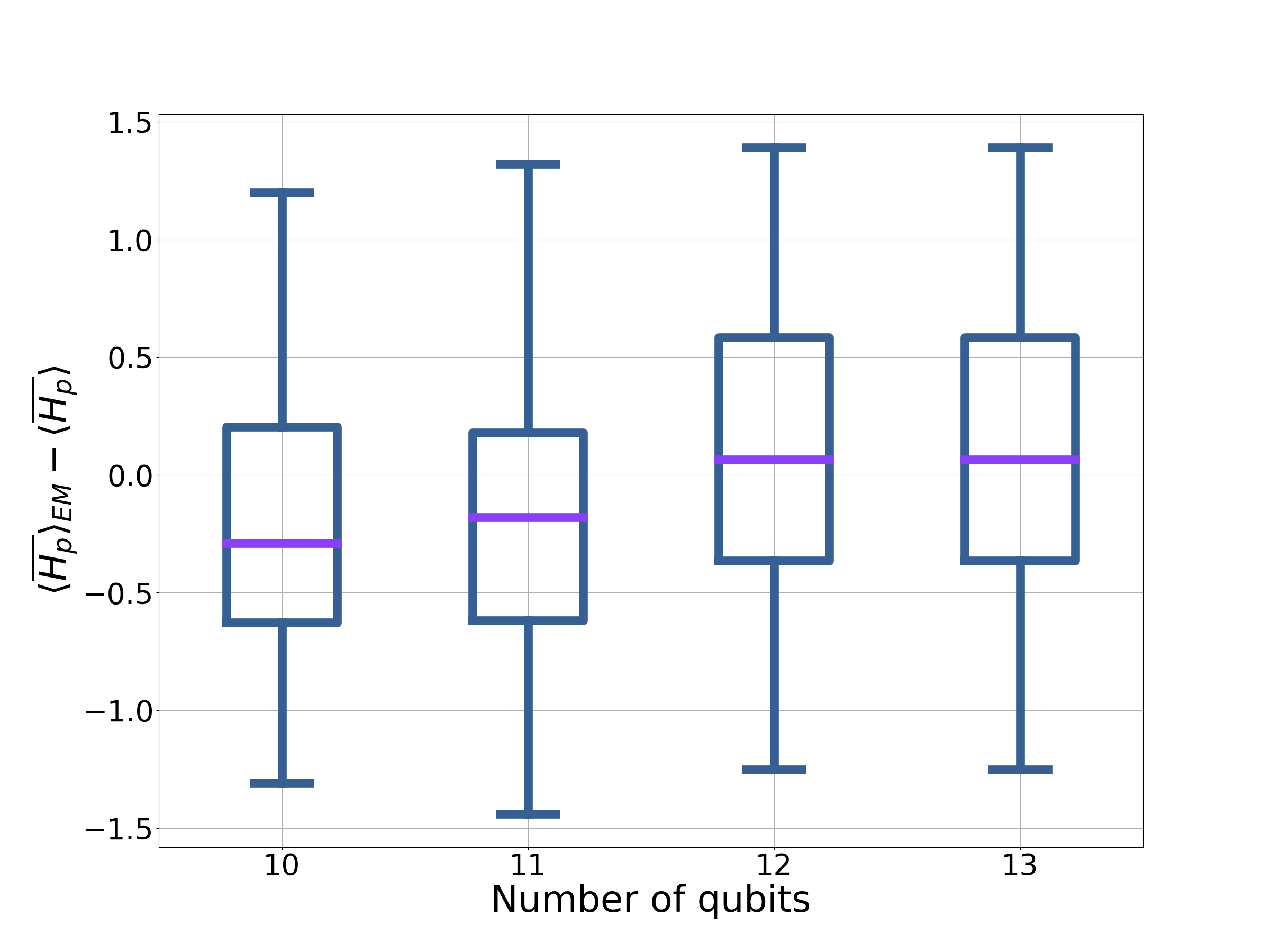}
    \caption{The difference between the prediction of $\langle H_p \rangle$ with an EMG DOS using Eq.\ \ref{eq:hppred} ($\langle H_p \rangle_{EM}$) and the true value. For each instance the value of $\gamma$ has been chosen such that $\langle H_p \rangle_{EM}$ is minimised. Each bar shows one hundred binomial graph instances.}
    \label{fig:emgHperr}
\end{figure}

\begin{figure}
    \centering
    \includegraphics[width=0.48\textwidth]{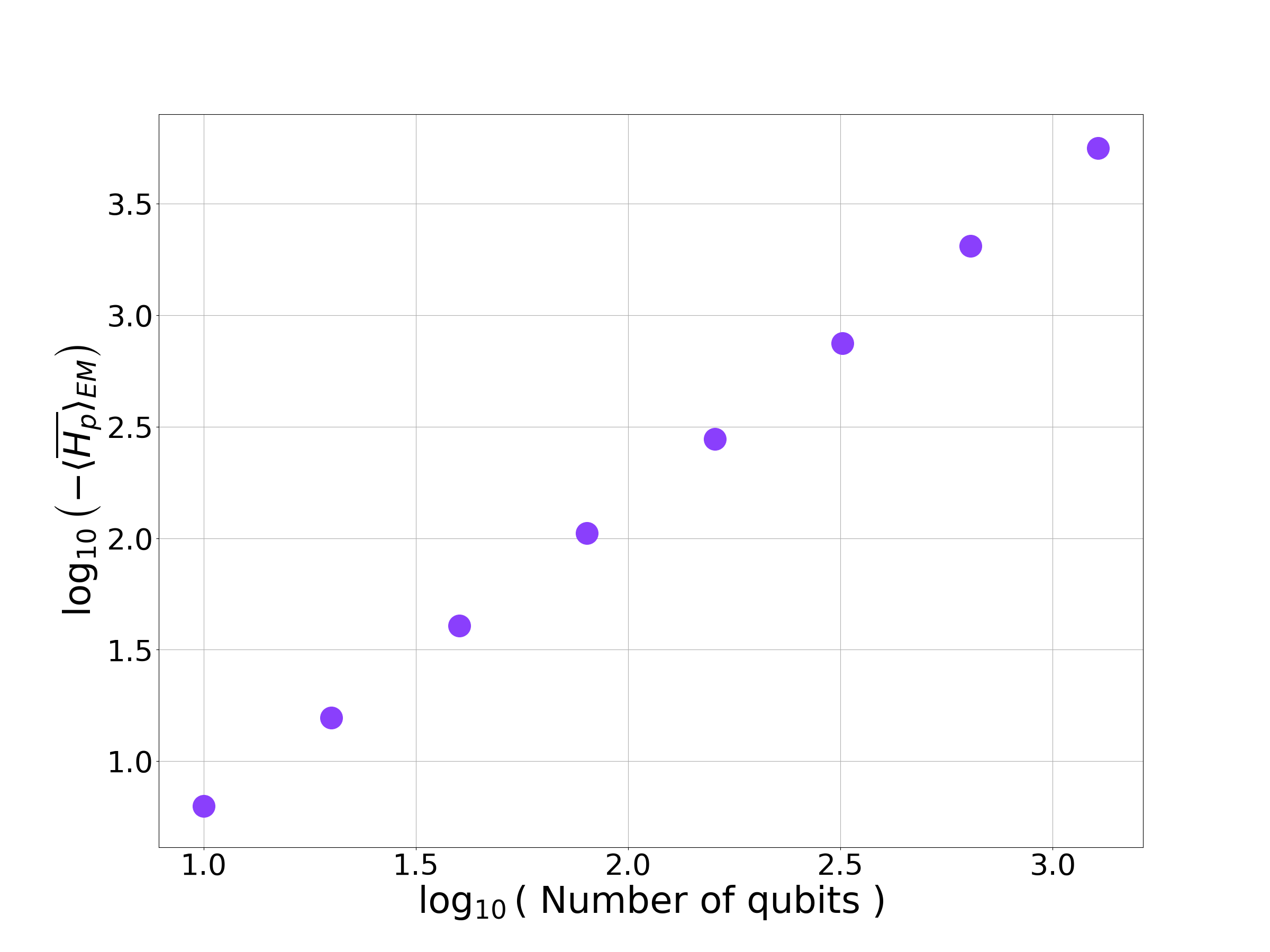}
    \caption{Estimated scaling of the CTQW for binomial graphs assuming thermalisation and an EMG DOS. Each data point shows the median performance of one hundred binomial graph instances.}
    \label{fig:emgHp}
\end{figure}

Fig.\ \ref{fig:emgHperr} shows the error in the performance between the true value of $\langle \bar{H_p} \rangle$ and the predicted value from the EMG approximation, for the $\gamma$ predicted to give the best possible performance from the EMG DOS. The error is relatively small for these small problem sizes. Given the tractability of Eq.\ \ref{eq:hppred}, and evidence of small errors, it is possible to estimate the performance for larger problem sizes. This is shown in Fig.\ \ref{fig:emgHp}.

By treating closed-system CTQWs as thermalising systems we have shown they thermalise. From here we have been able to provide useful heuristic choices for optimising $\gamma$ and to make practical predictions for $\langle \bar{H_p} \rangle$. For further discussion of the performance of CTQWs see Appendix \ref{app:per}. We have also provided an alternative approach to understanding CTQW that might be applicable to other combinatorial optimisation problems. 

\section{Multi-stage quantum walks}
\label{sec:MSQW}

\subsection{The set-up}
So far we have focused on time-independent Hamiltonians. In this section we introduce time-dependence into the CTQW, by implementing multi-stage quantum walks (MSQWs). In this case $\gamma$ is increased monotonically in a step-wise fashion. With each step of $\gamma$ the system is allowed to reach equilibrium. For instance an $l$-stage quantum walk would have the following schedule:
\begin{equation}
    \gamma(t)=\begin{cases*}
    \gamma_1 \text{ for } 0\leq t<t_1\\
    \gamma_2 \text{ for } t_1 \leq t<t_2\\
    \cdots\\
    \gamma_k \text{ for } t_{k-1} \leq t<t_k\\
    \cdots\\
    \gamma_l \text{ for } t_{l-1} \leq t\\
    \end{cases*}
\end{equation}
with $\gamma_l>\dots \gamma_k\dots>\gamma_2>\gamma_1$. This schedule can be seen as being motivated by approaches like Quantum Annealing \cite{Alb18}.
For each step, while $\gamma$ is held constant energy is conserved. Therefore we can modify Eq.\ \ref{eq:betaFix} to be:
\begin{equation}
    \label{eq:betaFix_MSQW}
    \Tr \left[H_{QW}^{(k)} \rho_{\beta}\left(H_{QW}^{\left(k\right)}\right)\right]=\bra{\psi\left(t_{k-1}\right)}H_{QW}^{(k)}\ket{\psi \left( t_{k-1}\right)},
\end{equation}
where $H_{QW}^{(k)}$ is the Hamiltonian during the $k^{\text{th}}$ stage of the quantum walk (i.e. with $\gamma_k$) and $\ket{\psi \left(t\right)}$ is the state-vector associated with the MSQW.
\begin{figure}
   \centering
    \includegraphics[width=0.48\textwidth]{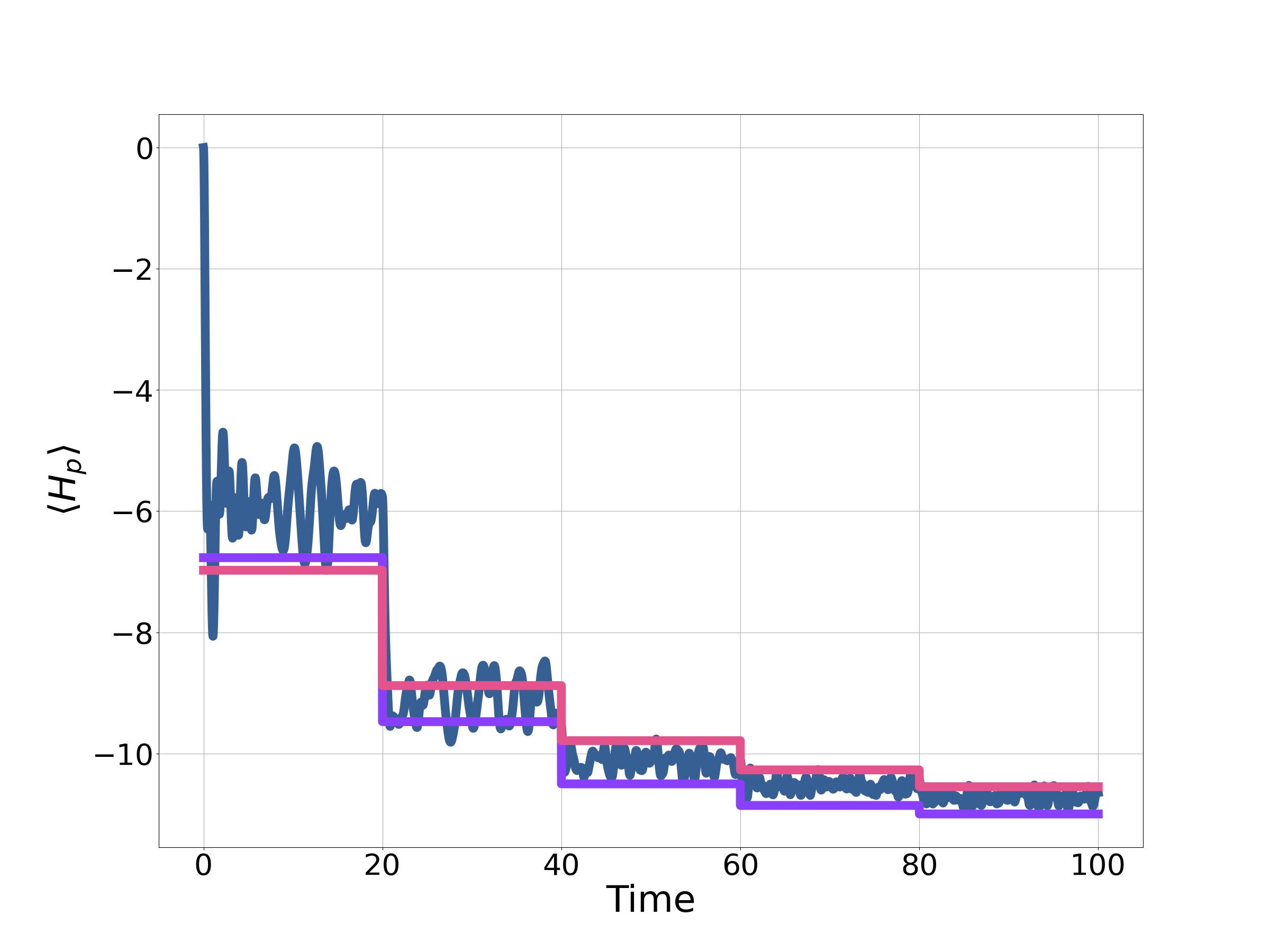}
    \caption{A five-stage CTQW on a 12-qubit binomial graph. The purple line is the numerical prediction assuming that the MSQW is well modelled by a CTQW. The pink line shows the prediction from Eq.\ \ref{eq:hppred} and modelling the DOS as an EMG distribution.}
    \label{fig:MSQWan}
\end{figure}
Fig.\ \ref{fig:MSQWan} shows a five-stage CTQW for a 12-qubit binomial graph. Each stage increases $\gamma$ by 0.5, starting with 0.5. Each stage has a duration of twenty units of time. With each stage $\langle H_p \rangle$ (the solid blue line) improves and quickly tends to an approximate steady-state. Solving Eq.\ \ref{eq:betaFix_MSQW} to find the temperature and associated performance gives the purple line in the figure.  For this instance, we have good qualitative and reasonable quantitive agreement between the thermal prediction and the Schr\"odinger equation.  During the first stage of the MSQW, the thermal prediction is quite far from the prediction of the Schr\"odinger equation, suggesting the state is far from thermal. This is to be expected as $\gamma$ is small and the driver Hamiltonian dominates, breaking the assumption that the DOS is continuous. Despite this, as $\gamma$ is increased the system thermalises. Perhaps most interestingly, as $\gamma$ is further increased the state remains thermal despite the problem Hamiltonian becoming more dominant. The inverse temperatures associated with this MSQW are $\beta=1.21, 0.72, 0.53, 0.41, 0.32$. This corresponds to heating the system, suggesting that despite the final stage providing the best performance, it might be the distribution easiest to simulate  classically \cite{Cro20}. Given the thermal behaviour, it might be reasonable to assume that the techniques developed earlier in this paper for CTQW can be applied to MSQW.

\subsection{Making predictions}

In Sec.\ \ref{sec:QWTL} the tools for making predictions about CTQWs were developed. Here we apply these tools to the MSQW. Here we only use the EMG distribution to model the DOS, since this also captures the Gaussian model.

To make predictions Eq.\ \ref{eq:betaFix_MSQW} needs to be approximated to find $\beta$. Since finding $\ket{\psi(t)}$ is likely to be numerically intractable for large systems we approximate Eq.\ \ref{eq:betaFix_MSQW} by:
\begin{align}
    \label{eq:lhsMSQW}
    \bra{\psi\left(t_{k-1}\right)}&H_{QW}^{(k)}\ket{\psi \left( t_{k-1}\right)} \nonumber \\
    =\bra{\psi\left(t_{k-1}\right)}&H_d+\gamma_k H_p\ket{\psi \left( t_{k-1}\right)} \nonumber \\
     =\bra{\psi\left(t_{k-1}\right)}&H_d+\gamma_{k-1} H_p\ket{\psi \left( t_{k-1}\right)} \nonumber\\
    &+\left(\gamma_k-\gamma_{k-1}\right)\bra{\psi\left(t_{k-1}\right)} H_p\ket{\psi \left( t_{k-1}\right)} \nonumber \\
    \approx \langle H_{QW}^{(k-1)}& \rangle_{\left(k-1\right)} +\left(\gamma_k-\gamma_{k-1}\right)\langle \bar{H_p} \rangle_{\left(k-1\right)},
\end{align}
where $\langle \cdot \rangle_{k-1}$ denotes the expectation during the $(k-1)^{\text{th}}$ stage. Therefore, we can estimate Eq.\ \ref{eq:betaFix_MSQW} by recursively evaluating  Eq.\ \ref{eq:lhsMSQW} using Eq.\ \ref{eq:hppred} to calculate $\langle \bar{H_p} \rangle_{\left(k-1\right)}$. In Fig.\ \ref{fig:MSQWan} the solid pink line shows the prediction, using Eq.\ \ref{eq:lhsMSQW}, to good agreement.

\begin{figure}
   \centering
    \includegraphics[width=0.48\textwidth]{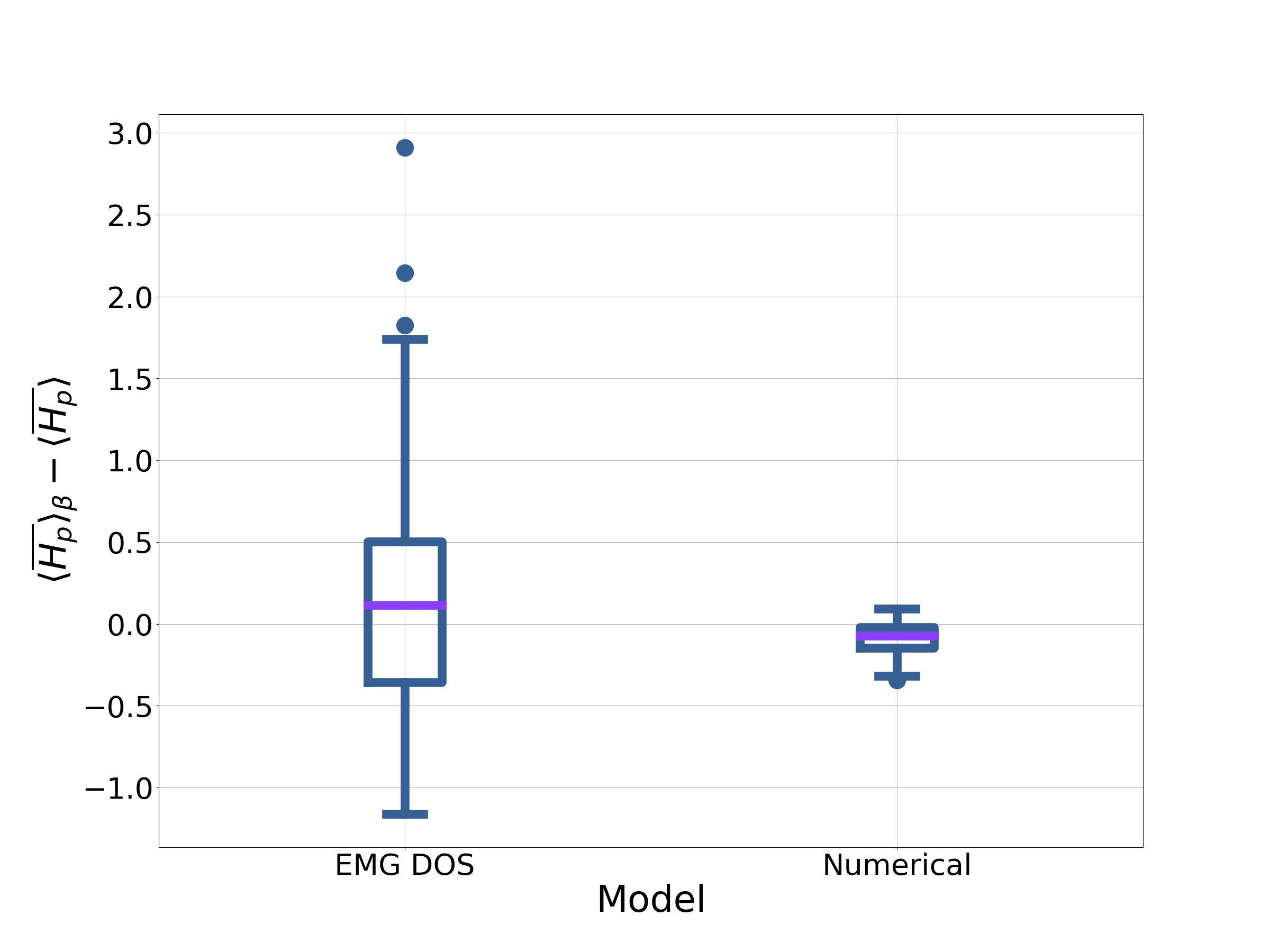}
    \caption{The difference between the stationary state calculated from the Schr\"odinger equation and a thermal state for five-stage CTQW on 100 12-qubit binomial graphs. ``EMG DOS'' refers to the prediction with the energy fixed by Eq.\ \ref{eq:lhsMSQW}. The numerical approach gives the prediction from using the numerically determined energy, numerically fixing the temperature and calculating $\langle H_p \rangle$.}
    \label{fig:MSQWerr}
\end{figure}

In Fig.\ \ref{fig:MSQWerr} we consider the final value of $\langle \bar{H_p} \rangle$  for a hundred five-stage CTQW on 12-qubit binomial graphs compared to the analytical prediction (i.e. Eq.\ \ref{eq:lhsMSQW}) and the numerical prediction assuming thermalisation. The schedule  is the same as the schedule used in Fig.\ \ref{fig:MSQWan}. Consider first the numerical prediction. Assuming thermalisation, the correct final steady state for all the problem instances is predicted within an error of 0.5. In contrast the EMG DOS model only achieves this error bound for approximately 50\% of the instances.  This approach suffers from cumulative errors in $\langle H_p \rangle$ at each stage of the MSQW (e.g. Eq.\ \ref{eq:lhsMSQW}), which can result in large errors in the prediction.

Here we have numeric evidence for MSQW exhibiting thermalisation. The EMG DOS model struggles to capture the performance for all instances, as well as the time-independent case. This is due to the difficulty in determining the energy of the system. To improve this approach a better model for the DOS or prediction of $\langle H_p \rangle$ needs to be developed, especially for the small $\gamma$ case.

\section{Implications for gate-based implementation}
\label{sec:QWF}
\begin{figure}
    \centering
    \includegraphics[width=0.48\textwidth]{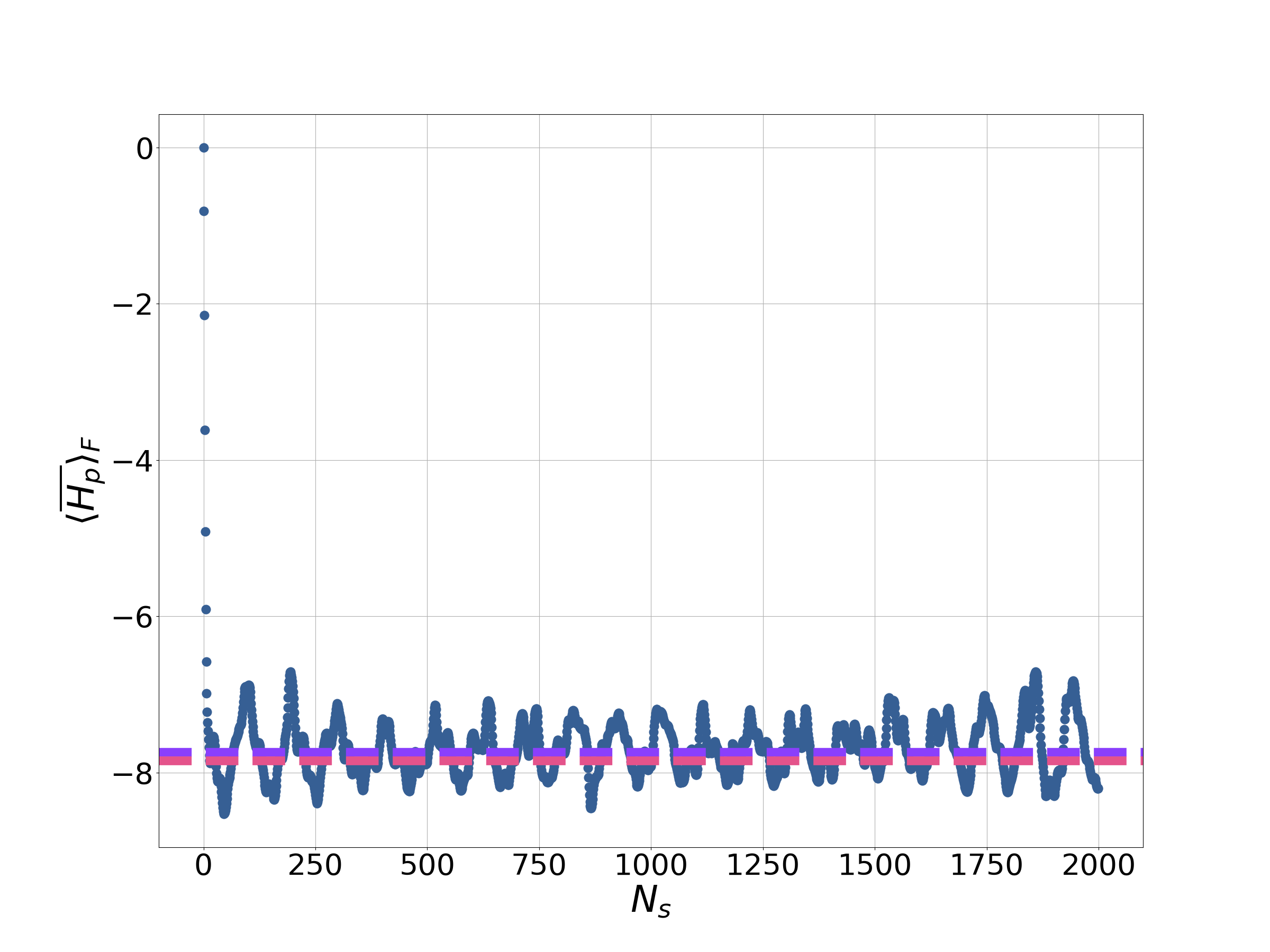}
    \caption{A Floquet system simulating a CTQW for a 12-qubit binomial graph, with $\gamma=1$ and $\tau=0.1$. The dashed purple line shows the time averaged value (i.e. Eq.\ \ref{eq:floqss}). The dashed pink line shows the prediction from Eq.\ \ref{eq:hppred}, assuming an EMG DOS. }
    \label{fig:floq_inst}
\end{figure}

\begin{figure}
     \centering
     \includegraphics[width=0.48\textwidth]{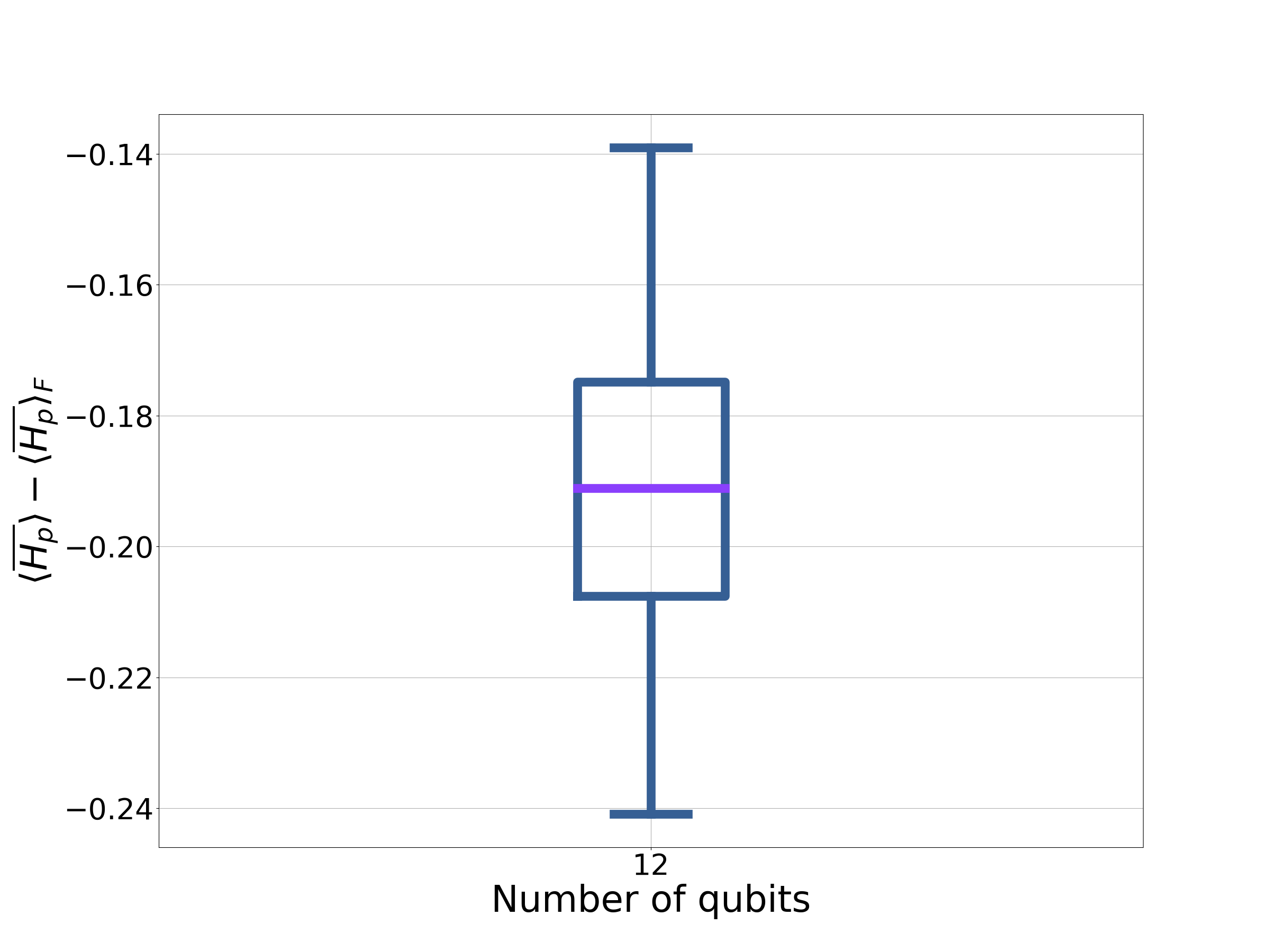}
     \caption{The reduced performance resulting from implementing the CTQW as a Floquet system for one-hundred 12-qubit binomial graphs. For the Floquet systems $\tau=0.2$. For all the problem instances $\gamma=1$.}
     \label{fig:red_per_f}
 \end{figure}

One approach for realising a CTQW might be to simulate it on a gate-based quantum computer. The simplest approach to achieve this is to Trotterise the evolution \cite{nielsen_chuang_2010}, such that the Hamiltonian alternates between $H_d$ and $H_p$. The resulting state-vector is prepared by the quantum computer:
\begin{equation}
    \label{eq:trot}
    \ket{\psi_{F}(t)}=\left(e^{-i \tau H_d/2}e^{-i \gamma \tau H_p/2}\right)^{N_s}\ket{+},
\end{equation}
where $t=N_s \tau$ and $N_s$ is is an integer. Here $N_s$ acts a proxy for circuit depth. Setting $\tau$ presents a trade-off between minimising gate-depth with a large $\tau$, and accuracy of the simulation that requires $\tau$ to be small. Alternatively, Eq.\ \ref{eq:trot} could be viewed as a QAOA \cite{far14} schedule with each step having the same set of angles. In the previous sections the Hamiltonians have been held constant for extended intervals in time. This has provided a local conserved quantity, resulting in a finite-temperature thermal state. The Trotterised system is not time-independent and energy is not conserved. In this section we investigate Eq.\ \ref{eq:trot}. Since the Hamiltonian is periodic in time, it is an example of a Floquet system (a very brief introduction can be found in Appendix \ref{app:flo}).

To analyse the long time behaviour of the Floquet system we evaluate:
\begin{equation}
    \label{eq:floqss}
    \langle \bar{H_p} \rangle_F=\sum_{\epsilon_\alpha=\epsilon_\beta} c_\alpha c_\beta^*\bra{\phi(\tau)_\beta}H_p\ket{\phi(\tau)_{\alpha}},
\end{equation}
where $\ket{\phi(t)_\alpha}$ are the Floquet modes, $\epsilon_\alpha$ the corresponding quasi-energies, and $c_\alpha=\bra{\phi(0)_\alpha}\ket{+}$ \cite{GRI98}. This is illustrated in Fig.\ \ref{fig:floq_inst} for a 12 qubit binomial graph with $\tau=0.1$ and $\gamma=1$. The solid blue line shows the Schr\"odinger equation. Within a few intervals of $\tau$ the system has already thermalised to $\langle \bar{H_p} \rangle_F=-7.74$. This is shown by the dashed purple line (i.e.\ the time averaged value from Eq.\ \ref{eq:floqss}).  The performance of the Floquet system  is marginally worse than the value of $\langle \bar{H_p} \rangle=-7.79$  for the corresponding CTQW. Indeed Fig.\ \ref{fig:red_per_f} shows the difference between $\langle \bar{H_p} \rangle_F$ and the corresponding value for a CTQW for one-hundred binomial qubit instances. For the Floquet instances $\tau=0.2$. For each problem instance the CTQW provided the better performance.

\begin{figure}
     \centering
     \includegraphics[width=0.48\textwidth]{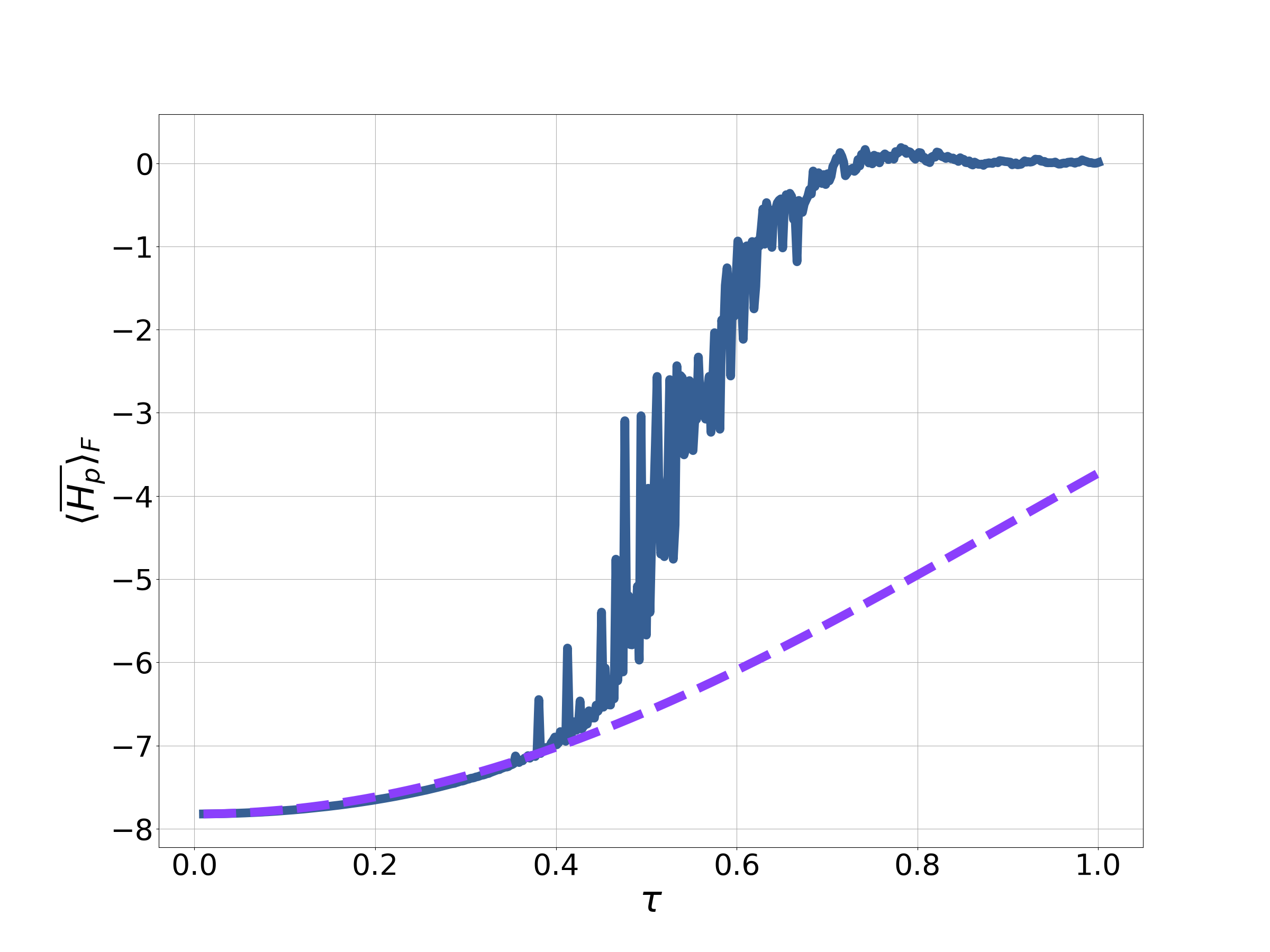}
     \caption{The steady state value for a single 12-qubit binomial graph with $\gamma=1$ as $\tau$ is increased. The blue line shows the prediction from Eq.\ \ref{eq:floqss}. The dashed purple line shows the prediction using the time-independent Hamiltonian shown in Eq.\ \ref{eq:floqConst}.}
     \label{fig:floq_sweep}
 \end{figure}

Depending on the size of $\tau$, we might expect two things to occur for $\langle \bar{H_p}\rangle_F$ \cite{Ued20}. For small $\tau$ the system does not have time to thermalise, so $\ket{\psi_{F}(t)}$ mimics the behaviour of a CTQW. For large $\tau$, the system thermalises at each step. Since energy is not conserved, the system tends to the infinite temperature Gibbs state, resulting in $\langle \bar{H_p} \rangle_F=0$. This is shown by the blue line in Fig.\ \ref{fig:floq_sweep} for a single 12-qubit binomial graph as $\tau$ is increased.

\subsection{The thermal model}
 \begin{figure}
     \centering
     \includegraphics[width=0.48\textwidth]{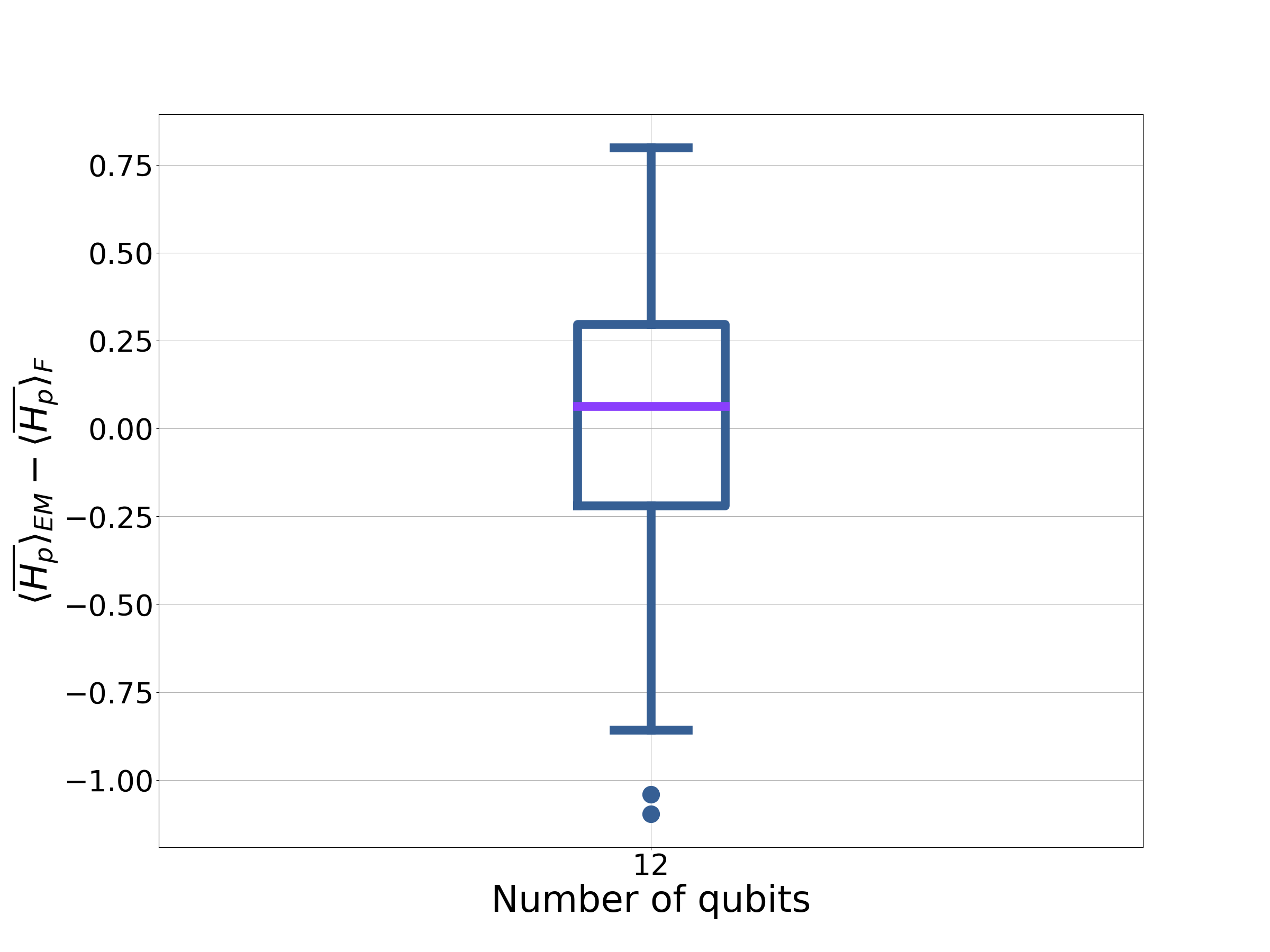}
     \caption{The difference between the true steady state and the steady state predicted assuming the system is well modelled by a time-independent Hamiltonian with a EMG distribution. The figure shows the results for one hundred 12-qubit binomial graph instances with $\gamma=1$ and $\tau=0.2$.  }
     \label{fig:floq_err_many}
 \end{figure}

In this section we provide an approximate time-independent model to capture the stroboscopic behaviour of the Floquet system  assuming the Trotter-step is small. Previous works have focused on using a Magnus expansion \cite{DAle13,DAle14,Laz14}, for ease of computation here we take an alternative approach.

The stroboscopic behaviour is captured  by the single period unitary:
\begin{equation}
    U_F(\tau)=e^{-i \tau H_d/2}e^{-i \gamma \tau H_p/2},
\end{equation}
from here it is a trivial exercise to recover the associated Hamiltonian:
\begin{align}
    H_F(t)=&i U_F^{\dagger}(t) \frac{\dd}{\dd t} U_F(t)\\
    =&\frac{1}{2} \left[U_p^\dagger (t) H_d U_p(t) +\gamma H_p \right]
\end{align}
where $U_p(t)=e^{-iH_p \gamma t/2}$.  Assuming $\tau$ is small, making the approximation $H_F(t)\approx H_F(\tau/2)$ gives a time-independent Hamiltonian. We can also neglect the constant scaling of the Hamiltonian that is not going to change the long time behaviour. Hence we are left with the following time-independent Hamiltonian:
\begin{equation}
   \label{eq:floqConst}
   \tilde{H}_F=U_p^\dagger (\tau/2) H_d U_p(\tau/2) +\gamma H_p 
\end{equation}
From here we can follow the steps outlined in Sec.\ \ref{sec:QWTL}  for the CTQW. Note that
\begin{align}
   \Tr \tilde{H}_F^k(t) =& \Tr\left[U_p^\dagger (\tau/2) H_d U_p(\tau/2) + \gamma H_p \right]^k \nonumber\\
    =& \Tr\left[ H_d +\gamma H_p \right]^k \nonumber \\
   =&  \Tr  H_{QW}^k ,
\end{align}
so we can use the moments for the DOS derived for the CTQW. Therefore any change in performance must come from the energy of the initial state:
\begin{equation}
\label{eq:enCon}
\bra{+}\tilde{H}_F\ket{+}=-\sum_{k=1}^n \cos^{\deg(k)}(\gamma \tau/4),
\end{equation}
where $\deg(k)$ is the degree of the $k^{\text{th}}$ node. Notably, for binomial graphs (or any graph where $\deg(k)$ increases with $n$) as $n$ goes to infinity, $\bra{+}\tilde{H}_F\ket{+}$ goes to zero. This corresponds to an infinite temperature thermal state. Consequently, in the thermodynamic limit, under the above assumptions, any attempt to approximate a CTQW with a Floquet system will fail. In contrast, regular graphs will have more stability. The numerical prediction for thermalising under $\tilde{H_F}$ can be seen in Fig.\ \ref{fig:floq_sweep} (dashed purple line) for a single 12-qubit binomial graph. The numerical prediction agrees well with the true numerical value of $\langle \bar{H_p} \rangle_F$ (solid blue line) up until the cross-over to an infinite temperature Gibbs state. 

As established in the earlier section on CTQWs, they correspond to Gibbs states. Therefore, we use the same model as for the CTQW with a modified energy constraint (i.e. Eq.\ \ref{eq:enCon}). This is shown by the dashed pink line in Fig.\ \ref{fig:floq_inst}, for a single 12-qubit binomial graph to good agreement. The results for one hundred 12-qubit binomial graphs with $\tau=0.2$ and $\gamma=1$ can be seen in Fig.\ \ref{fig:floq_err_many}.  As shown in the figure the median error between the numerically calculated value of $\langle \bar{H_p} \rangle_F$  and the analytic value, assuming an EMG DOS and thermalisation, is close to zero. 

\begin{figure}
     \centering
     \includegraphics[width=0.48\textwidth]{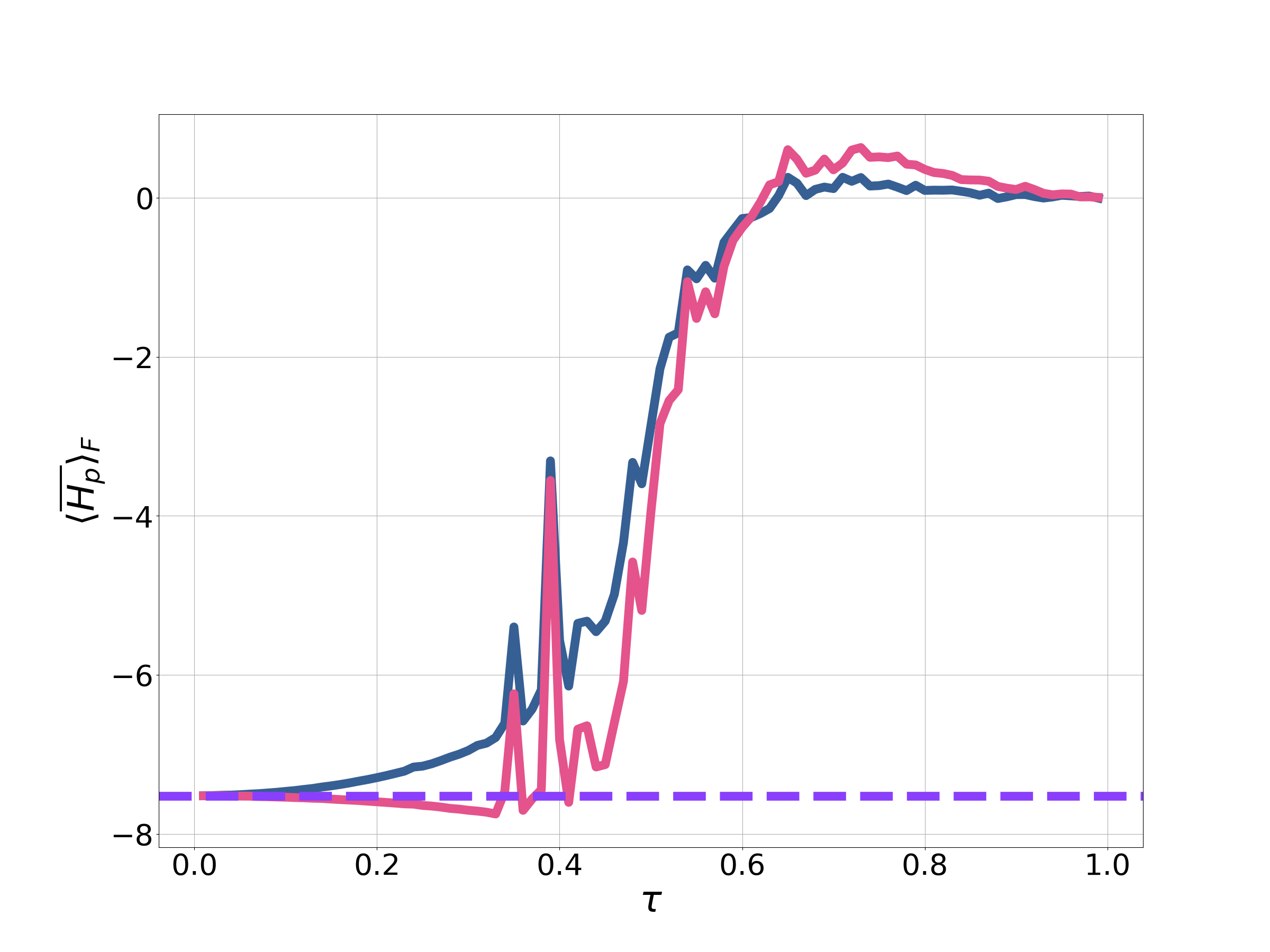}
     \caption{Improved performance from changing the initial state.The dashed purple line shows the steady-state value for a CTQW. The solid blue line shows the steady state of the Floquet system with initial state $\ket{+}$. The solid pink line shows the steady state of the Floquet system with initial state $U_p^\dagger(\tau/2)\ket{+}$.}
     \label{fig:floq_rot}
 \end{figure}

Perhaps most usefully, this thermal model provides some intuition on how mitigate Trotter errors for this system. Essentially, we wish to start in the effective ground-state of the Floquet system. Eq.\ \ref{eq:floqConst}, suggests this might be achieved by changing the initial state from $\ket{+}$ to $U_p^\dagger(\tau/2)\ket{+}$.  This is illustrated for a single 12-qubit binomial graph in Fig.\ \ref{fig:floq_rot} with $\gamma=1$.  The dashed purple line shows the steady-state value for a CTQW. The solid blue shows the steady state of the Floquet system with the initial state being the $\ket{+}$ state. The pink line shows the same Floquet system with the modified initial state $U_p^\dagger(\tau/2)\ket{+}$. Indeed, for the short time steps, the modified initial state better matches the CTQW. Indeed it even provides a slight advantage for a small interval in $\tau$.

\section{Conclusion}

Throughout this paper we have attempted to understand the performance of CTQWs outside of what is classically simulable for non-integrable models.  For the short-time limit, we were able to exploit the effective low dimensionality to demonstrate that the performance can be characterised by underlying graph properties, such as the number of edges in the graph.

For the steady-state we conjectured that the system thermalises. This means despite the state-vector consisting of $2^n$ complex amplitudes, for a given $\gamma$, the value of $\langle H_p\rangle$ will depend on one real number, the energy. This provides some insight into the computational mechanisms  involved in CTQWs.

Assuming thermalisation,  classical statistical physics provides an alternative route to understanding CTQWs for MAX-CUT on a broad range of graphs within a closed-system setting. By associating the unitary dynamics with a temperature, it provides an alternative route to achieve the same value of $\langle \bar{H_p} \rangle$ , either through dissipative dynamics or classical simulation.  Here we have introduced an EMG distribution to account for some of the frustration in the model to make analytic predictions.  By exploiting this model we were able to find reasonable estimates for the optimal choice of $\gamma$, that utilised properties of the underlying graph. Importantly, we were able to make predictions far away from what it is easy to directly and completely simulate. We have also demonstrated some preliminary results for MSQWs.

Further to this we were able to extend these results to Floquet systems as an approach to implementing CTQWs. For all the problems considered the Floquet system performed worse than the corresponding CTQW with the same $\gamma$. Our results predict that in the thermodynamic limit, MAX-CUT on binomial graphs will thermalise to an infinite temperature Gibbs state, independent of the step size. This is consistent with known work on thermalisation in Floquet systems \cite{Ued20}. The Floquet systems considered  in this paper can be thought of as a QAOA \cite{far14} ansatz with restricted angle choices.  For QAOA we wish to avoid the system thermalising to an infinite temperature Gibbs state, hence it would be prudent to avoid a Floquet ansatz. Further to this it is reasonable to assume large step sizes might be detrimental for high-depth QAOA circuits. It is has been observed that a QAOA circuit with randomly chosen step sizes also thermalises to an infinite temperature state \cite{Du22}, suggesting that thermalisation is not solely the consequence of the discrete time symmetry of the Floquet system. How far the results of this paper might extend to other NISQ algorithms such as QAOA and Quantum Annealing is an ongoing area of research. 

\section*{Acknowledgements}
We gratefully acknowledge Itay Hen, Michael Kolodrubetz, Matthew Cooper, Henry Chew, Natasha Feinstein, Leon Guerrero and Pias Tubsrinuan for inspiring discussion and helpful comments.  Robert Banks would like to acknowledge Orbyts initiative for providing the opportunity to collaborate with an outstanding team of students at Newham Collegiate Sixth Form Centre. Many thanks to Serena Maugeri, Abbie Bray, and William Dunn for making this possible and the Mathematical and Physical Sciences faculty at UCL through the ``Take bold action for inclusion 2022-2023'' grant for funding it. The authors acknowledge the use of the UCL Myriad High Performance Computing Facility (Myriad@UCL), and associated support services, in the completion of this work. This work was supported by the Engineering and Physical Sciences Research Council through the Centre for Doctoral Training in Delivering Quantum Technologies [grant number EP/S021582/1]. Dan Browne was supported by the Engineering and Physical Sciences Research Council [grant numbers EP/S005021/1 and EP/T001062/1]. Paul Warburton was supported by the Engineering and Physical Sciences Research Council [grant number EP/W00772X/2]. For the purpose of open access, the author has applied a Creative Commons Attribution (CC BY) licence to any Author Accepted Manuscript version arising.

\bibliographystyle{quantum}
\bibliography{references}

\onecolumn\newpage
\appendix

\section{Details for the short time limit}
\label{app:tor}

\subsection{Calculating the torsion}
Here we provide the details for the short time limit for CTQW. Firstly, we calculate the torsion, and the related quantity curvature. The curvature is a local measure of how much the quantum evolution deviates from a geodesic. These are conserved quantities valid for the whole evolution. We then show how this leads to short time predictions on $\langle H_p \rangle$ for these problems in Sec.\ \ref{sec:ste_deets}.

From Laba et al.\ \cite{Lab17} the curvature of the wave-function, for a time-independent Hamiltonian is given by:
\begin{equation}
    \mathcal{C}=\langle \Delta H^4\rangle - \langle \Delta H^2\rangle^2,
\end{equation}
where $\Delta H= H-\langle H \rangle$. The torsion is given by:
\begin{equation}
    \mathcal{T}=\langle \Delta H^4\rangle - \langle \Delta H^2\rangle^2-\frac{\langle \Delta H^3\rangle^2}{\langle \Delta H^2\rangle},
\end{equation}
Therefore, calculating the torsion and curvature reduces to evaluating $\langle \Delta H^j \rangle$, for $j=2,3,4$. Since these are all conserved quantities, we can take the expectation with respect to the initial state, $\ket{+}$. The rest of this section details the rather tedious steps required to find these expectation values. Those willing to take the expression for $\mathcal{T}$ at face-value are invited to skip to Sec.\ \ref{sec:ste_deets}. 

The first step is to find $\langle H \rangle$:
\begin{align*}
    \langle H \rangle & = \bra{+} H_d+\gamma H_p \ket{+}\\
    & = -n,
\end{align*}
since each $Z_iZ_j$ in $H_p$ will flip qubits $i$ and $j$ to the minus state. Consequently, only $H_d$ contributes to this expectation value.

Consider now:
\begin{align*}
    \langle \Delta H^2 \rangle & = \langle H^2 \rangle-\langle H \rangle^2\\
    & = \langle H_d^2 \rangle + \gamma^2 \langle H_p^2 \rangle+ \gamma \langle H_p H_d+H_d H_p \rangle-\langle H \rangle^2\\
    &=n^2+\gamma^2\kappa_2+0+0-n^2\\
    &=\gamma^2 \kappa_2,
\end{align*}
where $\kappa_2=\abs{E}$ is the number of edges in the graph.  It was pointed out by Gnatenko et al.\ \cite{Gna22}, that $\langle H_p^2 \rangle$ is equal to the number of edges in the graph. To see why this is the case, we can write out this term explicitly to give:
\begin{equation*}
    \langle H_p^2 \rangle = \sum_{i_1,j_1,i_2,j_2} \bra{+} Z_{i_1}Z_{j_1}Z_{i_2}Z_{j_2} \ket{+},
\end{equation*}
where the sum is over the ordered pairs $ \left(i_k,j_k\right)$, for $k=1,2$. The only possible way a term in the sum can be non-zero is if $i_1=i_2$ and $j_1=j_2$. Therefore the number of non-zero terms corresponds to the number of edges in the graph. The rest of the terms in $\langle \Delta H^2 \rangle$ are relatively trivial to calculate. 

Extending the above logic, Gnatenko et al.\ \cite{Gna22} demonstrated that:
\begin{equation}
    \label{eq:Hp3}
    \langle H_p^3 \rangle= 6 \kappa_3
\end{equation}
where $\kappa_3$ is the number of triangles in the graph, and
\begin{equation}
    \label{eq:hp4}
    \langle H_p^4 \rangle= \kappa_2+3\kappa_2\left(\kappa_2-1\right)+24 \kappa_4,
\end{equation}
where $\kappa_4$ is the number of squares in the graph.

Using the above results we can calculate $\langle \Delta H^3 \rangle$ and $\langle \Delta H^4 \rangle$. Starting with $\langle \Delta H^3 \rangle$:
\begin{equation}
    \langle \Delta H^3 \rangle=\langle H^3 \rangle -3 \langle H^2 \rangle \langle H \rangle +2 \langle H \rangle ^3.
\end{equation}

Calculating $\langle H^3 \rangle$ gives:
\begin{align*}
    \langle H^3 \rangle &=\langle \left(H_d+\gamma H_p \right)^3\rangle\\
    &=\langle H_d^3+\gamma^2 H_d H_p^2+\gamma^2 H_p^2H_d+\gamma^3 H_p^3+\gamma^2 H_p H_d H_p \rangle\\
    &=-n^3-2n\gamma^2\kappa_2+6\gamma^3 \kappa_3 +\gamma^2 \langle H_p H_d H_p\rangle,
\end{align*}
where we have neglected to write out terms that are trivially zero and made use of Eq.\ \ref{eq:Hp3}. To find $\langle \Delta H^3 \rangle$ it just remains to calculate $\langle H_p H_d H_p\rangle$:
\begin{equation}
   \label{eq:HpHdHp}
   \langle H_p H_d H_p\rangle=-\sum_{i_1,j_1,i_2,j_2,k} \bra{+} Z_{i_1}Z_{j_2}X_kZ_{i_2}Z_{j_2}\ket{+}.
\end{equation}
The $X_k$ in each term is not going to change which terms are non-zero. It will only introduce a minus sign if $k$ is $i_2$ or $j_2$. Hence $\langle H_p H_d H_p\rangle$ is equal to:
\begin{align*}
   &=-\sum_{i_2,j_2,k} \bra{+} Z_{i_2}Z_{j_2}X_kZ_{i_2}Z_{j_2}\ket{+}\\
   &=-\sum_{i_2,j_2\atop k\neq i_2,j_2}\bra{+} Z_{i_2}Z_{j_2}Z_{i_2}Z_{j_2}\ket{+}+\sum_{j_2} \bra{+} Z_{k}Z_{j_2}Z_{k}Z_{j_2}\ket{+}+\sum_{i_2}\bra{+} Z_{i_2}Z_{k}Z_{i_2}Z_{k}\ket{+}\\
  &=-(n-2)\kappa_2+\kappa_2+\kappa_2\\
  &=-(n-4)\kappa_2.
\end{align*}
Alternatively, we can argue as follows: in the absence of the $X_k$ term, Eq.\ \ref{eq:HpHdHp} would give $- n \kappa_2$. Accounting for the edge cases where $k=i_2$ or $k=j_2$ gives:
\begin{equation}
   \langle H_p H_d H_p\rangle=-(n-4)\kappa_2.
\end{equation}
Combining all of the above we are left with:
\begin{equation}
    \langle \Delta H^3 \rangle = 2 \gamma^2 \left(2 \kappa_2+3 \gamma \kappa_3 \right).
\end{equation}
To find $\mathcal{C}$ and $\mathcal{T}$, it remains to find $\langle \Delta H^4 \rangle$.
\begin{equation*}
    \langle \Delta H^4 \rangle=\langle H^4\rangle-4 \langle H \rangle \langle H^3\rangle + 6\langle H \rangle^2 \langle H^2\rangle  -4 \langle H \rangle^3 \langle H\rangle+\langle H \rangle^4
\end{equation*}
Evaluating $\langle H^4\rangle$ gives:
\begin{multline*}
    \langle H^4 \rangle =\langle H_d^4+\gamma^2\left( H_d^2 H_p^2+H_dH_pH_dH_p+ H_dH_p^2H_d
    +H_pH_d^2H_p+H_pH_dH_pH_d+H_p^2H_d^2\right)\\
    +\gamma^3 \left(H_dH_p^3+ H_p^2H_dH_p+ H_p^3H_d+H_pH_dH_p^2\right)+\gamma^4H_p^4 \rangle.
\end{multline*}
Again, we have neglected to write out the terms that evaluate to zero. Most of the terms in this expansion we have already evaluated, or are simple to evaluate. We are left with $\langle H_p H_d^2 H_p \rangle$, $\langle H_pH_dH_p^2 \rangle$, $\langle H_p^2H_dH_p \rangle$, and $\langle H_p^4\rangle$, the last of which can be evaluated with Eq.\ \ref{eq:hp4}. Turning now to $\langle H_p H_d^2 H_p \rangle$:
\begin{equation*}
    \langle H_p H_d^2 H_p \rangle=\sum_{i_1,j_1,i_2,j_2,k_1,k_2} \bra{+} Z_{i_1}Z_{j_1}X_{k_1}X_{k_2}Z_{i_2}Z_{j_2}\ket{+}.
\end{equation*}
Again the presence of $X$ in the terms of the sum is not going to change which terms evaluate to be non-zero. Fixing the edge $(i_2,j_2)$, if either one of the $X$ terms coincides with this edge and the other $X$ term does not, it contributes $-1$ to the sum; there are $4(n-2)$ ways of this happening. If both $X$ terms do not coincide with this edge then the term contributes 1; there are $(n-2)^2$ ways of this happening. Finally if both edges coincide, this edge contributes 1; there are four ways of this happening. The result is:
\begin{equation*}
    \langle H_p H_d^2 H_p \rangle=(n-4)^2 \kappa_2.
\end{equation*}

Thus, it remains to evaluate  $\langle H_pH_dH_p^2 \rangle$ and  $\langle H_p^2H_dH_p \rangle$. Note that:
\begin{equation*}
    \langle H_pH_dH_p^2 \rangle^*=\langle H_p^2H_dH_p \rangle.
\end{equation*}
To evaluate $\langle H_pH_dH_p^2 \rangle$, consider $\langle H_p^3 \rangle$. As before, the introduction of $H_d$ is not going to change which terms in the sum are non-zero, if we were to write out a similar sum for this term, as in Eq.\ \ref{eq:HpHdHp}. Therefore, $\langle H_pH_dH_p^2 \rangle$ is going to depend on the number of triangles in the graph. The introduction of $H_d$ is going to result in sign flips to some terms. The result is 
\begin{equation}
    \langle H_pH_dH_p^2 \rangle=-6\kappa_3(n-4).
\end{equation}

Finally, we can assemble all the expectation values to find
\begin{equation}
    \langle \Delta H^4 \rangle= \gamma^2\left[-2 \kappa_2(\gamma^2-8)+3\gamma^2\kappa_2^2+24 \gamma(2\kappa_3+\gamma \kappa_4)\right].
\end{equation}
The resulting expression for the curvature is:
\begin{equation}
    \mathcal{C}=2\gamma^2\left[ -\kappa_2\left(\gamma^2-8\right)+\gamma^2\kappa_2^2+12\gamma \left(2 \kappa_3+\gamma \kappa_4\right) \right].
\end{equation}
and for the torsion:
\begin{equation}
    \mathcal{T}=2\gamma^4\left[\kappa_2\left(\kappa_2-1\right)-\frac{18 \kappa_3^2}{\kappa_2} + 12 \kappa_4 \right].
\end{equation}
Note that the torsion is zero for a two-qubit system, with one edge, as expected from the spin-flip symmetry in the problem. The torsion is also zero for a ring of three qubits.

\subsection{Short-time estimate for the performance}
\label{sec:ste_deets}

In the previous section we calculated the torsion. The torsion measures how much the wave-function deviates from a two-dimensional space in a time $\Delta t$. The error from deviating from this space is given by \cite{Gna22}:
\begin{equation}
    \varepsilon_{2D}=\mathcal{T}\Delta t^4.
\end{equation}
with $\varepsilon_{2D}=0$ corresponding to remaining in the two-dimensional subspace and $\varepsilon_{2D}=1$ corresponding to departing the subspace. If $\varepsilon_{2D} \ll 1$, then the two-dimensional approximation holds well. That is for times $t\ll \/\mathcal{T}^{-1/4}$, we can apply a two-level approximation.

At $t=0$ the state of the system is $\ket{\psi \left(t=0\right)}=\ket{+}$. Integrating the system forward some infinitesimal time $\delta t$, gives  $\ket{\psi \left(t=\delta t\right)}=\left(I-iH\delta t\right)\ket{+}$.

Applying the Gram-Schmidt procedure to the two vectors, $\ket{\psi \left(t=0\right)}$ and $\ket{\psi \left(t=\delta t\right)}$, gives the orthonormal basis for the subspace:

\begin{align}
    \ket{e_0}&=\ket{+}\\
    \ket{e_1}&=\Delta H \ket{+}/\sqrt{\langle \Delta H^2 \rangle},
\end{align}
where the expectation is with respect to the $\ket{+}$ state.
Writing out $H$ in this subspace is:
\begin{equation}
    H^{(2D)}=
    \begin{pmatrix}
         \langle H \rangle & \frac{\langle H \Delta H\rangle}{\sqrt{\langle \Delta H^2 \rangle}}\\
         \frac{\langle \Delta H H\rangle}{\sqrt{\langle \Delta H^2 \rangle}} & \frac{\langle \Delta H H \Delta H \rangle}{\langle \Delta H^2 \rangle}
    \end{pmatrix},
\end{equation}
which evaluates to
\begin{align}
    H^{(2D)}&=
    \begin{pmatrix}
         -n & \sqrt{\gamma^2 \kappa_2}\\
         \sqrt{\gamma^2 \kappa_2} & 2\left(2+\frac{3\gamma \kappa_3}{\kappa_2}\right)-n
    \end{pmatrix}\\
    &\\
    &=\sqrt{\gamma^2 \kappa_2} X-\left(2+\frac{3\gamma \kappa_3}{\kappa_2}\right) Z,
\end{align}
neglecting the uninteresting term proportional to the identity. 

The problem Hamiltonian, within this space is:
\begin{equation}
    H_p^{(2D)}=
    \begin{pmatrix}
         \langle H_p\rangle & \frac{\langle H_p \Delta H\rangle}{\sqrt{\langle \Delta H^2 \rangle}}\\
         \frac{\langle \Delta H H_p\rangle}{\sqrt{\langle \Delta H^2 \rangle}} & \frac{\langle \Delta H H_p \Delta H \rangle}{\langle \Delta H^2 \rangle}
    \end{pmatrix},
\end{equation}
evaluating this gives:
\begin{equation}
    H_p^{(2D)}=
    \begin{pmatrix}
         0& \sgn\left(\gamma\right)\sqrt{\kappa_2}\\
         \sgn\left(\gamma\right)\sqrt{\kappa_2} &  \frac{6 \kappa_3}{\kappa_2}
    \end{pmatrix},
\end{equation}
where $\sgn\left(\gamma\right)$ denotes the sign of $\gamma$.

Having now explicitly evaluated the relevant operators in this subspace, it remains to calculate the expectation of the problem Hamiltonian, which is:

\begin{equation}
    \label{eq:Hp2D}
    \langle H_p^{(2D)} (t) \rangle=-4\gamma \kappa_2\frac{\sin^2 \omega t}{\omega^2},
\end{equation}
where 
\begin{equation}
    \omega^2=\gamma^2 \kappa_2+\frac{\left(2 \kappa_2+3 \gamma \kappa_3\right)^2}{\kappa_2^2}.
\end{equation}

\section{Spin-flip symmetry and the ETH}
\label{app:sfs}
In Sec.\ \ref{sec:QWMC} we pointed out MAX-CUT has a spin-flip symmetry. To recap the notation, given a CTQW for MAX-CUT with Hamiltonian $H$, then $\left[H,G\right]=0$, where:
\begin{equation}
    G=\prod_{i=1}^n X_i.
\end{equation}
Typically, the system starts in the $\ket{+}$ state so the evolution is restricted to the plus one eigenspace of $G$ for the entire evolution.

Despite this restriction, in Sec.\ \ref{sec:QWTL}, we used the `complete' Hamiltonian as opposed to the Hamiltonian projected onto the correct symmetry sector to apply the ETH. This was numerically verified in Sec.\ \ref{sec:therm_check}. We justify this choice on the following grounds \cite{Ess16}:
\begin{enumerate}
    \item $G$ is a global operator and corresponds to a global symmetry.
    \item We are interested in calculating the expectation values of local interactions only.
\end{enumerate}

It is well established within the ETH literature that local symmetries can prevent thermalisation \cite{Ess16}. In contrast the spin-flip symmetry is a global symmetry. In general, there are a large number of globally conserved quantities for evolution under a constant Hamiltonian. If $H$ belongs to a Hilbert Space $\mathcal{H}$ with dimension $\dim \mathcal{H}$, then there are at least $\dim \mathcal{H}$ globally conserved quantities \cite{Ess16}. To see this, let the projector onto each eigenstate of $H$ be denoted by $P_k$ with $k=1,\dots, \dim \mathcal{H}$, then \cite{Ess16}:
\begin{align}
    \left[H ,P_k\right]&=0\\
    \left[P_k,P_j\right]&=0.
\end{align}
Hence $\langle P_k \rangle$ are globally conserved quantities. Despite the large number of conserved quantities, we generally do not expect these global symmetries to have much impact on the dynamics. 

Though the above argument suggests we should be careful when considering the role of globally conserved quantities, it fails to consider the specific case of the spin-flip symmetry which we know restricts dynamics to the positive eigenspace of $G$ for a CTQW. Here we emphasise again that we are only interested in local quantities. A locally measurable consequence of the spin-flip symmetry is that:
\begin{align*}
\langle Z_i(t) \rangle =&\bra{+} U^\dagger Z_i(t) U \ket{+}\\
=&\bra{+} G U^\dagger Z_i(t) U G \ket{+}\\
=&\bra{+} U^\dagger G Z_i(t) G U \ket{+}\\
=&-\bra{+} U^\dagger G Z_i(t) G U \ket{+}\\
=&-\langle Z_i(t) \rangle,
\end{align*}
where we have used the initial state being an eigenstate of $G$ and $\left[U,G\right]=0$, consequently $\langle Z_i(t) \rangle=0$.

Since $\langle Z_i(t) \rangle$ is a local observable, this should agree with the result if we were to replace the unitary evolution with a Gibbs distribution:
\begin{equation}
    \langle Z_i(t) \rangle_{\beta} \propto \Tr\left(Z_i e^{-\beta H}\right).
\end{equation}
As $\left[H,G\right]=0$ this is also zero. Thus the thermal-state recovers the correct expectation value for any local observable that flips sign under conjugation by $G$ (i.e., $GO_LG=-O_L$). 

That is not to say that the spin-flip symmetry has no role in calculating expectation values. Consider the case where the initial state
\begin{equation}
    \ket{\psi_i}=\cos\theta/2 \ket{\varphi_+}+e^{-i \phi} \sin\theta/2 \ket{\varphi_-},
\end{equation}
is a linear superposition of two states, where $G\ket{\varphi_+}=\ket{\varphi_+}$  and $G\ket{\varphi_-}=-\ket{\varphi_-}$.  Then 
\begin{multline}
    \bra{\psi_i}U^\dagger O_L U \ket{\psi_i}= \cos^2\theta/2 \bra{\varphi_+} U^\dagger O_L U \ket{\varphi_+}
    +\sin^2\theta/2 \bra{\varphi_-} U^\dagger O_L U \ket{\varphi_-}\\
    +e^{-i \phi} \cos \theta/2 \sin\theta/2 \bra{\varphi_+}U^\dagger O_L U\ket{\varphi_-}+c.c.
\end{multline}
If $O_L$ also commutes with $G$ then (as is the case with $H_p$) it follows that:
\begin{align} 
    \bra{\varphi_+}U^\dagger O_L U\ket{\varphi_-}=&-\bra{\varphi_+}GU^\dagger O_L UG\ket{\varphi_-}\\
    =&-\bra{\varphi_+}U^\dagger O_L U\ket{\varphi_-},
\end{align}
hence $\bra{\varphi_+}U^\dagger O_L U\ket{\varphi_-}=0$ and 
\begin{equation}
    \bra{\psi_i}U^\dagger O_L U \ket{\psi_i}= \cos^2\theta/2 \bra{\varphi_+} U^\dagger O_L U \ket{\varphi_+}
    +\sin^2\theta/2 \bra{\varphi_-} U^\dagger O_L U \ket{\varphi_-}.
\end{equation}

However, we have conjectured each (and numerically investigated) that each of the above matrix elements can be represented by replacing each unitary evolution with a thermal state. That is to say, 
\begin{equation}
    \bra{\psi_i}U^\dagger O_L U \ket{\psi_i}\approx \cos^2\theta/2 \frac{\Tr{O_L e^{-\beta_+ H}}}{\Tr{ e^{-\beta_+ H}}}\\
    +\sin^2\theta/2 \frac{\Tr{O_L e^{-\beta_- H}}}{\Tr{ e^{-\beta_- H}}},
\end{equation}
therefore the system would need to be assigned two temperatures for each symmetry sector.

Throughout this paper we have tacitly assumed we are in the paramagnetic phase, justified by the associated high temperatures. Work on spontaneous symmetry breaking and the ETH can be found in \cite{Fra15,Fra16}.

\section{Calculating the moments of the density of states}
\label{sec:mom}
To better understand the density of states (DOS) it is possible to calculate moments of the distribution. We can use these as fitting parameters to the model for the DOS. We are interested in calculating the moments of the distribution produced by the eigenenergies of the CTQW Hamiltonian for the graph $G=(V,E)$:
\begin{equation}
    H_{QW}=H_d+\gamma H_p,
\end{equation}
where:
\begin{align}
    H_d=&-\sum_{i=1}^n X_i\\
    H_p=&\sum_{(i,j)\in E} Z_iZ_j,
\end{align}
where $n=\abs{V}$ is the number of nodes in the graph. Note that the Hamiltonian is constructed of Pauli matrices, which are traceless \cite{nielsen_chuang_2010}.

Denoting the eigenenergies of $H_{QW}$ as $E_k$, then the mean of the distribution of eigenenergies is given by:
\begin{equation}
    \mu=\frac{1}{2^n}\sum_k E_k=\Tr H_{QW},
\end{equation}
and since the Hamiltonian is traceless, this evaluates to zero.

Repeating this approach for the variance gives:
\begin{align*}
   \sigma^2&=\frac{1}{2^n}\sum_k E_k^2\\
   &=\frac{1}{2^n}\Tr H_{QW}^2\\
   &=\frac{1}{2^n}\Tr \left(H_d^2+2\gamma H_d H_p +\gamma^2 H_p^2 \right)\\
   &=n+0+\gamma^2 \kappa_2,
\end{align*}
where $\kappa_2$ is the number of edges in the graph. Between the penultimate and final line, we have used that only terms that are equal to the identity will contribute to the trace. 

The same approach can be used to find the third moment:
\begin{align*}
   \frac{1}{2^n}\sum_k E_k^3 &=\frac{1}{2^n}\Tr H_{QW}^3\\
   &=\frac{1}{2^n}\Tr \left(\gamma^3 H_p^3\right)\\
   &=6 \gamma^3 \kappa_3,
\end{align*}
where $\kappa_3$ is the number of triangles in the graph. Further details can be seen in \cite{Gna22} or Appendix \ref{app:tor}.
The fourth moment also follows from the above logic:
\begin{align*}
   \frac{1}{2^n}\sum_k E_k^4 &=\frac{1}{2^n}\Tr H_{QW}^4\\
   &=\frac{1}{2^n}\Tr \left(H_d^4 +4\gamma^2 H_d^2 H_p^2 +2\gamma^2 H_d H_p H_d H_p +\gamma^4 H_p^4\right)\\
   &=\left(n^2+2(n^2-n)\right)+4 \gamma^2 n \kappa_2 +2 \gamma^2 (n-4)\kappa_2+\gamma^4 \left(\kappa_2+3\kappa_2(\kappa_2-1) +24 \kappa_4\right).
\end{align*}
From here it is straightforward to calculate the skewness and excess kurtosis. In theory, this process of moments could be continued to higher orders, incorporating loops of greater lengths. However, as shown above, even by the fourth order this becomes cumbersome.

\section{Estimating the temperature}
\subsection{Gaussian density of states}
\label{sec:norm}
In Sec.\ \ref{sec:therm_check} we numerically verified that CTQW were well approximated by a thermal state $\rho_{th}\left(\beta\right)$, with the temperature fixed by:
\begin{equation}
    \Tr{H_{QW} \rho_{th}\left(\beta\right)}=-n,
\end{equation}
in agreement with the ETH.  In this section we provide the details for estimating $\beta$ and associated quantities, under the following assumptions:
\begin{enumerate}
    \item \textbf{Continuity}: the energy levels can be modelled as a continuous variable.
    \item \textbf{Normality}: the density of states is well modelled by a normal distribution.
\end{enumerate}

Under these assumptions, we can take the density of states to be:
\begin{equation}
    g(E)\dd E = \frac{1}{\sqrt{2 \pi \sigma^2}}e^{-\frac{E^2}{2 \sigma^2}} \dd E,
\end{equation}
where $\sigma^2=\Tr\left(H_{QW}^2\right)/2^n=n+\gamma^2 \kappa_2$. The density of states has been normalised to one. Calculating the partition function gives:
\begin{align*}
    \mathcal{Z}&=\int_{-\infty}^\infty e^{-\beta E} g(E)\dd E\\
    &=\frac{1}{\sqrt{2 \pi \sigma^2}} \int_{-\infty}^\infty e^{-\frac{E^2}{2 \sigma^2}} e^{-\beta E} \dd E\\
    &=e^{\beta^2 \sigma^2/2}.
\end{align*}
Using $\langle E \rangle =-\partial \ln \mathcal{Z} /\partial \beta $, gives:
\begin{equation}
    \langle E \rangle =-\beta \sigma^2.
\end{equation}
Setting this equal to $-n$ gives
\begin{equation}
    \label{eq:bguessapp}
    \beta=\frac{n}{n+\gamma^2 \kappa_2}.
\end{equation}
This is our estimate for $\beta$, in the thermodynamic limit. From here the entropy follows trivially from Eq.\ \ref{eq:ent}. Calculating $\langle H_p \rangle$ from Eq.\ \ref{eq:hppred} gives:
\begin{equation}
    \langle H_p \rangle_{est}=-\beta \gamma \kappa_2.
\end{equation}
Substituting Eq.\ \ref{eq:bguessapp} into the above equation gives Eq.\ \ref{eq:hpnorm}. This can then be minimised to find the optimal choice of $\gamma$.

\subsection{Exponentially modified Gaussian density of states}
\label{sec:EMG}

The steps for treating the EMG DOS are exactly the same as for the Gaussian distribution. Therefore in this section we only demonstrate how to fit the EMG and calculate the associated partition function.

The EMG distribution is a convolution of a Gaussian distribution with an exponential distribution \cite{FEL94}.  Denoting the exponential distribution as:
\begin{equation}
    f(x)=
    \begin{cases}
        0 & \text{ for } x<0 \\
        \lambda e^{-\lambda x} & \text{ for } x\geq0 
    \end{cases}
\end{equation}
and the normal distribution as 
\begin{equation}
    g(x)=\frac{1}{\sqrt{2\pi\nu^2}}e^{-\frac{\left(x-m\right)^2}{2 \nu^2}},
\end{equation}
then the exponentially modified Gaussian distribution is given by:
\begin{align}
    h(x)&=f(x)*g(x) \nonumber\\
    &=\int_{-\infty}^{\infty} \dd y \, f(y) g(x-y)\nonumber \\
    &=\frac{\lambda}{2}e^{\frac{\lambda}{2}\left(2m+\lambda \nu^2-2x\right)}\erfc \left(\frac{m+\lambda \nu^2 -x}{\sqrt{2 \nu^2}} \right).
\end{align}
Through application of the convolution theorem \cite{RHB06} it is straightforward to write down the  moment generating function (and hence partition function) from the moment generating functions of the exponential and Gaussian distributions. The resulting partition function in terms of the fitting parameters $m$, $\nu^2$ and $\lambda$ is:
\begin{equation}
    \mathcal{Z}\left(\beta\right)=\left(1+\frac{\beta}{\lambda}\right)^{-1}e^{-m \beta +\frac{1}{2}\nu^2 \beta^2}
\end{equation}
To find the fitting parameters we fit the moments of $h(x)$ to the moments associated with $H_{QW}$. Utilising the moment generating function to find the mean ($\mu$), variance ($\sigma^2$) and skew ($s$) of $h(x)$ gives:
\begin{align}
    \mu&=m+\frac{1}{\lambda}\\
    \sigma^2&=\nu^2+\frac{1}{\lambda^2}\\
    s&=\frac{2}{\nu^3 \lambda^3} \left(1+\frac{1}{\nu^2 \lambda^2}\right)^{-3/2}.
\end{align}
The above equations can be inverted to find the fitting parameters in terms of the mean, variance and skew associated with $H_{QW}$. These properties can be extracted from $H_{QW}$ by calculating $\Tr{H_{QW}^k}$ for $k=1,2,3$. The method can be found in Appendix \ref{sec:mom}. From here the same steps as Appendix \ref{sec:norm} can be followed to derive predictions for the temperature, entropy, $\langle H_p \rangle$ and optimal choices of $\gamma$

\newpage
\section{Derivation of Eq.~\ref{eq:hppred}.}
\label{app:hp_reason}

Using Eq.~\ref{eq:hppred} removes the need for matrix exponentiation, hence here we discuss in greater detail the assumptions behind this equation.

We assume that the system is well modelled by a Gibbs state and consider the function
\begin{equation}
    \Omega=e^{\beta H_{QW}} \partial_\gamma e^{-\beta H_{QW}},
\end{equation}
which we differentiate with respect to $\beta$, to get:
\begin{align}
    \partial_\beta \Omega&=e^{\beta H_{QW}}H_{QW} \partial_\gamma e^{-\beta H_{QW}}-e^{\beta H_{QW}} \partial_\gamma \left(H_{QW} e^{-\beta H_{QW}}\right) \nonumber\\
    &=-e^{\beta H_{QW}} \partial_\gamma\left(H_{QW}\right) e^{-\beta H_{QW}}\nonumber\\
    &=-e^{\beta H_{QW}} H_p e^{-\beta H_{QW}}.
\end{align}
Now through successive differentiation or by simply applying the well-known result for writing the right-hand-side in terms of nested commutators \cite{hall2000} we conclude:
\begin{equation}
    \Omega=-\beta H_p -\frac{\beta^2}{2}\left[H_{QW},H_p\right]-\frac{\beta^3}{3!}\left[H_{QW},\left[H_{QW},H_p\right]\right]+\dots
\end{equation}
 where $[\cdot, \cdot]$ denotes the commutator. Tidying this expression up using the notation:
 \begin{equation*}
     [H_{QW}^{(k)}, H_p]=
     \begin{cases}
         H_p & \text{if } k=0\\
         \underbrace{[H_{QW},\cdots[H_{QW},[H_{QW}}_{\text{k times}}, H_p]]\cdots] & \text{otherwise}.
     \end{cases}
 \end{equation*}
     
gives:
\begin{equation}
    \Omega=-\sum_{k=1}^\infty\frac{\beta^k}{k!}\left[H_{QW}^{(k-1)},H_p\right].
\end{equation}
Acting on both sides with $e^{-\beta H_{QW}}$ and taking the trace gives:

\begin{equation}
    \label{eq:ass_29}
    \Tr{\partial_\gamma e^{-\beta H_{QW}}}=
    -\sum_{k=1}^\infty\frac{\beta^k}{k!} \Tr{\left[H_{QW}^{(k-1)},H_p\right]e^{-\beta H_{QW}}}.
\end{equation}
\\
Evaluating the terms in the sum for which $k>1$:
\begin{align*}
    &\Tr{\left[H_{QW}^{(k-1)},H_p\right]e^{-\beta H_{QW}}}\\
    &=\sum_j \bra{E_j}\left[H_{QW}^{(k-1)},H_p\right]e^{-\beta H_{QW}}\ket{E_j}\\
    &=\sum_j \bra{E_j}\left[ H_{QW},\left[H_{QW}^{(k-2)},H_p\right] \right] \ket{E_j}e^{-\beta E_j}\\
    &=\sum_j E_j \bra{E_j}   \left(\left[H_{QW}^{(k-2)},H_p\right]- \left[H_{QW}^{(k-2)},H_p\right]\right) \ket{E_j}e^{-\beta E_j}\\
    &=0,
\end{align*}
where $\ket{E_j}$ denotes the eigenvectors of $H_{QW}$. Therefore, once we assume the state is well modelled by a Gibbs distribution it follows that:
\begin{equation}
    \Tr{\partial_\gamma e^{-\beta H_{QW}}}=-\beta \Tr{H_p e^{-\beta H_{QW}}}.
\end{equation}
Commuting through the trace with the partial derivative and dividing both sides by the partition function gives Eq.~\ref{eq:hppred}.

\section{The performance of CTQWs for MAX-CUT}
\label{app:per}

\begin{figure}
     \centering
     \begin{subfigure}[b]{0.48\textwidth}
         \centering
         \includegraphics[width=\textwidth]{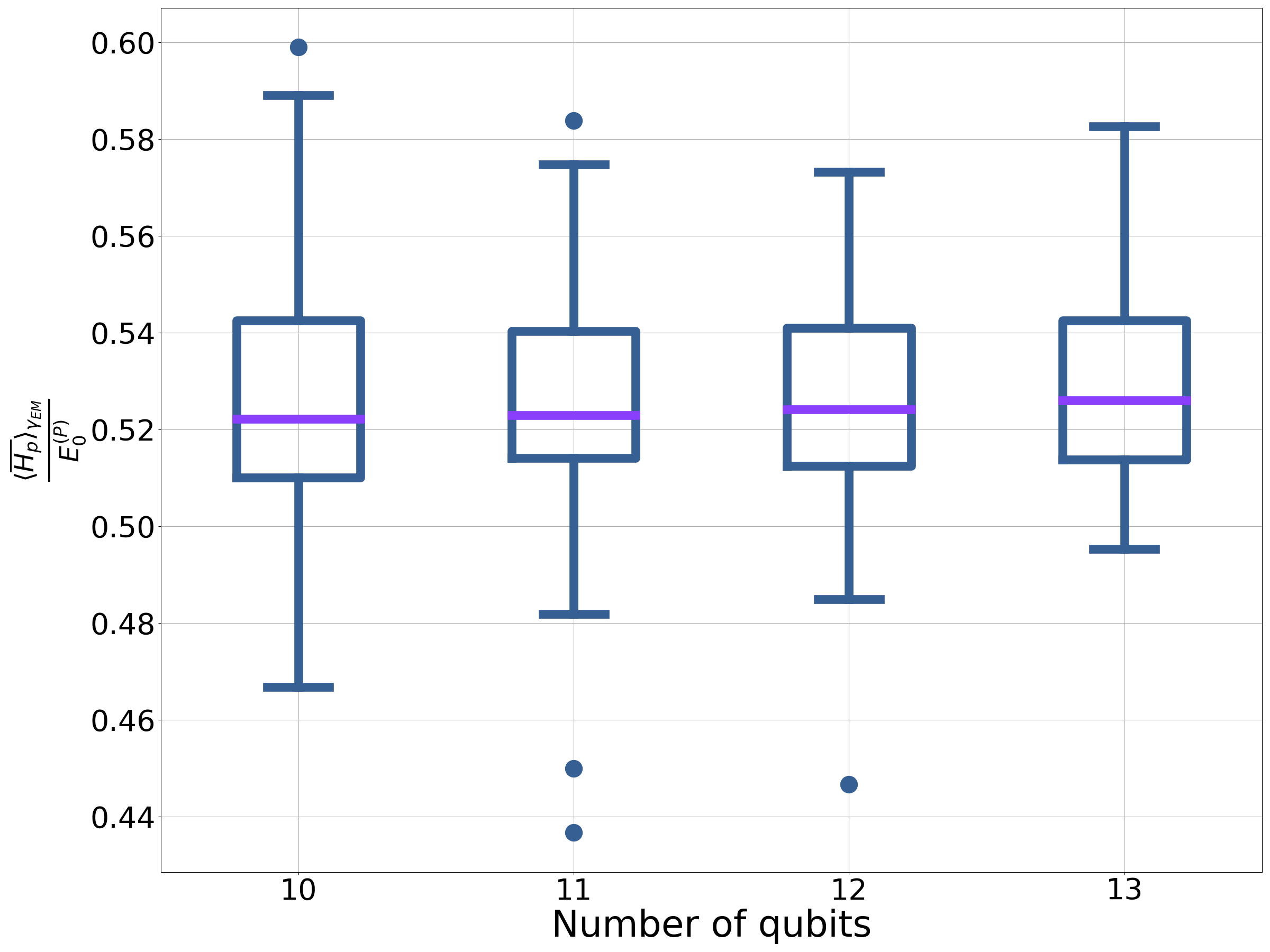}
         \caption{Binomial graphs}
         \label{fig:Rand_per}
     \end{subfigure}
     \hfill
     \begin{subfigure}[b]{0.48\textwidth}
         \centering
         \includegraphics[width=\textwidth]{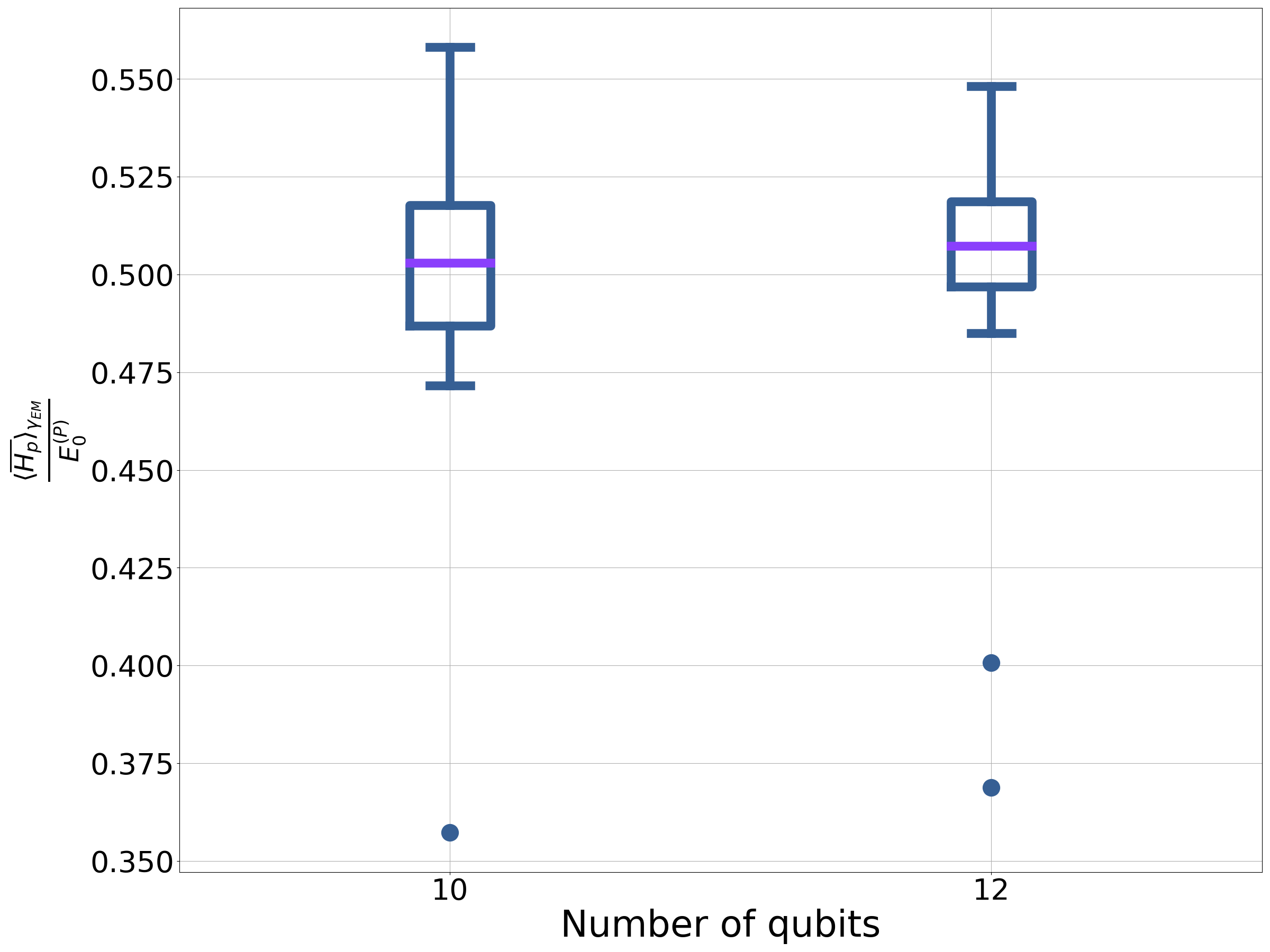}
         \caption{Three-regular graphs}
         \label{fig:Reg_per}
     \end{subfigure}
        \caption{The performance (i.e., $\langle \bar{H_p} \rangle$) of the CTQW for the optimal $\gamma$ predicted by the EMG DOS compared to the ground state energy of the problem Hamiltonian}
        \label{fig:per_app}
\end{figure}

For the problem instances considered in Sec.~\ref{sec:QWTL} we consider how the value of $\langle \bar{H_p} \rangle$ evaluated at the $\gamma$ predicted by the EMG DOS compares to the ground state energy of $\langle H_p \rangle$, (i.e.\ $E_0^{(P)}$). The results can be seen in Fig.~\ref{fig:per_app}. For the small problem sizes considered the performance seems largely independent of system size. This perhaps is unsurprising for a high temperature Gibbs state with a local Hamiltonian \cite{Wolf08,Kliesch_2018}. We might expect different parts of the system to be weakly correlated, hence the CTQW is optimising locally and does not scale with problem size.

In practice one is perhaps less interested in how the average of the energy distribution of $\langle H_p\rangle$ compares with the absolute minimum, particularly if considering CTQWs as an exact solver. In such a case one might be more interested in the ground-state probability or time-to-solution. This requires some notion of run-time not discussed in this paper and will be the subject of future investigations.

\section{A brief introduction to Floquet systems}
\label{app:flo}
In this appendix we provide a very brief introduction to Floquet systems, with the aim of providing sufficient background to understand Sec.\ \ref{sec:QWF}. This introduction draws from \cite{GRI98}.

In this paper we have used Floquet system to refer to a closed quantum system with a time periodic Hamiltonian. A closed system will evolve under the Schr\"odinger equation:
 \begin{equation}
     i \frac{\dd}{\dd t} \ket{\psi(t)}=H(t)\ket{\psi(t)}.
 \end{equation}
We are interested in Hamiltonians that are time-periodic, such that:
 \begin{equation}
     H(t)=H(t+\tau),
 \end{equation}
where $\tau$ is the period of the Hamiltonian. The Floquet theorem then tells us that there exist solutions to the Schr\"odinger equation of the form 
\begin{equation}
    \label{eq:floq_mode}
    \ket{\psi(t)}=e^{-i\epsilon_{\alpha}t}\ket{\phi_{\alpha}(t)},
\end{equation}
where $\epsilon_{\alpha}$ are termed quasi-energies and $\ket{\phi_{\alpha}(t)}$ the Floquet modes. The Floquet modes are time periodic, $\ket{\phi_{\alpha}(t)}=\ket{\phi_{\alpha}(t+\tau)}$. The quasi-energies are time-independent but are only uniquely defined up to multiples of $2\pi/\tau$.  This is reminiscent of Bloch's theorem \cite{simon2013oxford} for spatially periodic crystals. 

Noting that
\begin{equation}
    U(t+\tau,t)\ket{\psi(t)}=\ket{\psi(t+\tau)}\\
\end{equation}
where $U(t+\tau,t)$ is the unitary associated with a single period. Substituting Eq.\ \ref{eq:floq_mode} into the above equation gives:
\begin{align}
    U(t+\tau,t)e^{-i\epsilon_{\alpha}t}\ket{\phi_{\alpha}(t)}=e^{-i\epsilon_{\alpha}(t+\tau)}\ket{\phi_{\alpha}(t+\tau)}\\
    =e^{-i\epsilon_{\alpha}\tau}e^{-i\epsilon_{\alpha}t}\ket{\phi_{\alpha}(t)}.
\end{align}
Hence $\ket{\phi_{\alpha}(t)}$ is an eigenvector of a single period unitary with eigenvalue $e^{-i\epsilon_{\alpha}\tau}$. Therefore, by studying the single period unitary we can construct the wave function at all times, including the long-time average in Eq. \ref{eq:floqss}.
\end{document}